\documentclass[twocolumn]{aastex63}
\usepackage{natbib}
\usepackage{lineno}

\newcommand{\logM}{log($M_\star/M_\odot$) \,}

\newcommand{\recomma}{$\mathrm{r_e}$}

\newcommand{\sersic}{S\'{e}rsic \,}

\usepackage{amsmath} 

\usepackage[encapsulated]{CJK}

\usepackage{rotating}

\usepackage{float}
\usepackage{textcomp}
\usepackage{makecell}

\begin{document}

\date{\today}


\shorttitle{Early Dense Core Formation at $z_{spec}=3.97$}

\shortauthors{Setton et al.}

\title{UNCOVER NIRSpec/PRISM Spectroscopy Unveils Evidence of Early Core Formation in a Massive, Centrally Dusty Quiescent Galaxy at $z_{spec}=3.97$}

\author[0000-0003-4075-7393]{David J. Setton}\thanks{Email: davidsetton@princeton.edu}\thanks{Brinson Prize Fellow}
\affiliation{Department of Astrophysical Sciences, Princeton University, 4 Ivy Lane, Princeton, NJ 08544, USA}

\author[0000-0002-3475-7648]{Gourav Khullar}
\affiliation{Department of Physics and Astronomy and PITT PACC, University of Pittsburgh, Pittsburgh, PA 15260, USA}

\author[0000-0001-8367-6265]{Tim B. Miller}
\affiliation{Center for Interdisciplinary Exploration and Research in Astrophysics (CIERA), Northwestern University, 1800 Sherman Ave, Evanston IL 60201, USA}

\author[0000-0001-5063-8254]{Rachel Bezanson}
\affiliation{Department of Physics and Astronomy and PITT PACC, University of Pittsburgh, Pittsburgh, PA 15260, USA}

\author[0000-0002-5612-3427]{Jenny E. Greene}
\affiliation{Department of Astrophysical Sciences, Princeton University, 4 Ivy Lane, Princeton, NJ 08544, USA}

\author[0000-0002-1714-1905]{Katherine A. Suess}
\altaffiliation{NHFP Hubble Fellow}
\affiliation{Kavli Institute for Particle Astrophysics and Cosmology and Department of Physics, Stanford University, Stanford, CA 94305, USA}

\author[0000-0001-7160-3632]{Katherine E. Whitaker}
\affiliation{Department of Astronomy, University of Massachusetts, Amherst, MA 01003, USA}
\affiliation{Cosmic Dawn Center (DAWN), Denmark}

\author[0000-0002-0243-6575]{Jacqueline Antwi-Danso}
\altaffiliation{Banting Postdoctoral Fellow}
\affiliation{David A. Dunlap Department of Astronomy \& Astrophysics, University of Toronto, 50 St George Street, Toronto, ON M5S 3H4, Canada}
\affiliation{Dunlap Institute for Astronomy \& Astrophysics, University of Toronto, 50 St George Street, Toronto, ON M5S 3H4, Canada}
\affiliation{Department of Astronomy, University of Massachusetts, Amherst, MA 01003, USA}

\author[0000-0002-7570-0824]{Hakim Atek}
\affiliation{Institut d'Astrophysique de Paris, CNRS, Sorbonne Universit\'e, 98bis Boulevard Arago, 75014, Paris, France}

\author[0000-0003-2680-005X]{Gabriel Brammer}
\affiliation{Cosmic Dawn Center (DAWN), Niels Bohr Institute, University of Copenhagen, Jagtvej 128, K{\o}benhavn N, DK-2200, Denmark}

\author[0000-0002-7031-2865]{Sam E. Cutler}
\affiliation{Department of Astronomy, University of Massachusetts, Amherst, MA 01003, USA}

\author[0000-0001-8460-1564]{Pratika Dayal}
\affiliation{Kapteyn Astronomical Institute, University of Groningen, 9700 AV Groningen, The Netherlands}

\author[0000-0002-1109-1919]{Robert Feldmann} \affiliation{Department of Astrophysics, University of Zurich, Winterthurerstrasse 190, 8057 Zurich, Switzerland}

\author[0000-0001-7201-5066]{Seiji Fujimoto}\altaffiliation{NHFP Hubble Fellow}
\affiliation{Department of Astronomy, The University of Texas at Austin, Austin, TX 78712, USA}

\author[0000-0001-6278-032X]{Lukas J. Furtak}\affiliation{Physics Department, Ben-Gurion University of the Negev, P.O. Box 653, Be’er-Sheva 84105, Israel}

\author[0000-0002-3254-9044]{Karl Glazebrook}
\affiliation{Centre for Astrophysics and Supercomputing, Swinburne University of Technology, PO Box 218, Hawthorn, VIC 3122, Australia}

\author[0000-0003-4700-663X]{Andy D. Goulding}
\affiliation{Department of Astrophysical Sciences, Princeton University, 4 Ivy Lane, Princeton, NJ 08544, USA}

\author[0000-0002-5588-9156]{Vasily Kokorev}
\affiliation{Kapteyn Astronomical Institute, University of Groningen, 9700 AV Groningen, The Netherlands}

\author[0000-0002-2057-5376]{Ivo Labbe}
\affiliation{Centre for Astrophysics and Supercomputing, Swinburne University of Technology, Melbourne, VIC 3122, Australia}

\author[0000-0001-6755-1315]{Joel Leja}
\affiliation{Department of Astronomy \& Astrophysics, The Pennsylvania State University, University Park, PA 16802, USA}
\affiliation{Institute for Computational \& Data Sciences, The Pennsylvania State University, University Park, PA 16802, USA}
\affiliation{Institute for Gravitation and the Cosmos, The Pennsylvania State University, University Park, PA 16802, USA}

\author[0000-0002-0463-9528]{Yilun Ma (\begin{CJK*}{UTF8}{gbsn}马逸伦\ignorespacesafterend\end{CJK*})}
\affiliation{Department of Astrophysical Sciences, Princeton University, 4 Ivy Lane, Princeton, NJ 08544, USA}

\author[0000-0001-9002-3502]{Danilo Marchesini}
\affiliation{Department of Physics \& Astronomy, Tufts University, MA 02155, USA}

\author[0000-0003-2804-0648 ]{Themiya Nanayakkara}
\affiliation{Centre for Astrophysics and Supercomputing, Swinburne University of Technology, PO Box 218, Hawthorn, VIC 3122, Australia}

\author[0000-0002-9651-5716]{Richard Pan}
\affiliation{Department of Physics \& Astronomy, Tufts University, MA 02155, USA}

\author[0000-0002-0108-4176]{Sedona H. Price}
\affiliation{Department of Physics and Astronomy and PITT PACC, University of Pittsburgh, Pittsburgh, PA 15260, USA}

\author[0000-0002-9337-0902]{Jared C. Siegel}
\affiliation{Department of Astrophysical Sciences, Princeton University, 4 Ivy Lane, Princeton, NJ 08544, USA}
\affiliation{NSF Graduate Research Fellow}

\author[0009-0007-1787-2306]{Heath Shipley}
\affiliation{Department of Physics, Texas State University, San Marcos, TX 78666, USA}

\author[0000-0003-1614-196X]{John R. Weaver}
\affiliation{Department of Astronomy, University of Massachusetts, Amherst, MA 01003, USA}

\author[0000-0002-8282-9888]{Pieter van Dokkum}
\affiliation{Department of Astronomy, Yale University, New Haven, CT 06511, USA}

\author[0000-0001-9269-5046]{Bingjie Wang (\begin{CJK*}{UTF8}{gbsn}王冰洁\ignorespacesafterend\end{CJK*})}
\affiliation{Department of Astronomy \& Astrophysics, The Pennsylvania State University, University Park, PA 16802, USA}
\affiliation{Institute for Computational \& Data Sciences, The Pennsylvania State University, University Park, PA 16802, USA}
\affiliation{Institute for Gravitation and the Cosmos, The Pennsylvania State University, University Park, PA 16802, USA}

\author[0000-0003-2919-7495]{Christina C.\ Williams}
\affiliation{NSF’s National Optical-Infrared Astronomy Research Laboratory, 950 North Cherry Avenue, Tucson, AZ 85719, USA}

\submitjournal{ApJ}

\begin{abstract}

We report the spectroscopic confirmation of a massive ($\log(M_\star/M_\odot)=10.34 \pm_{0.07}^{0.06}$), HST-dark ($m_\mathrm{F150W} - m_\mathrm{F444W} = 3.6$) quiescent galaxy at $z_{spec}=3.97$ in the UNCOVER survey. NIRSpec/PRISM spectroscopy and a non-detection in deep ALMA imaging surprisingly reveals that the galaxy is consistent with a low ($<$10 $M_\odot \ \mathrm{yr^{-1}}$) star formation rate despite evidence for moderate dust attenuation. The F444W image is well modeled with a two component \sersic fit that favors a compact, $r_e\sim200$ pc, $n\sim2.9$ component and a more extended, $r_e\sim1.6$ kpc, $n\sim1.7$ component. The galaxy exhibits strong color gradients: the inner regions are significantly redder than the outskirts. Spectral energy distribution models that reproduce both the red colors and low star formation rate in the center of UNCOVER 18407 require both significant ($A_v\sim1.4$ mag) dust attenuation and a stellar mass-weighted age of 900 Myr, implying 50\% of the stars in the core already formed by $z=7.5$. Using spatially resolved annular mass-to-light measurements enabled by the galaxy's moderate magnification ($\mu=2.12\pm_{0.01}^{0.05}$) to reconstruct a radial mass profile from the best-fitting two-component \sersic model, we infer a total mass-weighted $r_\mathrm{eff} = 0.72 \pm_{0.11}^{0.15}$ kpc and log$(\Sigma_\mathrm{1 kpc} \ [\mathrm{M_\odot/kpc^2}]) = 9.61 \pm_{0.10}^{0.08}$. The early formation of a dense, low star formation rate, and dusty core embedded in a less attenuated stellar envelope suggests an evolutionary link between the earliest-forming massive galaxies and their elliptical descendants. Furthermore, the disparity between the global, integrated dust properties and the spatially resolved gradients highlights the importance of accounting for radially varying stellar populations when characterizing the early growth of galaxy structure.

\end{abstract}



\section{Introduction} \label{sec:intro}

The centers of the most massive quiescent galaxies in the local universe show evidence of old ($>10$ Gyr) stellar populations \citep{Trager2000, Thomas2005,Sanchez-Blazquez2006b, Graves2007, Greene2013, McDermid2015} and compact structures \citep[e.g.,][]{Shen2003, Trujillo2004,Cappellari2013}. This finding indicates that their star forming progenitors formed the bulk of their massive, dense cores in the first few Gyr of cosmic time before rapidly quenching their star formation. To link this population of red and dead galaxies to their high-z counterparts, numerous studies have attempted to connect compact star forming, sub-millimeter galaxies, post-starburst, and quiescent populations near cosmic noon to their lower-z descendants \citep[e.g.,][]{Swinbank2006,VanDokkum2015, Ma2015, Barro2017, Suess2019a, Suess2021} and to quantify the rate of rapid quenching across cosmic time \citep[e.g.,][]{Wild2016, Pattarakijwanich2016, Rowlands2018a, Belli2019, Park2022, Setton2023,Gould2023,Park2024}. The emerging picture is that the most massive galaxies in the universe formed a significant portion of their stellar mass in the very early universe ($z\gtrsim3$) in dense, compact regions that evolve into the cores of present-day ellipticals \citep[e.g.,][]{Bezanson2009}, but precisely how early this formation occurs remains an open question. As such, there has been a concerted effort to find and characterize quiescent galaxies at high-redshift to understand when they form and how they quench their star formation.

Prior to the launch of the James Webb Space Telescope (JWST), the majority of these efforts were focused on using ground-based medium-band NIR photometry to select for strong breaks at $z>3$ \citep[e.g.,][]{Whitaker2011,Straatman2014, Straatman2016,Zaidi2024}. These surveys discovered systems with a wide range of SED shapes, hinting at variety in the age and dust content of quiescent systems at this crucial epoch. However, essentially all successful spectroscopic follow-up programs of $z>3$ quiescent galaxies have observed young, blue ``post-starburst" spectra \citep[e.g.,][]{Marsan2015,Glazebrook2017,Schreiber2018a,Schreiber2018b,Tanaka2019, Forrest2020b, Saracco2020, Valentino2020, DeugenioC2020,Kalita2021,Man2021,Tanaka2023,Kakimoto2024}. 

In the absence of selection effects, the discovery of exclusively recently-quenched galaxies at $\gtrsim3$ would indicate that the first quiescent galaxies formed the majority of their stellar mass between $z$=3-5 before rapidly quenching. However, continuum observations of post-starburst systems, which are brighter at fixed stellar mass due to their younger ages, at these redshifts were straining even the largest ground-based observing facilities. It has therefore been suggested that some of the apparent lack of old, quiescent systems at $z\gtrsim3$ may be biased by the inability of ground-based facilities to robustly discover older stellar populations with higher mass-to-light ratios \citep{Forrest2020b,Antwi-Danso2023b}. As such, a hidden red-and-dead population that formed in the first $\sim$1 Gyr of cosmic time \citep[as suggested by star formation history and chemical measurements of lower-z systems that appeared maximally old, e.g.,][]{Thomas2005, Kriek2016} could still be found in deeper observations.

Early JWST data has quickly confirmed this suggestion, with photometric \citep[e.g.,][]{Carnall2023a, Valentino2023,Alberts2023} and spectroscopic \citep[e.g.,][]{Carnall2023b,Nanayakkara2024,Glazebrook2024,deGraaff2024, Carnall2024} observations at $z>3$ finding massive, quiescent galaxies with old ages that could only have resulted from extremely early-onset star formation, with star formation history fitting of deep spectroscopic samples at $z\sim$2 providing archeological evidence of similar populations \citep{Slob2024, Park2024}. Characterizing such early-forming systems has become especially important given that high-redshift studies are turning up candidate examples of rapid mass growth in the $z>7$ universe \citep[][]{Labbe2023, Boyett2023, Tacchella2023}--the star formation histories and structures of the earliest-forming quiescent galaxies provide an important constraint on how quickly galaxies can have assembled. As such, the detailed study of massive, old quiescent galaxies, in addition to younger co-eval post-starburst galaxies \citep[e.g.,][]{Belli2023,Deugenio2023}, in the first $\sim2$ Gyr of cosmic time, promises to illuminate the physical drivers of rapid mass assembly and quenching in the early universe.

In this work, we present a unique example of JWST-identified early forming galaxy--UNCOVER 18407\footnote{We note that 18407 is an ID based on an internal photometric catalog that was not part of the public UNCOVER data releases, but which was used for MSA designs that will be presented in Price et al. in preparation. Throughout this work we will refer to this source by its MSA ID for consistency with that work. However, its public UNCOVER DR3 ID is 29930. For ease, we also note the right ascension and declination (J2000): (3.56275, -30.39095)}--a moderately lensed ($\mu~\sim2$) massive (\logM$\sim10.4$) quiescent galaxy at $z_{spec}=3.97$ that was observed in seven band NIRCAM imaging and with deep PRISM spectroscopy as part of the UNCOVER Cycle\,1 Treasury Program \citep{Bezanson2022b}. In contrast with the majority of spectroscopically confirmed quiescent sources at high-redshift, UNCOVER 18407 has an extremely red ($m_\mathrm{F150W} - m_\mathrm{F444W}\sim3.6$) spectral energy distribution, suggesting either an old stellar population, significant dust attenuation, or some combination of the two. We leverage the high-spatial resolution of JWST --- complete with the added bonus from the lensing boost of Abell 2744 \citep{Furtak2023a} --- to apply state-of-the-art methods \citep[e.g.,][]{Suess2019a, Akhshik2020, Jafariyazani2020, Miller2023,Akhshik2023} to perform the first spatially resolved analysis of an HST-dark quiescent galaxy.

This paper is laid out as follows. In Section \ref{sec:data}, we briefly describe the UNCOVER Treasury Program \citep[PIs: Labbé and Bezanson, JWST-GO-2561][]{Bezanson2022b} and the imaging and spectroscopy we will analyze to constrain the stellar populations and structure of UNCOVER 18407. In Section \ref{sec:analysis}, we describe the inference of the imaging, photometric, and spectroscopic data to infer the star formation history and structural parameters of this galaxy. In Section \ref{sec:results}, we measure the spatially resolved and global properties of UNCOVER 18407. In Section \ref{sec:discussion}, we discuss the implications of our discovery of a massive, dusty, compact quiescent system at $z_{spec}=3.97$. Finally, in Section \ref{sec:conclusions}, we summarize our results and propose future avenues to build statistical samples of dusty, old quiescent galaxies at high-z to understand their cosmological significance.

Throughout this work, we adopt the best-fit cosmological parameters from the WMAP 9 year results \citep{Hinshaw2013}: $H_0 = 69.32 \ \mathrm{km \ s^{-1} \ Mpc^{-1}}$, $\Omega_m = 0.2865$, and $\Omega_\Lambda = 0.7135$, utilize a Chabrier initial mass function \citep{Chabrier2003}, and quote AB magnitudes. All effective radii ($r_e$) values measured denote the semi-major axis of the best-fitting 2D \sersic ellipse, and are corrected for the effect of lensing. Unless otherwise mentioned, all star formation rates and stellar masses are also corrected for the effect of lensing.

\section{Data} \label{sec:data}

\begin{figure*}
    \centering
    \includegraphics[width=\textwidth]{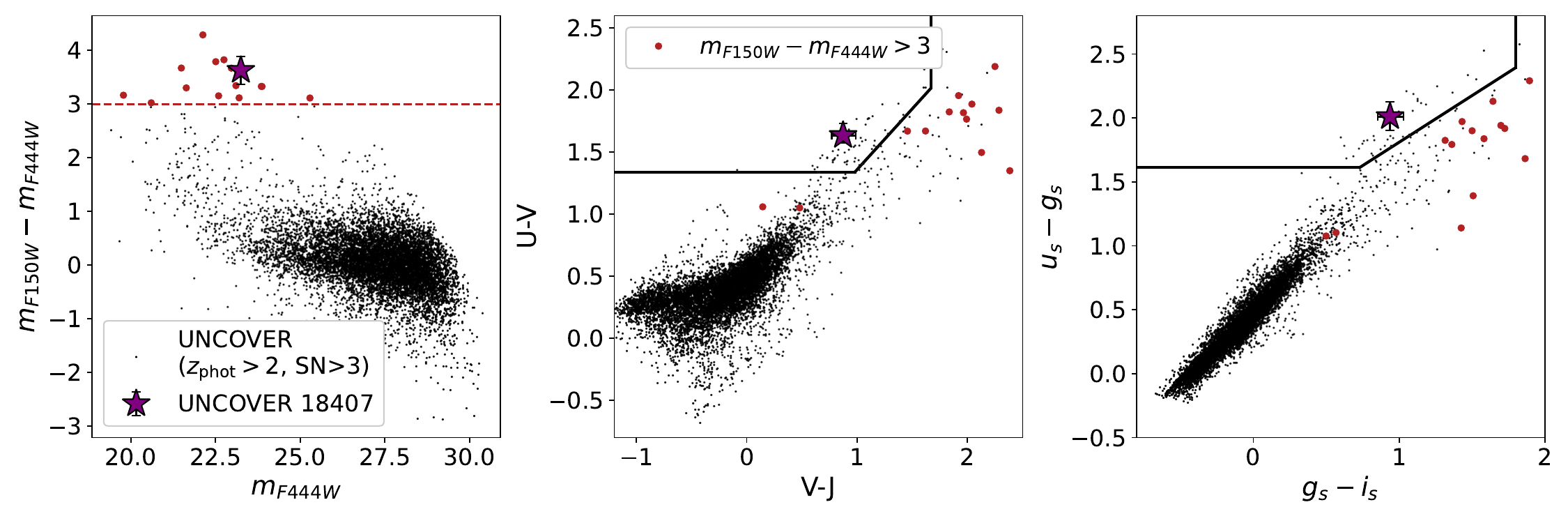}
    \caption{Observed $m_{F444w}$ versus $m_{F150W} - m_{F444W}$ (left, \citealt{Weaver2024}), rest frame $U-V$ versus $V-J$ colors (center), and rest frame $u_s - g_s$ and $g_s - i_s$ colors \citep[right, using the marginalized colors from][]{WangB2024_uncoverpops} for the UNCOVER sample ($z_\mathrm{phot}>2$ and signal-to-noise in F150W and F444W $>3$, black). Points with $m_{F150W} - m_{F444W} > 3$ are indicated as red, and UNCOVER 18407 is indicated as a purple star. The quiescent $UVJ$ box and $ugi_s$ boxes are shown as in \cite{Williams2009} and \cite{Antwi-Danso2023a} respectively. UNCOVER 18407 was targeted for its unique combination of an extremely red color and quiescent photometric solution. \label{fig:selection}}
\end{figure*}

\begin{figure*}
    \centering
    \includegraphics[width=\textwidth]{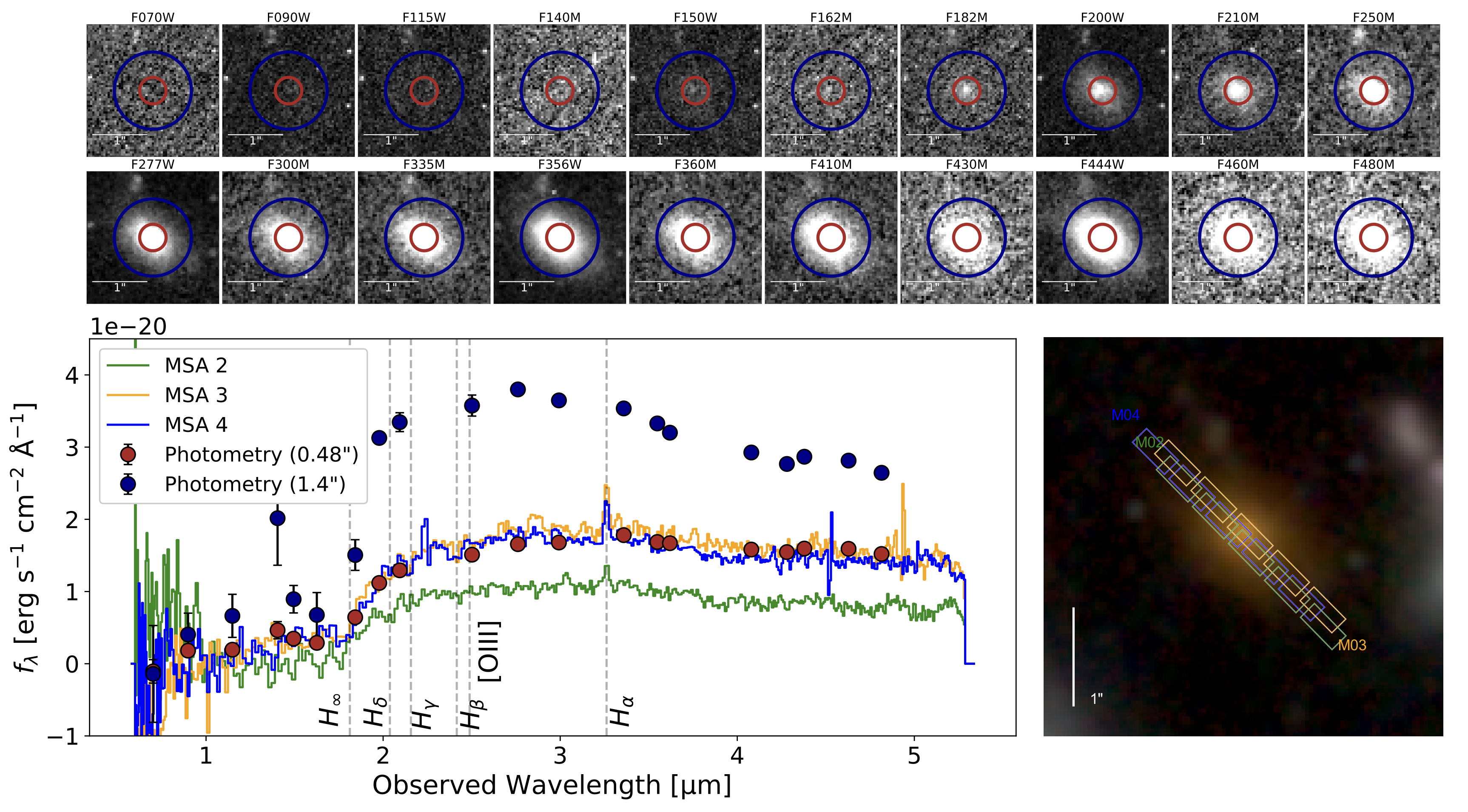}
    \caption{\textbf{Top:} NIRCAM UNCOVER \citep{Bezanson2022b} and MegaScience \citep{Suess2024} imaging in increasing order of wavelength for our target, with neighboring sources masked following \citep{Weaver2024}. The 0.48" (red) and 1.4" (dark blue) diameter apertures are also indicated as circles. \textbf{Bottom:} (Left): The extracted spectra from the three visits to UNCOVER 18407 (pre-photometric flux calibration, MSA2: green, MSA3: orange, MSA4: blue), along with the central 0.48" (red) and global 1.4" (dark blue) diameter photometry. The SED shape in the central region and in the observed spectra is considerably flatter redward of the break than in the global photometry. Key Balmer series features are labeled with dashed lines. Weak $H_\alpha$ emission is detected in all three exposures (including MSA2, which misses the direct center of the galaxy), and the Balmer series is otherwise seen in absorption only. (Right): An RGB image of the source with the microshutter arrays labeled. Because the source is extended, we utilize a global background subtraction from empty slits. \label{fig:spectrum}}
\end{figure*}

\subsection{Imaging and Photometry}

All JWST/NIRCam images used in this work were obtained as a part of the UNCOVER Treasury Program \citep[PIs: Labbé and
Bezanson, JWST-GO-2561][]{Bezanson2022b}, which took seven-band (NIRCam F115W, F150W, F200W, F277W, F356W, F410W, and F444W with AB point source depths 29-30) imaging of the Abell 2744 cluster, and Medium Bands, Mega Science \citep[PI: Suess, JWST-GO-4111][]{Suess2024}, which obtained NIRCam F070W and F090W broadband imaging in addition to all eleven medium bands (F140M, 162M, 182M, F210M, F250M, F300M, 335M, 360M, 430M, 460M, and F480M). The full reductions of the imaging are described in \cite{Bezanson2022b} and \cite{Suess2024}, and we utilize the v7.2 imaging mosaics hosted on the DAWN JWST Archive \citep{Valentino2023}\footnote{https://dawn-cph.github.io/dja/imaging/v7/}. 

The construction of photometric catalogs based on a combination of the full JWST/UNCOVER mosaics and archival HST imaging is described in \cite{Weaver2024}. In this work, we utilize the DR3 catalog that consists of photometry extracted using circular apertures on PSF-matched images with apertures selected to be closest to the total size of the galaxy in the segmentation map. We also adopt rest frame colors measured using photometric SED fits as detailed in \cite{WangB2024_uncoverpops}, which fits all available photometry in the catalog using the {\texttt{Prospector-$\beta$}} model \citep{Wang2023}. To account for the lensing of the foreground Abell 2744 cluster, we correct for magnifications calculated with the internal lensing model with updated spectroscopic redshifts following the methods of \citet{Furtak2023a}.

\subsection{Spectroscopic Targeting of UNCOVER 18407}

In addition to seven-band NIRCAM imaging, the JWST UNCOVER Treasury Program includes NIRSpec/PRISM spectroscopy ($0.6 \mu \mathrm{m}<\lambda < 5.3 \mu \mathrm{m}$) of $\sim$700 sources selected for a variety of science cases. Two stated goals of the UNCOVER program are the study of obscured/old ``HST-dark" sources \citep[e.g.,][]{Nelson2022, Barrufet2023,PerezGonzalez2023,Barrufet2024} and the study of the first quiescent galaxies at high-redshift; UNCOVER 18407 is a unique source within the survey that was targeted for neatly meeting both the above criteria.

In Figure \ref{fig:selection}, we show the full UNCOVER dataset with $z_\mathrm{phot}>2$ and signal-to-noise in F150W and F444W $>3$ (black points) in observed $m_\mathrm{F150W} - m_\mathrm{F444W}$ versus $m_\mathrm{F444W}$ color magnitude space (left), in addition to rest-frame $UVJ$ (center) and $(ugi)_s$ (right) plots with commonly applied quiescent selection criteria from \cite{Williams2009} and \cite{Antwi-Danso2023a}, respectively. We additionally highlight the 15 galaxies with $m_\mathrm{F150W} - m_\mathrm{F444W}$ (red), effectively selecting the reddest bright sources in the sample that would have been undetectable with the Hubble Space Telescope. The vast majority of these galaxies are best fit in \cite{WangB2024_uncoverpops} as dusty star forming galaxies at $z\gtrsim2.5$, as suggested by their positions in the upper right in the rest frame UVJ and ugi plots. The dusty star-forming nature of these sources is supported by detections in ALMA 1.2 mm dust continuum observations in several cases \citep{Fujimoto2023a, MunozArancibia2023}, with morphological analyses presented in \cite{Price2023}.

However, one galaxy, UNCOVER 18407 (shown as a star), is unique in that it lies in the quiescent regions of $UVJ$ and $(ugi)_s$ space. In the \citet{WangB2024_uncoverpops} catalog, it is best fit with $z_\mathrm{phot}=4.06\pm_{0.15}^{0.13}$, log($M_\star/M_\odot$)$=10.29\pm_{0.05}^{0.06}$, log(sSFR [$\mathrm{\mathrm{yr^{-1}}}$])$=-11.03\pm_{1.12}^{0.64}$, and a mass weighted age of $0.92\pm_{0.12}^{0.08}$ Gyr, marking it as potentially one of the earliest-forming quiescent galaxies ever identified. The best fit model of this galaxy also suggests significant dust attenuation ($A_v=0.49\pm_{0.14}^{0.17}$). The confluence of these factors with its magnification ($\mu\sim2$) made it a high priority for spectroscopic follow up to confirm its quiescent and old nature. A full presentation of the photometric identification of quiescent galaxies in the UNCOVER field will be presented in Khullar et al. in preparation.

UNCOVER 18407 was observed in three of the seven micro-shutter array (MSA) configurations of the UNCOVER program between July 31 and August 2, 2023. MSA designs were optimized to target a wide range of target types based on the required depth of observation and source priority. Because it is extended, UNCOVER 18407 was assigned five slitlets, and observations were conducted with a \texttt{2-POINT-WITH-NIRCam-SIZE2} dither pattern. The total exposure times in each of the MSA2, MSA3, and MSA4 exposures that targeted UNCOVER 18407 were 2.6 hr, 2.3 hr, and 4.4 hr, respectively. 

\subsection{Spectroscopic Reduction}

The spectroscopic reduction for each of the individual MSA exposures of UNCOVER 18407, which will further be described in Price et al. in preparation, is as follows. Stage 2 data products were downloaded from MAST reduced with msaexp \citep[v0.6.10][]{Brammer2022}. msaexp masks artifacts (including snowballs), and corrects for 1/f noise. Individual slits are identified and WCS solutions are applied and data are flat fielded. Because the source is extended, we do not elect to background subtract using vertically shifted spectra, but instead utilize empty slits to avoid self-subtraction. 
The spectra are best fit by msaexp with $z_{spec}=3.97$, corresponding to a magnification $\mu=2.12\pm_{0.01}^{0.05}$ using the models in \cite{Furtak2023a}. Throughout this work, we correct all physical measurements of mass, size, and star formation rate for the effect of this magnification.

In Figure \ref{fig:spectrum}, we show the extracted spectra for the three MSA configurations that targeted UNCOVER 18407, in addition to photometry from two physical aperture diameters: 1.4" (purple) and 0.48" (red). Additionally, we present an RGB image of our target with the 5-slit MSA configuration for each exposure colored as in the spectrum. The spectra exhibit a range in normalization due to differences in alignment over the galaxy; for example, MSA2 is off-center enough that it misses the core of the galaxy between two shutters, and as such is significantly fainter than the other two exposures that sample that part of the galaxy. However, all three exposures, regardless of their orientation, exhibit strong Balmer breaks and weak $H\alpha$ emission (see Section \ref{subsec:halpha}), lending strong credence to the quiescent photometric solution for UNCOVER 18407.

Our primary science goal using the spectrum is to place limits on the star formation rate of UNCOVER 18407 via the $H\alpha$ line strength. We therefore elect to coadd only the red end of the spectrum to create a stacked one-dimensional spectrum of the $H\alpha$/[NII]/[SII] complex. Additionally, it is clear in Figure \ref{fig:spectrum} that the MSA2 slit is missing the core of the galaxy, as evidenced by the significantly lower flux in the spectrum. As such, we elect to combine only MSA3 and MSA4 in our coadd.

To calibrate the spectrum, we fit a 2nd order polynomial to synthetic photometry measured from the MSA3 and MSA4 spectra in addition to the UNCOVER DR2 0.48" diameter aperture photometry \cite{Weaver2024}, which visually provides the best match to the normalization of the MSA3 and MSA4 spectra, for the F200W, F277W, F356W, F410W, and F444W imaging, in addition to the medium bands spanning that same wavelength range. We do the same to synthetic photometry in the same bands measured on each of the individual MSA exposures, and we multiply the observed spectra and error vectors by the ratio of these polynomials. We then take the inverse variance weighted average of the MSA3 and MSA4 spectra to produce the flux calibrated data product that we will utilize for this analysis.

\section{Measuring Galaxy Properties} \label{sec:analysis}

\begin{figure*}
    \centering
    \includegraphics[width=\textwidth]{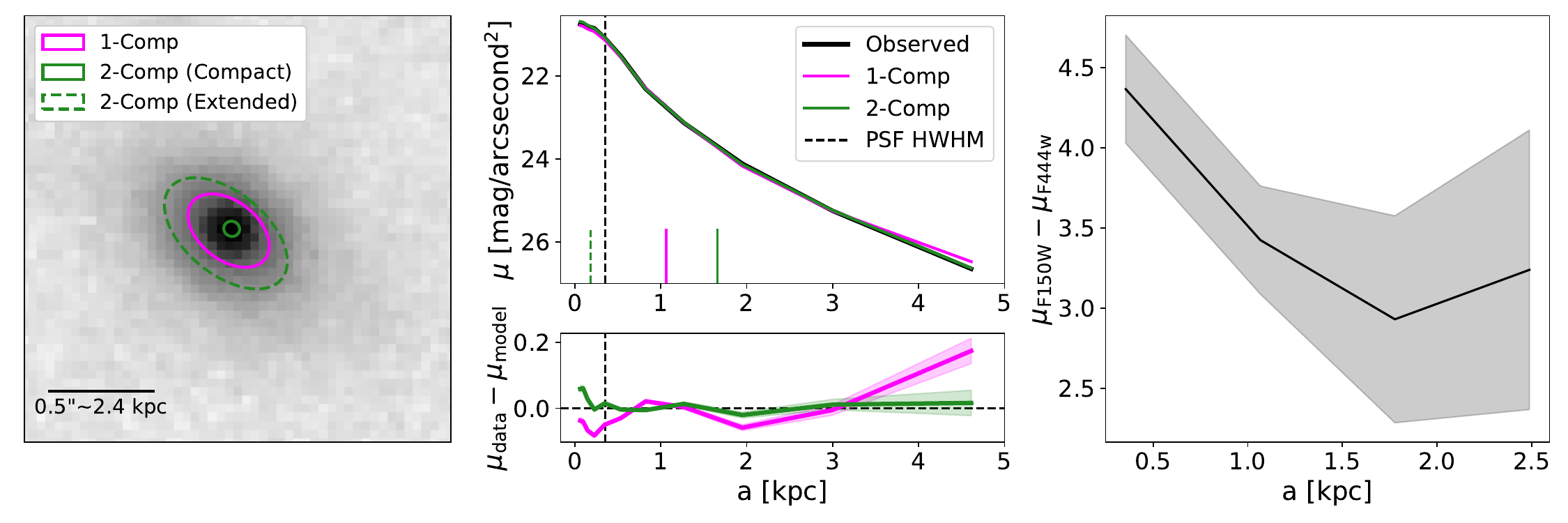}
    \caption{(Left): The F444W image of UNCOVER 18407. The plotted ellipses show the single (magenta) and two (green, component 1 solid, component 2 dashed) best fitting \sersic models to this image at \recomma, accounting for the PSF smearing of the axis ratio as described in Section \ref{subsec:annularfit}. (Center): The 1D surface brightness profiles of the observed galaxy (black) and the best fitting PSF-convolved one- and two- component models, along with residuals. We indicate the PSF half-width-half-maximum in the F444W image as a dashed vertical line. The two component model clearly does a better job at capturing the exponentially declining wings that are missed by the single component fit. The effective radii of the \sersic components are labeled, with solid and dashed lines again corresponding to the compact and extended components of the two component model. (Right): The F150W-F444W color gradient, measured using the best fitting sersic ellipse on the F444W PSF-convolved images. The two component structure can be seen in a peaky central red excess that flattens outside the central kpc. \label{fig:sersic}}
\end{figure*}

\begin{deluxetable*}{cccccc}
    \tablecaption{Best fitting \sersic parameters to the F444W image \label{tab:structure}}
    \tablehead{\colhead{\bf{Fit Type}} & \colhead{mag} & \colhead{$r_e$} & \colhead{n} & \colhead{b/a} & \colhead{$\chi^2$} \\
    \colhead{} & \colhead{[AB]} & \colhead{[kpc]} & \colhead{} & \colhead{} & \colhead{} \\ }
    \startdata
         \hline
         \bf{Sersic:} & $ 22.98 \pm_{ 0.01 }^{ 0.01 }$ & $ 1.07 \pm_{ 0.03 }^{ 0.03 }$ & $ 4.66 \pm_{ 0.15 }^{ 0.16 }$ & $ 0.58 \pm_{ 0.01 }^{ 0.01 }$ & 6888 \\
         \bf{Sersic+Sersic:} &  &  &  & & 6173 \\
         Extended & $ 23.55 \pm_{ 0.07 }^{ 0.06 }$ & $ 1.63 \pm_{ 0.07 }^{ 0.08 }$ & $ 1.71 \pm_{ 0.18 }^{ 0.17 }$ & $ 0.54 \pm_{ 0.01 }^{ 0.01 }$ &  \\
         Compact & $ 24.19 \pm_{ 0.11 }^{ 0.11 }$ & $ 0.18 \pm_{ 0.02 }^{ 0.02 }$ & $ 2.87 \pm_{ 0.96 }^{ 0.84 }$ & $ 0.71 \pm_{ 0.07 }^{ 0.06 }$ &  \\
         \bf{Sersic+PSF:} &  &  &  & & 6240 \\
         Sersic & $23.24 \pm_{ 0.02 }^{ 0.02 }$ & $ 1.3 \pm_{ 0.03 }^{ 0.03 }$ & $ 2.65 \pm_{ 0.13 }^{ 0.15 }$ & $ 0.56 \pm_{ 0.01 }^{ 0.01 }$ &  \\
         PSF & $ 24.96 \pm_{ 0.05 }^{ 0.05 }$ &  &  & &  \\
    \enddata
    \tablecomments{The best fitting structural parameters to the F444W image of UNCOVER 18407, as discussed in Section \ref{subsec:structure}. The physical effective radii presented are corrected for the effect of gravitational lensing by dividing by $\sqrt{\mu}$. }
\end{deluxetable*}

The differing shapes of the spectral energy distribution within different apertures \citep[see Figure \ref{fig:spectrum},][]{Weaver2024} hints at radially varying stellar populations or dust attenuation that we can study due to the excellent resolution and broad wavelength coverage of NIRCam in conjunction with the $\sim\sqrt{2}$ boost from the gravitational lensing of Abell 2744. We therefore elect to allow radial variation in the stellar populations by adopting and modeling annular photometry following the geometry of the best-fitting \sersic models.  In this section, we demonstrate the extraction of annular photometry, our approach to radial stellar population fitting with priors informed by the weak observed $H\alpha$ emission.

\subsection{Light-Weighted Structure} \label{subsec:structure}

To measure the structure of UNCOVER 18407, we use \texttt{pysersic} \citep{Pasha2023} to perform \sersic fits to the 2D image of the galaxy. Because the galaxy clearly exhibits strong color gradients (see Figure \ref{fig:spectrum}), in addition to a traditional single component \sersic fit, we also perform two component modeling analysis, one with two \sersic profiles and another with a \sersic and point source component. For these, we allow the central locations of the components to vary independently. To ensure accurate sampling, for each model we re-parameterize the posterior following the procedure in ~\citet{Hoffman2019} using a Block Neural Auto regressive flow.~\citep{DeCao2020} We then sample the posterior using a No U-turn sampler with two chains 1000 warm-up and 1000 sampling steps~\citep{Hoffman2014,Phan2019}. We ensure that the results of the sampling are robust by computing the $\hat{r}$ and effective sample size metrics which are $<1.01$ and $>300$ for all variables in all of the models~\citep{Vehtari2021}. In our fitting, we adopt the empirical PSF measured in \cite{Weaver2024} and smooth all intrinsic profiles to the resolution of the F444W image before fitting. The results of this fitting are presented in Table \ref{tab:structure}. Using $\chi^2$ as a goodness-of-fit metric, it is clear that the single \sersic model is disfavored relative to the two component models. The difference between the two \sersic and \sersic plus PSF models is more marginal, given how compact the central component is. 

In Figure \ref{fig:sersic}, we demonstrate visually the preference for a two component model by showing the F444W image of the galaxy, with ellipses overlaid showing the best fitting galaxy shapes at $r_e$ for the single component fit (purple) and the two component fits (green, compact as solid and extended as dashed). We correct the axis ratio for the effect of the PSF as:

\begin{equation}
    q_\mathrm{corr} = \sqrt{\frac{(q r_e)^2 + r_\mathrm{PSF \ HWHM}^2}{r_e^2 + r_\mathrm{PSF \ HWHM}^2}}
\end{equation}

following \cite{Suess2019a} and using the most compact effective radius as our benchmark $r_e$, and $r_\mathrm{PSF \ HWHM} = 0.0725$". In the middle panel of Figure \ref{fig:sersic}, we show the one-dimensional surface brightness profile (extracted using the best fitting compact component in the two component model) of the galaxy in black, and of the one- and two- component \sersic models again as purple and green respectively. The two component model, which consists of a $\sim200$ pc, \sersic n$\sim$2.9 compact profile and a $\sim1.6$ kpc, $n\sim$1.7 extended profile, is better able to capture both the compact galaxy core and the extended exponential envelope, whereas the single \sersic fit compromises by fitting a $\sim$1 kpc $n\sim4.5$ profile that overshoots the wings and undershoots the galaxy center, as evidenced by the residuals. We note that the two components are physically offset from one another by $\sim$1 pixel (0.04"$\sim$0.2 kpc); while this difference is well within the PSF FWHM, it is a potential hint at a disturbed morphology or the presence of strong dust lanes. In the third panel, we show the F150W - F444W surface brightness profiles (corresponding to just blueward of the break and redward of $H\alpha$ at $z_{spec}=3.97$) measured again using the geometry of the compact \sersic profile on images that have been matched to the F444W PSF \citep{Weaver2024}, demonstrating that the color of UNCOVER 18407 is significantly redder in the central kpc than it is in the outskirts. We note again that the two component \sersic model is only marginally preferred over the \sersic+point source model. However, given that the two \sersic fit favors a resolved central component with a clear preference for slight intrinsic elongation, we carry forward with this profile as our benchmark for annular extraction, noting that none of our conclusions are strongly impacted by the choice to extract photometry with elliptical versus circular annuli.

\subsection{$H\alpha$ and FIR Star Formation Rate Constraints} \label{subsec:halpha}

In order to constrain the total star formation rate, we require measurements of the $H\alpha$ line luminosity. However, at the resolution of the NIRSpec/PRISM, $H\alpha$ and the [NII] doublet are completely blended \textbf{and appear as a single line with a joint rest-frame equivalent width of 21 \AA}. As such, in order to measure the line fluxes and associated errors, we jointly fit the spectrum at $0.63 \ \mu m <\lambda_\mathrm{rest}< 0.68 \ \mu m$, modeling $H\alpha$ along with the [NII] and [SII] doublets as Gaussians with tied $\sigma$ (which, in this narrow region of wavelength, is a nuisance parameter that accounts for both the instrumental resolution and any intrinsic line spread) and a single free redshift. We fix the flux ratio of the [NII] doublets at 1:3, and allow the ratio of [SII]$\lambda6716/\lambda6731$ to vary from 0.4375 to 1.4484 following \cite{Sanders2016}. 

\begin{figure}
    \centering
    \includegraphics[width=0.4\textwidth]{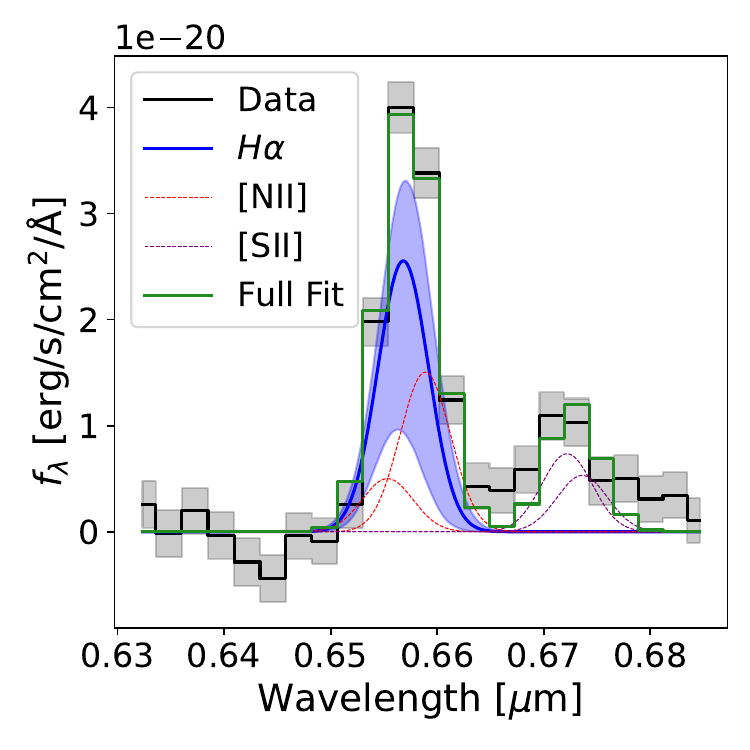}
    \caption{The stellar continuum-subtracted (see Section \ref{subsec:halpha}) spectrum of UNCOVER 18407 (black) in the region surrounding $H\alpha$/[NII]/[SII], along with the best fit (green). Also shown is the 1$\sigma$ confidence interval for $H\alpha$ (blue) along with the best fitting $H\alpha$, [NII], and [SII] lines. At the PRISM resolution, we are not able to rule out a line that is totally dominated by $H\alpha$.}
    \label{fig:halpha}
\end{figure}

In order to account for the continuum, including $H\alpha$ absorption, before fitting for the lines we fit a simple stellar population with a free dust law, convolved to the wavelength dependent PRISM resolution, to the flux-calibrated spectrum redward of 2 $\mu$m. We mask $H\alpha$ in this fit but otherwise leave the spectrum free, and we double the error vector to account for uncertainties in the small scale flux calibration. We subtract the best fitting continuum model from our spectrum, and we then utilize \texttt{emcee} \citep{Foreman-Mackey2013} to fit the covariant posteriors of the line strengths.

We show the results of this fitting in Figure \ref{fig:halpha}, with the shaded region surrounding the best fit on $H\alpha$ indicating the $1\sigma$ constraints on that line ($f_{H\alpha}=5.8\pm_{3.4}^{3.9} \times 10^{-19}$ $\mathrm{erg/s/cm^2}$ after accounting for the magnification). At the NIRSpec/PRISM resolution, we are not able to deblend $H\alpha$ and [NII] effectively, and the line is consistent with being essentially totally dominated by $H\alpha$. As such, we adopt the 3$\sigma$ $H\alpha$ luminosity (flux), $2.1\times10^{41} \ \mathrm{erg/s}$ ($1.32 \times 10^{-18} \ \mathrm{erg/s/cm^2}$), corresponding to a star formation rate of 1.15 $M_\odot \ \mathrm{yr^{-1}}$ with the assumption of $5.5\times10^{-42} \ \frac{M_\odot \ \mathrm{yr^{-1}}}{\mathrm{erg/s}}$ \citep{Leitherer1999} after correcting for magnification, as the limiting dust-unobscured star formation rate contained within the slit. Under the assumption of $A_v\sim1-3$ magnitudes and a \cite{Kriek2016} dust law with a dust index of 0, this corresponds to a limit of $\sim2.5-13 \ M_\odot \ \mathrm{yr^{-1}}$.

We note that the $H\alpha$/[NII]/[SII] complex is the only region of the spectrum of UNCOVER 18407 with significantly detected emission lines. We see no evidence of strong [OIII], even in the MSA configurations that most strongly overlap with the center of the galaxy. However, given the strong measured color gradients that could arise from centrally concentrated dust, we cannot conclusively rule out a dust-obscured AGN, and therefore cannot conclusively ascribe all $H\alpha$ flux to star formation. The lack of ability to rule out AGN ionization is further bolstered by our constraints on line ratios of the detected features; we measure log($\mathrm{[NII_{6583}]}/H_\alpha$)$=-0.07\pm_{0.54}^{0.54}$ and log($\mathrm{[SII_{6716,6731}]}/H_\alpha$)$=-0.21\pm_{0.22}^{0.37}$, spanning the full range of observed ratios in star forming galaxies and AGN \citep[e.g.,][]{Kewley2006}. This further motivates our choice to treat the $H\alpha$ star formation rate as an upper limit rather than a firm measurement.

The $H\alpha$ limits on the total star formation rate even under the assumption of negligible [NII] contribution are also corroborated by the ALMA non-detection of the galaxy in the DUALZ survey \citep{Fujimoto2023dualZ}. The 3$\sigma$ upper limit at 1.2-mm flux observed is 113 $\mu$Jy. Assuming a FIR SED with a modified black body ($T_\mathrm{dust}$ = 35K and spectral index = 1.8) and accounting for the magnification due to lensing, the ALMA non-detection places a 3$\sigma$ upper limit of $8.5 \ M_\odot \ \mathrm{yr^{-1}}$ on the galaxy star formation rate. This limit is quite sensitive to the assumed dust temperature, but in conjunction with the weak observed $H\alpha$, it provides a strong constraint that the total star formation rate for UNCOVER 18407 should not exceed $\sim10$ $M_\odot$yr$^{-1}$.



\subsection{Annular Stellar Population Fitting} \label{subsec:annularfit}

\begin{figure*}
    \centering
    \includegraphics[width=0.95\textwidth]{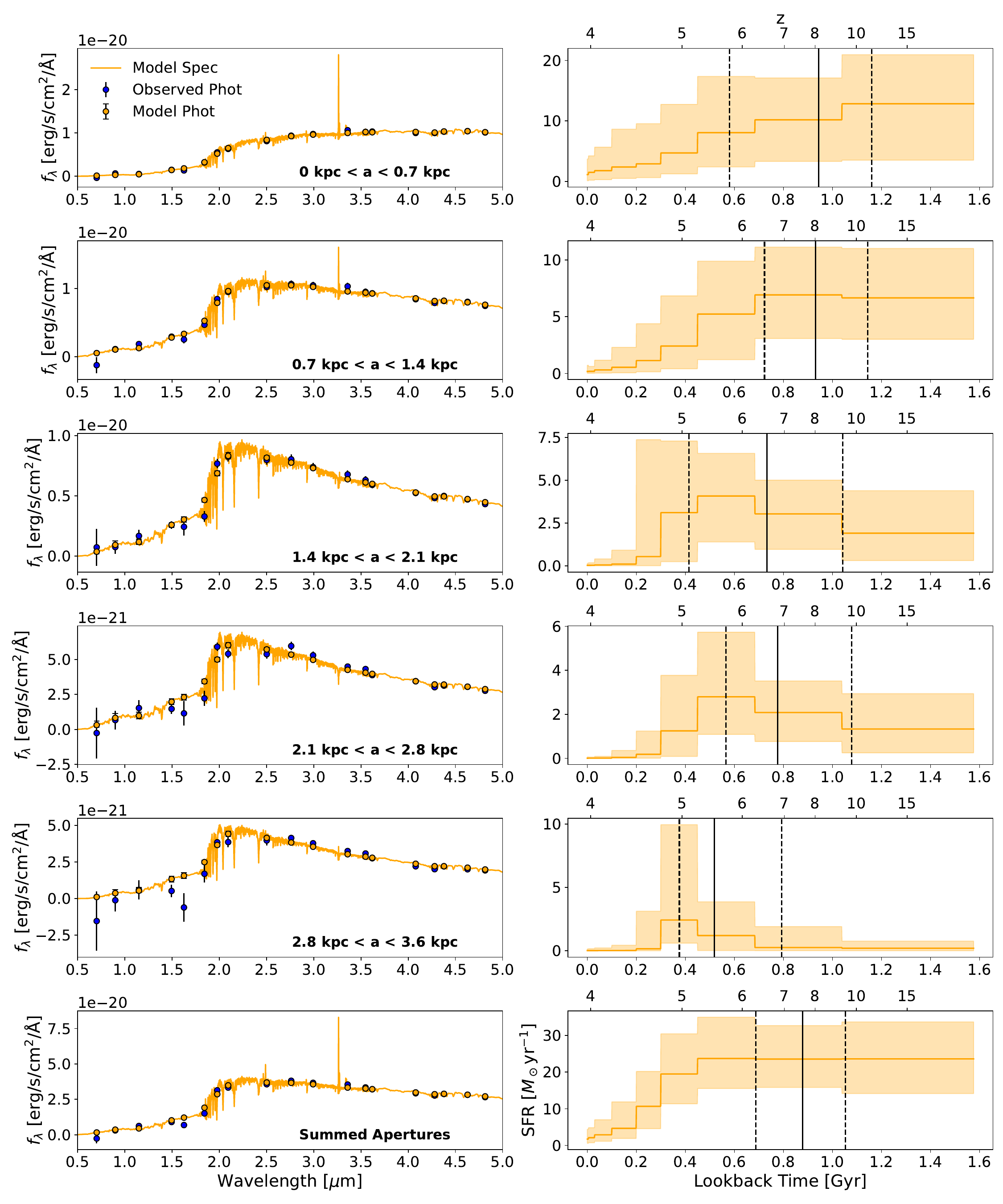}
    \caption{Annular prospector SED fits (left) and measured star formation histories (right) with $1\sigma$ confidence intervals shown as a shaded region. Labels indicating the physical radius of the aperture being fit, except in the final panel, which shows the sum of the five annular fits. The solid and dashed lines on the star formation histories show the median and $1\sigma$ constraints on the mass weighted age. The SED shape of the galaxy clearly becomes bluer at large physical radii, and the median mass weighted age correspondingly shifts to lower values, albeit with significant uncertainty. We note that while the SEDs are plotted in observed flux units, the star formation histories have been corrected for the effect of lensing.}
    \label{fig:annular_fits}
\end{figure*}

To further quantify the cause of these observed color gradients, we make use of the 7-band UNCOVER NIRCAM imaging to fit the radially varying spectral energy distribution of UNCOVER 18407. Using the best-fitting structural parameters from the compact component of the two-component \sersic fit and the corrected axis ratio as in Section \ref{subsec:structure}, we extract annular photometry in annular bins, with a minimum annulus size of 0.14" ($\sim$0.7 kpc), the PSF FWHM of F444W PSF-matched images \citep{Weaver2024}. We extract annuli in five bins, out to $\sim3.6$ kpc ($>$3$r_e$ in the single \sersic fit). We enforce a signal-to-noise threshold of 20 on our photometry to account for systematic uncertainties in the models and the flux calibration \citep[e.g.,][]{WangB2024_uncoverpops}. We perform our SED fits using 19/20 available bands, omitting F140M as it exhibits discrepant colors with F150W despite the lack of any strong features in that part of the spectrum blueward of the break (see Figure \ref{fig:spectrum}).

We utilize the non-parametric star formation histories in \texttt{Prospector} \citep{Johnson2017, Leja2017,Johnson2021} to fit the annular photometry in nine fixed bins of constant star formation. We set the widths of the nine star formation history bins in order of lookback time to 5 Myr, 25 Myr, 70 Myr, 100 Myr, 100 Myr, 150 Myr, and the final three bins are set to logarithmically fill the remaining time since the Big Bang, and, following the continuity prior in \cite{Leja2019}, the logarithmic difference between the SFR in neighboring bins is sampled over with a Student-T distribution prior centered at 0 with a width of 0.3 and $\nu=2$. We utilize the Flexible Stellar Population Synthesis (FSPS) stellar population synthesis models \citep{Conroy2009, Conroy2010}, the MILES spectral library \citep{Sanchez-Blazquez2006a} and MIST isochrones \citep{Choi2016, Dotter2016}. We assume a \cite{Chabrier2003} Initial Mass Function and fix the model redshift to the spectroscopic redshift. We assume the \cite{Kriek2013} dust law with a free $A_v$ and dust index (spanning [0,2.5] and [-1,0.4] respectively). Additionally, following \cite{Wild2020}, we fix the optical depth of attenuation around young ($<10^7$ yr) stars to be double the optical depth for the rest of the stars. We leave stellar metallicity as a free parameter, and sample linearly in the range $0.01Z_\odot<Z<2Z_\odot$. We include nebular emission in our fits, with the ionization parameter log(U) varying in the range [-4,-1] and the gas phase metallicity fixed to the stellar metallicity. Because at $z_{spec}=3.97$ even the reddest NIRCam/NIRSpec does not constrain wavelengths $\gtrsim1 \ \mu$m, we fix the shape of the IR SED following the \cite{Draine2007} dust emission templates, with $U_\mathrm{min}$ = 1.0, $\gamma_e$ = 0.01, and $q_\mathrm{PAH}$ = 2.0. For similar reasons, we do not fit for any IR AGN contribution. We sample our posteriors using the \texttt{dynesty} nested sampling package \citep{Speagle2020}.

\section{The spatially resolved structure of UNCOVER 18407} \label{sec:results}

\subsection{Can the core be maximally old and dust free?}

\begin{figure}
    \includegraphics[width=0.5\textwidth]{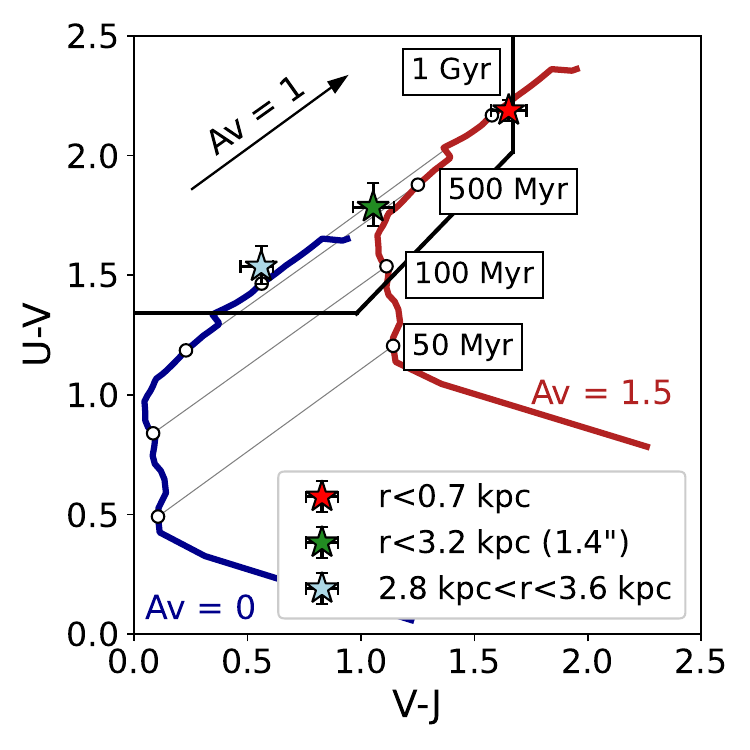}
    \caption{Rest frame U-V and V-J colors for the central 0.7 kpc (light red), the full galaxy (green, using the 1.4" aperture photometry from \cite{Weaver2024}), and the outer 2.8-3.6 kpc (light blue) for UNCOVER 18407. Also plotted are simple stellar populations evolutionary tracks with solar metallicity generated with \texttt{fsps} \citep{Conroy2013} for a dust free (dark blue) and $A_v=1.5$ (dark red) stellar population. The model tracks are labeled with circles at 50 Myr, 100 Myr, 500 Myr, and 1 Gyr. The black vector shows the effect of $A_v = 1$ for a 1 Gyr old population. The central regions of the galaxy cannot be produced by a dust-free, maximally old stellar population, and require significant dust to produce the observed colors, in contrast with the outer regions.}
    \label{fig:color_ssp}
\end{figure}

\begin{figure}
    \includegraphics[width=0.5\textwidth]{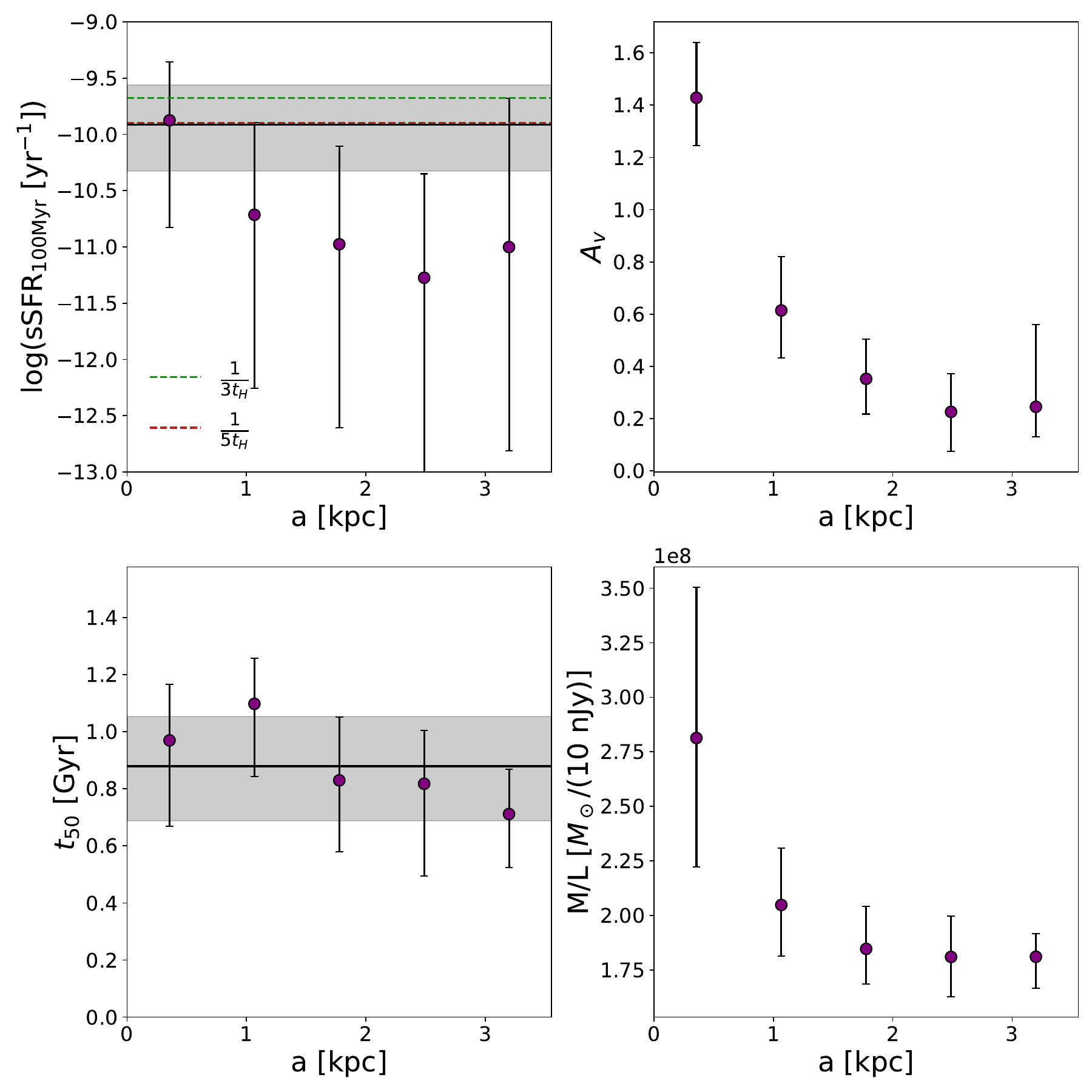}
    \caption{Photometric radial constraints on the specific star formation rate (top left, with the transition sSFR boundary $\frac{1}{3t_H(z=3.97)}$ from \cite{Tacchella2022} and the quiescence boundary $\frac{1}{5t_H(z=3.97)}$ from \cite{Pacifici2016} in green and red respectively), $A_v$ (top right), the mass-weighted age (bottom left, in units of lookback time), and the mass-to-light ratio (bottom right) for UNCOVER 18407. The error bars capture the 1$\sigma$ constraints on parameters. When relevant, the 1$\sigma$ constraints from the global star formation history are indicated as a grey shaded region around the median value. The fits indicate that observed color gradients largely result from centrally concentrated dust. We find very slight evidence of a negative age gradient, although the profile is consistent with flat at the $2\sigma$ level. All together, these results manifest in a declining mass-to-light gradient. \label{fig:radial}}
\end{figure}

Before delving into the marginalized radial stellar population trends, we first present a simple demonstration that the core of UNCOVER 18407 must contain significant dust based on rest frame colors. Given the age of the universe at $z=3.97$, there is only $\sim1.5$ Gyr over which any star formation could have occurred; using rest-frame colors, we posit the question of whether the extremely red SED of UNCOVER 18407 could result from a maximally old stellar population. In Figure \ref{fig:color_ssp} we show rest frame $UVJ$ colors \citep{Williams2009} for the galaxy center (red), the full galaxy (green), and the galaxy outskirts (light blue), with solar metallicity simple stellar populations generated with \texttt{fsps} shown for dust-free (dark blue) and $A_v=1.5$ (dark red) spanning 0 Gyr to 1.6 Gyr, the age of the universe at $z=3.97$. 

It is immediately evident that the center of these galaxy can \textit{only} result from a simple stellar population with significant extinction; the dust free curve does not reach high enough values in $U-V$ or $V-J$ to produce the observed central colors. In fact, to produce such high $U-V$ and $U-J$ colors, a dust free, simple stellar population would need to be $\sim$10 Gyr old, an order of magnitude older than is physically possible at this redshift. In contrast, the integrated galaxy $UVJ$ colors are still within the realm of possibility for a dust free SSP, albeit a quite old one, and cannot be produced in the presence of dust attenuation that is comparable to the galaxy center.

\subsection{Radial dust and age gradients}

The annular SED fitting allows us to fully probe the range of age and dust solutions that can produce the observed galaxy radial SEDs. The results of this fitting are shown in Figure \ref{fig:radial} as a function of the semi-major axis (corrected for lensing) in blue. It is clear that the color gradients in the SED of the galaxy are predominantly driven by highly concentrated dust with $A_v\sim1.4$, with the galaxy light significantly less attenuated outside the central kiloparsec. This color gradient, in light of the low-sSFR solutions at all radii, translates nearly directly to a mass-to-light gradient that indicates that the center of the galaxy hosts $\sim1.5$ as much mass per unit of light relative to the outer regions of the galaxy. There is significant uncertainty in this measurement, however, owing to the uncertainty in the precise amount of attenuation in the central regions.

In contrast with the clear gradient in $A_v$, gradients in the mass-weighted age are more ambiguous. 
We measure the slope in $t_{50}$ to be $-70 \pm_{100}^{110}$ Myr/kpc, and we are unable to rule out flat gradients due to the uncertainty in the mass weighted age measurements. We can, however, conclusively rule out any strong positive age gradients. There is some indication of radially declining specific star formation rate (averaged over the past 100 Myr in lookback time); while the galaxy is consistent with low rates of ongoing star formation at all radii, the star formation rate within the central 700 parsecs is constrained to be an average of $1.7\pm_{1.2}^{2.4}$ $M_\odot$yr$^{-1}$ over the 100 Myr before observation, corresponding to log(sSFR$_\mathrm{100 Myr}$ [yr$^{-1}$]) = $-9.7\pm_{0.8}^{0.5}$, slightly higher than at r=1-3 kpc but consistent at the 1$\sigma$ level.


\subsection{Global galaxy properties}

Ideally, in order to obtain global galaxy properties, one would jointly model the spectrum and photometry to increase constraining power on the total mass assembly of the galaxy. However, we note that the strong gradients measured in this section preclude the possibility of performing a combined spectrophotometric fit for this galaxy. Joint modeling of NIRSpec/PRISM spectra and photometry is already quite difficult due to the differential slit loss as a function of wavelength over a wide wavelength range \citep[e.g.,][]{deGraaff2023}, and with the added complication of color gradients, we are not confident we can robustly recover the shape of the spectrum and the corresponding photometric aperture after correcting for slit losses.

As such, we obtain global constraints on the star formation history of the galaxy by combining the annular SED fits to construct a total star formation history for the galaxy by summing the posteriors of each individual fit. This method of determining the total star formation history is preferred over a global SED fit due to the strong dust gradients, which are smoothed over when all photometry is summed. From this summed posterior, we measure the total galaxy SFR averaged over the 100 Myr before observation to be $2.8\pm_{1.7}^{3.5} \ M_\odot \ \mathrm{yr^{-1}}$, well in line with the ALMA $3\sigma$ limit of 8.5 $M_\odot \ \mathrm{yr^{-1}}$ and the total limits from $H\alpha$ (see Section \ref{subsec:halpha}). We constrain the total stellar mass within 3.5 kpc to be $\log(M_\star(<3.5 \mathrm{kpc})/M_\odot) = 10.35\pm_{0.06}^{0.06}$. In conjunction with the measured star formation rate, this corresponds to a specific star formation rate of $\log(\mathrm{sSFR_{100 Myr} [\mathrm{yr}^{-1}]}) = -9.9\pm_{0.5}^{0.4}$, $\sim1$ dex below the star forming main sequence at $z\sim4$ \citep[using the star forming main sequence in][scaled from \citealt{Speagle2014} under similar IMF assumptions]{Shapley2023}. Definitions of ``quenched" vary in the literature, especially at $z>3$ \citep[e.g.,][]{Pacifici2016,Tacchella2022}, but our constraints on the star formation history of UNCOVER 18407 show a strong preference for a declining star formation rate over the past $\sim500$ Myr (see Figure \ref{fig:annular_fits}).

We emphasize that there is a significant discrepancy between the constraints that come from our method of analyzing spatially resolved annuli and the constraints from fitting the entire galaxy SED with a single dust law. We perform an identical fit to the one performed in Section \ref{subsec:annularfit}, but now on the catalog photometry within a 1.4" aperture. This global fit measures a total stellar mass $\log(M_\star/M_\odot) = 10.34\pm_{0.07}^{0.06}$, which is quite consistent with the total mass estimate from the summed annuli. However, the dust and star formation properties of the galaxy differ significantly as a result of the averaging of dust attenuation. The global fit measures $A_v=0.66\pm_{0.17}^{0.18}$, almost a magnitude less than the attenuation we measure in the central annulus. Similarly, the star formation rate that we measure from the global fit is $1.2\pm_{1.1}^{3.3} \ M_\odot \ \mathrm{yr^{-1}}$, about a factor of 2 lower than the global star formation rate measured by summing the annular fits. This is a direct result of the ability for a dusty central region can hide significantly more star formation. While such a result was implied by the empirical $F150W-F444W$ and rest-frame $UVJ$ color gradients, only detailed modeling that marginalizes over the degeneracy between age, dust, and metallicity is capable of robustly correcting for such trends in mass and star formation rate estimates.


Using the combined posterior for the total star formation history of the galaxy, we also constrain the global mass weighted age to be $880\pm^{190}_{170}$ Myr. This corresponds to an age of the universe when 50\% of the galaxy's stellar mass was formed of $\sim700$ Myr after the Big Bang ($z_\mathrm{form} = 7.5 \pm_{1.3}^{1.8}$), suggesting that UNCOVER 18407 formed the majority of its stellar mass within the first Gyr of cosmic time. We measure the global $t_{90}$, the mass weighted age when UNCOVER 18407 formed 90\% of its stars, to be 360$\pm_{120}^{100}$ Myr, indicating that the galaxy has not been forming significant stellar mass in the past few hundred Myr. Such ages appear to broadly line up with the formation of the first quiescent galaxies with \logM$\sim$10.3, which appear as quenched quite close to $z\sim4$ in simulations \citep[e.g.,][]{Wellons2015, Hartley2023, Xie2024}.

\subsection{Constraints on the Dust and Gas Mass}

Because the spectrophotometric fit to the galaxy implies significant dust extinction, we also use the ALMA Band 6 non-detection \citep[see Section \ref{subsec:halpha},][]{Fujimoto2023dualZ} to constrain the dust mass under the assumption of cold dust ($T_\mathrm{dust}$ = 20K), given that star formation does not seem to be extreme from SED fitting and $H\alpha$ constraints. Using Equation 2 from \cite{Greve2012}, we obtain a 3$\sigma$ limit of log($M_\mathrm{dust}/M_\odot$)$<$7.7. Taking the ALMA $M_\mathrm{dust}$ limit and applying standard assumptions \citep[$\delta_{M_{H_2}/M_\mathrm{dust}}$=100, e.g.,][]{Scoville2016}, this corresponds to a gas mass limit of  log($M_\mathrm{H_2}/M_\odot$)$<$9.7, which in turn corresponds to an upper limit on the molecular gas fraction $M_\mathrm{H_2}/M_\star\lesssim20\%$. These limits are quite similar to those found in other quiescent systems at similar redshift \citep{Suzuki2022}. However, we note that this upper limit is quite sensitive to the assumed dust temperature \citep[e.g.,][]{Cochrane2022}.

Simulations that include predictive dust physics forecast considerable scatter in the dust-to-gas mass ratios of quiescent systems, with the wide range of values spanning four orders of magnitude becoming evident by $z=2$ and below \citep[][]{Whitaker2021b}. While very few quiescent systems exist at $z=3$ in the simulations, these early quenched galaxies instead appear to have more normal gas-to-dust mass ratios. So while dust masses can be extremely small in quiescent systems at $z=2$ and below \citep[e.g.,][]{Whitaker2021a,Caliendo2021}, it may be that the dust reservoirs at $z=3$ and above are instead still largely intact in the earliest quiescent galaxies and consistent with standard gas-to-dust mass ratios. With a sample of one galaxy, we cannot yet quantify how common this dusty phase of quiescence is. However, population studies using spatially resolved attenuation measurements as a proxy for dust richness provide an opportunity to constrain the evolution of this population as large-area JWST imaging surveys become public.

\section{Discussion} \label{sec:discussion}

\subsection{A old, centrally dusty, quiescent galaxy at $z_{spec}=3.97$}

The question of \textit{how} the most massive quiescent galaxies in the universe form is fundamentally linked to \textit{when} they form. In the pre-JWST landscape, the empirical findings pointed to a when of $\sim z=3-5$--the wide array of spectroscopic observations of quiescent galaxies at $z\gtrsim3$ indicated that the only populations of dead galaxies that existed at those redshifts were ``post-starburst" galaxies that shut down in the few hundred million Myr before observation \citep[e.g.,][]{Marsan2015, Glazebrook2017, Schreiber2018a,Schreiber2018b,Tanaka2019, Forrest2020b, Valentino2020, DeugenioC2020,Saracco2020,Kalita2021, Kubo2021, Tanaka2023,Kakimoto2024}. However, the presence of systems in the local Universe \citep[e.g.,][]{Thomas2005} and at $z\sim2$ \citep[e.g.,][]{Kriek2016,Beverage2023} with spectral features indicating earlier formation in conjunction with the ambiguous results of of following up ``old" galaxy candidates at $z>3$ with ground-based facilities \citep[e.g.,][]{Antwi-Danso2023b,Nanayakkara2024} suggested that a population of earlier forming quiescent galaxies existed just below the sensitivity of our best NIR facilities.

As with many aspects of galaxy evolution, the advent of JWST has confirmed this suspicion. Early photometric samples have revealed a previously-hidden population of massive, quiescent galaxies at $z>3$ with mass-weighted ages consistent with formation redshifts of $\sim$10 \citep[$\sim500$ Myr after the Big Bang, see][]{Carnall2023a}. A smaller number of high-z quiescent systems have also been the targets of follow-up spectroscopy, revealing formation times as early at $z\sim11$ \citep[e.g.,][]{Carnall2023b,Nanayakkara2024,Glazebrook2024,deGraaff2024, UrbanoStawinski2024,Carnall2024}, providing support to the argument that early universe star formation was more vigorous than previously thought \citep[e.g.,][]{Labbe2023, Boyett2023}. UNCOVER 18407 represents an additional confirmed example of early-onset star formation; the mass weighted age of the center of the galaxy is consistent with $z_\mathrm{form}=7.6 \pm_{1.3}^{1.8}$ ($700 \pm_{190}^{170}$ Myr after the Big Bang). 

\begin{figure}
    \centering
    \includegraphics[width=0.4\textwidth]{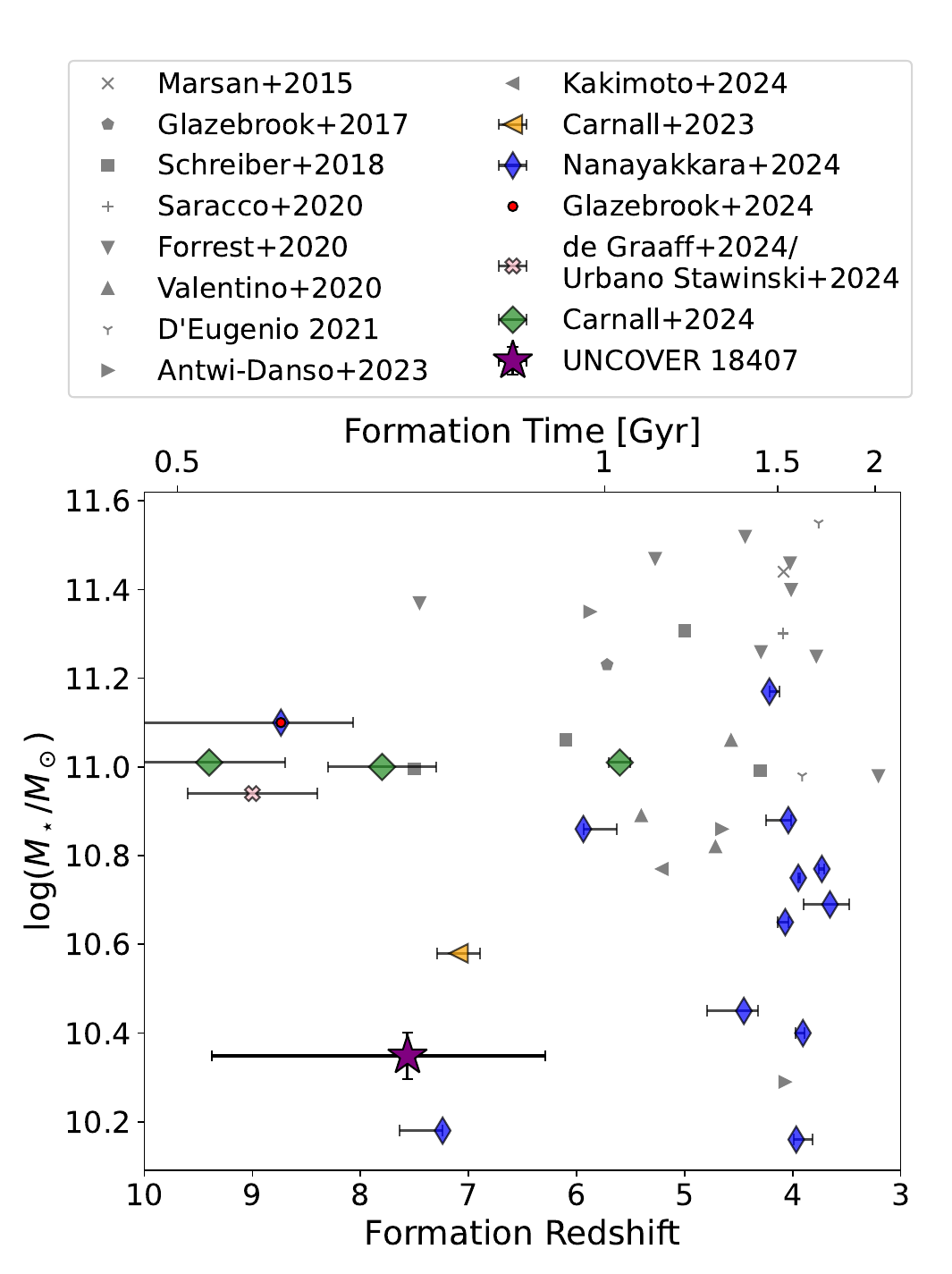}
    \caption{The stellar mass versus formation redshift (or time of formation) for UNCOVER 18407 (purple star) and other JWST-identified (color) and pre-JWST (grey) spectroscopically confirmed quiescent systems at $z>3$. We note a galaxy reported in \cite{Nanayakkara2024}, \cite{Glazebrook2024}, and \cite{Carnall2024} with two markers but show only the age and mass estimates from \cite{Nanayakkara2024} clarity. For objects that appear in both \cite{Schreiber2018a} and \cite{Nanayakkara2024}, we show only the latter up-to-date JWST measurements. UNCOVER 18407 is among the oldest quiescent systems ever identified, and as it was found in a very small field, it is likely that significant old and quiescent populations exist and have eluded prior detection.}
    \label{fig:mass_zform}
\end{figure}

We show this new post-JWST landscape of $z>3$ quiescent galaxies in Figure \ref{fig:mass_zform}, highlighting the new JWST samples (including UNCOVER 18407) as colored points and pre-JWST spectroscopically confirmed quiescent systems in grey. There is still considerable uncertainty in the formation redshift of UNCOVER 18407 owing to the coarse sensitivity of SED shape to age, especially in the presence of significant dust, but it is consistent at the $1\sigma$ level with forming the majority of its mass in the first Gyr of cosmic time. This formation time is consistent with the emerging picture of earlier-than-expected massive galaxy formation. However, UNCOVER 18407 adds another wrinkle to this literature, as while there has been clear evidence of spatially varying dust in photometric samples of HST-faint/dark galaxies \citep{PerezGonzalez2023}, no spectroscopically confirmed massive, quenching system at $z>3$ to date shows such clear indications of strong central dust extinction ($A_v>1$).

While it is commonly assumed that quiescent galaxies are dust poor, the presence of dust lanes in the centers of massive, local early type galaxies are quite often hosts to centrally concentrated dust \citep{Ebneter1985, Ebneter1988, VanDokkum1995, Tomita2000, Tran2001, Laine2003, Lauer2005,Whitaker2008, Davis2013,Boizelle2017}, indicating either that quiescent galaxy environments are able to retain significant leftover dust and gas for a few hundred Myr after quenching or that they are able to consistently replenish their central reservoirs. Such high attenuation is also seen in massive quiescent galaxies at $z>2$ with red SED coverage. Pre-JWST work leveraging Spitzer to probe the red SED of massive, dusty galaxies has identified dusty quiescent galaxies at $z>1$ \citep{Martis2019}. Similarly, \cite{Marchesini2014} finds that the typical $A_v$ in quiescent systems grows from $\sim0$ at $z=0$ to $\sim0.5$ at $z=2$ using ULTRAVISTA photometry, and many of the spectroscopically confirmed quiescent systems in \cite{Schreiber2018a} are fit with comparable $A_v$. With the benefit of the deep NIR imaging capability of JWST, \cite{VanDokkum2023} presents a massive lens that exhibits an integrated $UVJ$ color consistent with a dusty, quiescent solution, and four of the $z>3$ systems reported in \cite{Nanayakkara2024} have solutions with $A_v>1$. Such systems also appear to be common near dusty star forming galaxies; \cite{Alberts2023} reports that the quiescent companion in a massive star forming/quiescent pair located in a $z\sim3.7$ overdensity is likely dusty, and \cite{Kokorev2023a} reports an average $A_v\sim1$ in a quiescent system at $z=2.58$ located 9 kpc from an extremely dusty star forming galaxy. 

Additionally, spatially resolved studies of quiescent systems have turned up dusty sub-regions similar to those at the center of UNCOVER 18407; \cite{Miller2022} identifies two $UVJ$-selected quiescent galaxies at $z\sim2$ that exhibit remarkably similar $UVJ$ color gradients to UNCOVER 18407, and the GRISM spectroscopy lensed galaxies at similar redshifts in \cite{Akhshik2023} discovered a system with $A_v\sim4$ in a single spatial bin \citep[though notably, not at the galaxy center, both in the GRISM map and in direct dust detections, see][]{Morishita2022}. The fact that large samples of high-$A_v$ quiescent galaxies have yet to be catalogued is likely in large part a testament to the power of spatially resolved measurements; the vast majority of catalogs of quiescent systems at high redshift \citep[e.g.,][]{Schreiber2018a, Carnall2023a, Alberts2023} select based on star formation histories inferred from integrated spectral energy distributions. As demonstrated in Figure \ref{fig:color_ssp}, the integrated color that averages over the obscured core and unobscured outer regions is far less extreme in $UVJ$ space than the dusty core, and a fit to the global photometry of the galaxy that does not account for spatially varying dust/stellar populations results in constraints on the dust content of $A_v=0.66\pm_{0.17}^{0.18}$, a typical level of attenuation seen in global SED measurements of quiescent galaxies at $z\gtrsim2$ \citep{Marchesini2014}.


As such, we propose that it is likely that many previously observed $z>3$ quiescent galaxies, when imaged with JWST, may exhibit similar color gradients due to centrally concentrated dust. This inference that many high-z quiescent galaxies host central gas and dust, based on indirect attenuation measurements and an upper limit of log($M_\mathrm{dust}/M_\star$)$<-2.69$, squares with direct dust-continuum and CO measurements that find that a large variety in interstellar medium content in quiescent and post-starburst galaxies \citep[e.g.,][]{Sargent2015,French2015,Suess2017,Gobat2018,Whitaker2021a,Belli2021,Williams2021,Bezanson2022a,Smercina2022,Woodrum2022,Spilker2022,Otter2022,Wu2023,Donevski2023,Lee2024}. 

The origins of this central gas and dust remain uncertain, but there are strong links to physical associations with mergers in local galaxies \citep{Weaver2018}. As mergers appear to be physically associated with quenching \citep[e.g.,][]{Pawlik2016, Sazonova2021, Ellison2022, Verrico2023}, dynamical effects may be allowing these galaxies to sustain a centrally concentrated interstellar medium in the wake of shutdown; alternatively, recent wet mergers may be responsible for briefly re-enriching a previously depleted ISM that is quickly removed \citep[e.g.,][]{Akhshik2021}. Given that there is evidence for some small residual star formation in the galaxy core, it is plausible that any remaining ISM could be removed by this, but significantly deeper sub-mm observations would be needed to quantify the remaining ISM reservoirs given that quiescent galaxies at these redshifts are consistently undetected or weakly detected, even in stacks \citep{Schreiber2018b,Santini2019,Chworowsky2023}.

In the absence of deep sub-mm detections, we demonstrate that resolved stellar population studies present an opportunity to identify ISM-rich massive quiescent galaxies at cosmological distances by exploiting the sensitivity of imaging that spans the rest-optical to NIR to dust gradients. In future work, we will leverage the growing samples of spectroscopically confirmed quiescent systems to understand the diversity of ISM content in quiescent systems as a function of cosmic time and morphology.

\begin{figure*}
    \centering
    \includegraphics[width=\textwidth]{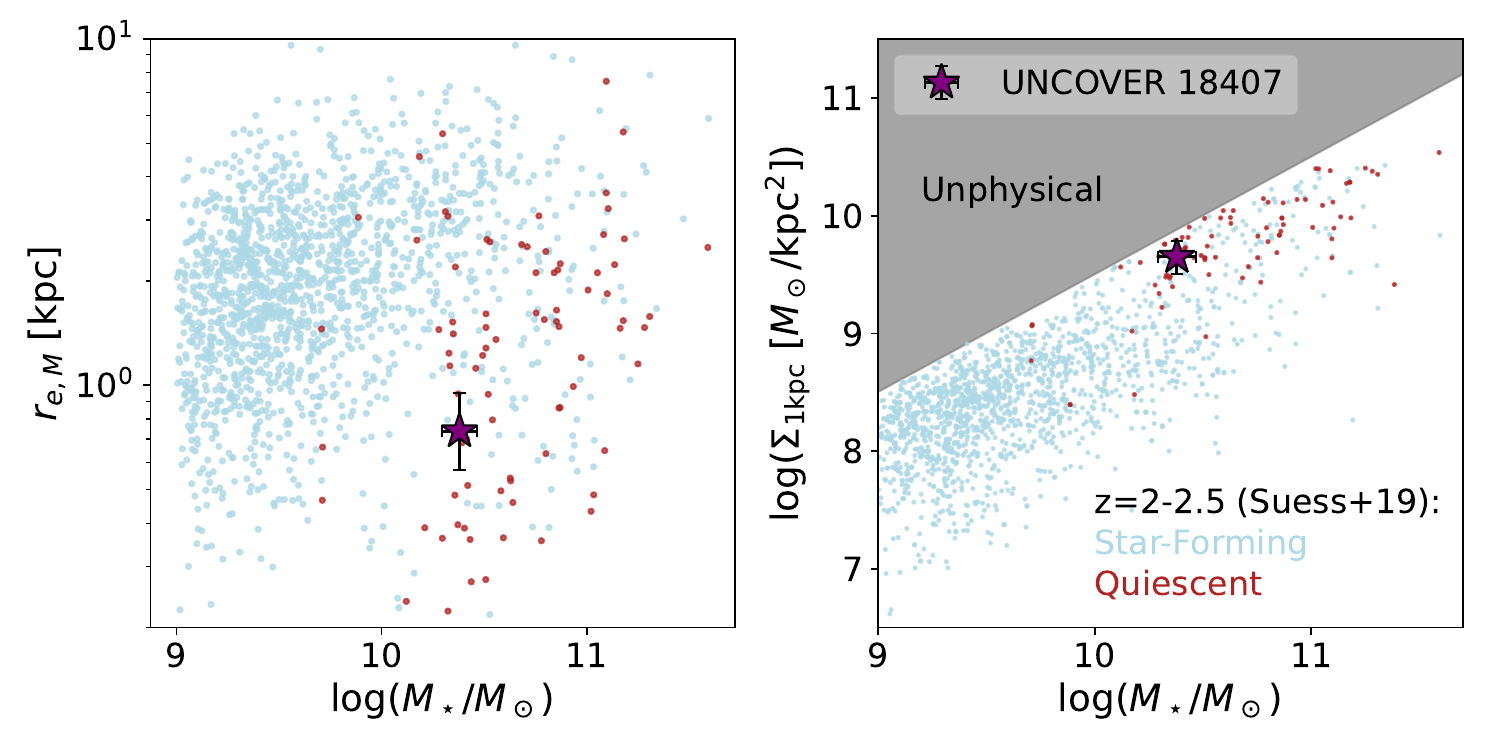}
    \caption{The mass-weighted effective radius (left) and the central kiloparsec surface density (right) of UNCOVER 18407 (dark blue) as compared to z=2-2.5 star-forming (light blue) and quiescent (red) galaxies, measured in \cite{Suess2019a}. The shaded grey ``unphysical" region denotes the space where a given density is unattainable at a given stellar mass. After accounting for the centrally concentrated dust, UNCOVER 18407 clearly shows structural signatures typical of quiescent galaxies at cosmic noon, with a dense central core that indicates intense star formation on sub-kpc scales in the first Gyr of cosmic time.} \label{fig:structures}
\end{figure*}

\subsection{Implications for dense core formation in the first Gyr of cosmic time} \label{subsec:dense}

It is observationally well-established that quiescent galaxies out to $z\sim3$ differ significantly from their co-eval star forming counterparts, whether in half light radius \citep[e.g.][]{VanDerWel2014,Mowla2019,Kawinwanichakij2021,Hamadouche2022,Ji2024}, half mass radius \citep{Suess2019a}, or the central mass density \citep[e.g.][]{Bezanson2009,Fang2013,VanDokkum2015,Barro2017,Mosleh2017,Whitaker2017,Suess2021}. The prevailing interpretation of these trends is that massive quiescent systems are the result of a quenching pathway resulting from rapid gas compaction \citep[e.g.,][]{Dekel2014, Zolotov2015, Tacchella2016}, leaving behind a dense core, with a more extended stellar disk component that quenches at later times \citep[e.g.,][]{Costantin2021,Costantin2022}. Direct evidence for this two component assembly and core formation on scales of $\sim$200 parcsecs have been seen as early as $z\sim7$ \citep{Baker2023}. 

However, inferences about galaxy growth based on galaxy sructure are complicated by the presence of color gradients, which bias structural measurements toward tracing younger stellar populations due to outshining \citep{Tortora2010, Wuyts2010, Suess2019a, Miller2023}. Because UNCOVER 18407 has broad rest-frame near-IR SED coverage and benefits from the mild lensing of Abell 2744 that results in a $\sim \sqrt{2}$ boost in angular resolution, it represents an excellent test of whether the stellar mass distribution of a quenched system that formed in the first Gyr of cosmic time exhibits compact structure after accounting for differential dust attenuation and stellar populations.

In order to measure the stellar mass distribution of UNCOVER 18407, we assume that the intrinsic M/L distribution of the galaxy follows a power law distribution:

\begin{equation}
    \mathrm{\log(M/L)} = \alpha \times \mathrm{\log(a)} + \beta
\end{equation}

following \cite{Suess2019a} where a is the semi-major axis of an ellipse following the distribution of the central component of the galaxy (see Table \ref{tab:structure}), $\alpha$ is a the slope of the power law, and $\beta$ is the intercept. We forward model the effect of the PSF on this power law by projecting the model onto a 2D grid following the aforementioned central component best fitting ellipses and convolving with the F444W PSF from \cite{Weaver2024}. We then measure the mean M/L in the PSF-corrected annuli defined in Section \ref{subsec:structure} to compare to our measured M/L profile. We fit the intrinsic power law with $\alpha$ and $\beta$ as free parameters using \texttt{emcee} \citep{Foreman-Mackey2013} to constrain the possible range in slopes allowed by the data. 

\begin{deluxetable}{ccc}
    \tablecaption{Mass-weighted structural measurements \label{tab:mass_weighted_structure}}
    \tablehead{\colhead{$\log(M_\star/M_\odot)$} & \colhead{$r_e$} & \colhead{$\log(\Sigma_\mathrm{1 kpc})$} \\ \colhead{} & \colhead{[kpc]} & \colhead{[$\log(M_\odot/\mathrm{kpc^2})$]}
    }
    \startdata
        $10.34\pm_{0.07}^{0.06}$ & $0.72 \ \pm_{0.11}^{0.15}$ & $9.61 \pm_{0.10}^{0.08}$ \\
    \enddata
    \tablecomments{The total stellar mass, the mass weighted effective radius, and the central surface mass density of UNCOVER 18407 measured as described in Section \ref{subsec:dense}. The physical radii presented here are corrected for the magnification by dividing by $\sqrt{\mu}$.}
\end{deluxetable}

As the uncertainty in the mass to light ratio is significantly larger than the uncertainty in the 2D structure of the two components of the galaxy, we then obtain a posterior in the intrinsic mass profile of the galaxy by multiplying the intrinsic 1D light profile (again measured using the best fitting compact ellipses of the central component applied to the best fitting two-component \sersic model) with the posterior in the measured intrinsic M/L profiles, after fixing the M/L within the central pixel to the value at 1 pixel (0.04") and the M/L outside the largest radius to the value in the outer annulus following \cite{Suess2019a}. We measure the total mass, the half mass radius, and the central surface mass density by integrating this new intrinsic mass profile. The results of this fitting are shown in Table \ref{tab:mass_weighted_structure}. 

In Figure \ref{fig:structures}, we show UNCOVER 18407 in mass-weighted size and $\Sigma_\mathrm{1 kpc}$ (as measured from the intrinsic mass profile after accounting for lensing) versus stellar mass as compared to star forming (blue) and quiescent (red) galaxies at z=2-2.5 as measured using 3D-HST images in \cite{Suess2019a}. It is clear from both panels that the structure of UNCOVER 18407 is comparably dense to $z\sim2$ quiescent systems, and has a mass distribution that is more compact than typical star forming galaxies at similar mass. Since the bulk of the stars in UNCOVER 18407 formed in the first $\sim$Gyr of cosmic time, the physical process that is responsible for the fairly rapid shutdown of star formation is acting in the extremely early universe; intense and compact star formation at physical scales $<1$ kpc that results in a quiescent core that may later grow from the inside out via minor mergers \citep[e.g.,][]{Bezanson2009, Belli2023}. 

The early formation of a dense stellar core suggests that the centers of galaxies like UNCOVER 18407 could evolve directly into local massive elliptical galaxies, which have central stellar ages that are consistent with such early formation \citep{Trager2000, Thomas2005,Sanchez-Blazquez2006b, Graves2007, Greene2013}, although the system would need to grow by $\sim$ an order of magnitude in total mass over the next 10 Gyr of cosmic time to resemble massive local elliptical galaxies. This kind of mass growth appears to be quite plausible based on cosmological simulations such as IllustrisTNG, where many of the earliest-forming quiescent systems at $z\sim4$ eventually merge with an even more massive companion and evolve into brightest cluster galaxies \citep{Hartley2023}.

Such early forming compact systems are suggested in cosmological simulations to be the products of secular, early dense star formation that results in a dense core with slightly older stellar populations in the center than the outskirts \citep[e.g.,][]{Wellons2015}. While the uncertainty in mass weighted age is too large to conclusively constrain radial age gradients, the observed SEDs are not inconsistent with stellar ages that are on the order of a few 100 Myr younger in the outskirts than in the dense, central region, suggesting that the quenching of star formation could be propagating outward from the center of the galaxy as suggested in \cite{Tacchella2015a}. 

The combination of the compact mass distribution and significant dust extinction in the central regions of UNCOVER 18407 is suggestive of the galaxy's progenitor as a compact, dusty star forming galaxy in the early universe. The early formation time and high mass of UNCOVER 18407 suggests that, if it were formed in a vacuum, it must have had sustained star formation at rates comparable to the most luminous $z\sim10$ galaxies \citep[e.g., ][]{Tacchella2023}. However, due to the hierarchical assembly of galaxies, it is not necessarily true (or likely) that UNCOVER 18407 was a single galaxy at $z\sim10$ when it was rapidly forming its stars \citep[e.g.,][]{Li2023}. That said, the star formation history and structural measurements of UNCOVER 18407 are consistent with it being the descendent of individual $z>7$ massive galaxy candidates that have been identified in early photometric samples; those systems exhibit extremely compact sizes at $\log(M_\star/M_\odot)\sim10$ \citep[][though their masses are quite uncertain, see \citealt{WangB2024_UB}]{Akins2023,Baggen2023}, and the formation of $\sim10^{10} M_\odot$ within a few hundred parsecs in the first Gyr of cosmic time is fully consistent with our measurements of UNCOVER 18407. 

Additionally, growing samples of massive, IR-luminous dusty star forming galaxies that may represent a key link between rapid star formation in the early universe and quiescence at $z\sim4$ have been spectroscopically confirmed at $z>5$ \citep{Williams2023, Nelson2023, McKinney2023,Smail2023}. The detailed study of the spatial scale of that star formation and the number densities of galaxies in the first Gyr of cosmic time and beyond \citep[e.g.,][]{Long2023} will allow for such populations to be linked to their dusty descendants such as UNCOVER 18407, as has been done with compact star forming galaxies near cosmic noon \citep[e.g.,][]{VanDokkum2015, Barro2017, Tadaki2020}.

\section{Conclusions} \label{sec:conclusions}

In this work, we present the spectrum and resolved photometry of UNCOVER 18407, a massive, lensed, extremely red galaxy at $z_{spec}=3.97$. We find the following:

\begin{itemize}
    \item The rest-frame 0.9 $\mu$m light distribution of UNCOVER 18407 is best modeled by a two-component \sersic fit, with a compact 200 parsec, $n\sim2.9$ core and a more extended 1.6 kpc, $n\sim1.7$ component. Annular photometry indicates that the galaxy exhibits a strong F150W-F444W color gradient that drops from 4.5 in the center to 3.0 in the outskirts (see Figure \ref{fig:sersic}).

    \item The weak (equivalent width $<21$ \AA) observed $H\alpha$ in UNCOVER 18407 confirms the quiescent $UVJ$ solutions. (see Figure \ref{fig:halpha}). While deblending $H\alpha$ and [NII] is difficult at the prism resolution, the 3$\sigma$ $H\alpha$ limit rules out star formation rates of 1 (13) $M_\odot \ \mathrm{yr^{-1}}$ under assumptions of $A_v$=0 (3) after flux calibrating the spectrum to the photometry of the central 1.2 kpc. This upper limit is corroborated by a FIR detection \citep[see][]{Fujimoto2023dualZ}, which constrains the star formation rate to be less than $\sim8.5$ $M_\odot \ \mathrm{yr^{-1}}$ at the 3$\sigma$ level.

    \item The observed central F150W-F444W and $UVJ$ colors of the galaxy can only be modeled by an old ($\sim$1 Gyr) stellar population with significant attenuation (see Figure \ref{fig:color_ssp}); the universe is simply not old enough at $z_{spec}=3.97$ for a maximally old dust-free solution to result in the observed colors, and star forming solutions violate the constraints on the star formation rate as measured from the weak $H\alpha$ emission. This implies that the core of UNCOVER 18407 is among the oldest stellar populations in the universe observed at such high redshift.
    
    \item Annular SED fits reveal that the color gradient in UNCOVER 18407 predominantly results from an $A_v$ gradient, with the central regions exhibiting a magnitude more extinction than the outskirts. Additionally, there are also hints of a negative age gradient, and we cannot entirely disentangle these trends due to age-dust degeneracy. The color gradients manifest in a mass-to-light gradient, suggesting that there is $\sim1.5\times$ as much mass-per-unit-light in the central $\sim$kpc as there is in the less-obscured outer regions (see Figure \ref{fig:radial}).

    \item The integrated star formation history of UNCOVER 18407 is consistent with formation of $\sim10^{10.4} M_\odot$ within the first Gyr of cosmic time. This object, in conjunction with the emerging JWST quiescent literature \citep[e.g.,][see Figure \ref{fig:mass_zform}]{Carnall2023b,Nanayakkara2024, Glazebrook2024}, suggests that old and dusty quiescent systems may be more common at high-redshift than previously anticipated.

    \item The mass-weighted size and surface density of UNCOVER 18407 are comparable to quiescent galaxies at cosmic noon, suggesting that it represents an early example of quenching after a period of compact star formation (see Figure \ref{fig:structures}). This also suggests that the growth in size of quiescent galaxies at fixed stellar mass as a function of redshift \citep[e.g.,][]{VanDerWel2014} is dominated by growth at $z<2$.
\end{itemize}

The discovery of this dense, old, and dusty massive quiescent galaxy at $z_{spec}=3.97$ opens questions as to how common such systems are. In order to connect UNCOVER 18407 to its progenitor population, we will need to perform searches over large volumes with well understood selection functions. While the precise redshift and the constraining power of the weak $H\alpha$ was useful for understanding the stellar populations in UNCOVER 18407, it is unlikely we will be able to build up well understood samples where all galaxies have spectroscopy due to the time intensive nature of performing pre-imaging followed up by targeted spectroscopy.

However, the fact that this source was identified as a potential dusty candidate in broadband imaging where the color gradient was already apparent is extremely promising for future searches. Future work can leverage the ability to select candidate quiescent galaxies in large-area imaging data to constrain how common strong color gradients are and what they can teach us about the physical processes that result in the formation of the first massive galaxies. In the coming years, a census of the quiescent population at $z>3$ will allow for a detailed study of this crucial phase of evolution for some of the earliest-forming galaxies in the universe.

\facilities{JWST (NIRCam, NIRSpec), ALMA}

\software{Astropy \citep{astropy2013, astropy2018, astropy2022}, \texttt{Dynesty} \citep{Speagle2020}, \texttt{grizli} \citep[\url{github.com/gbrammer/grizli}]{grizli},
Matplotlib \citep{Hunter:2007}, Flexible Stellar Population Synthesis \citep{Conroy2009, Conroy2010}, SEDPy \cite{sedpy2019}, Prospector \citep{Johnson2017, Leja2017, Johnson2021}, Pysersic \citep{Pasha2023}}

\acknowledgements

Support for this work was provided by The Brinson Foundation through a Brinson Prize Fellowship grant. This work is based in part on observations made with the NASA/ESA/CSA \textit{James Webb Space Telescope}. The data were obtained from the Mikulski Archive for Space Telescopes at the Space Telescope Science Institute, which is operated by the Association of Universities for Research in Astronomy, Inc., under NASA contract NAS 5-03127 for JWST. These observations are associated with JWST Cycle 1 GO program \#2561 and Cycle 3 GO program \#4111. The JWST data presented in this article were obtained from the Mikulski Archive for Space Telescopes (MAST) at the Space Telescope Science Institute. The specific observations analyzed can be accessed via \dataset[DOI]{https://doi.org/10.17909/qgpt-c913}. Support for program JWST-GO-2561 was provided by NASA through a grant from the Space Telescope Science Institute, which is operated by the Associations of Universities for Research in Astronomy, Incorporated, under NASA contract NAS5-26555. DS acknowledges helpful conversations with Desika Narayanan in the early stages of assessing how interesting this object is.

TBM was supported by a CIERA Postdoctoral Fellowship. JAD acknowledges funding from the Dunlap Institute for Astronomy \& Astrophysics. The Dunlap Institute is funded through an endowment established by the David Dunlap family and the University of Toronto. RP and DM acknowledge funding from JWST-GO-2561. HA is supported by CNES, focused on the JWST mission. H.A. also acknowledges support from the Programme National Cosmology and Galaxies (PNCG) of CNRS/INSU with INP and IN2P3, co-funded by CEA and CNES. PD acknowledge support from the NWO grant 016.VIDI.189.162 (``ODIN") and warmly thanks the European Commission's and University of Groningen's CO-FUND Rosalind Franklin program. VK acknowledges funding from the Dutch Research Council (NWO) through the award of the Vici Grant VI.C.212.036. The work of CCW is supported by NOIRLab, which is managed by the Association of Universities for Research in Astronomy (AURA) under a cooperative agreement with the National Science Foundation. K.G. and T.N. acknowledge support from Australian Research Council Laureate Fellowship FL180100060. S.F. acknowledges the support from NASA through the NASA Hubble Fellowship grant HST-HF2-51505.001-A awarded by the Space Telescope Science Institute, which is operated by the Association of Universities for Research in Astronomy, Incorporated, under NASA contract NAS5-26555. Some of the data products presented herein were retrieved from the Dawn JWST Archive (DJA). DJA is an initiative of the Cosmic Dawn Center, which is funded by the Danish National Research Foundation under grant No. 140.

\bibliography{all}

\begin{thebibliography}{}
\expandafter\ifx\csname natexlab\endcsname\relax\def\natexlab#1{#1}\fi
\providecommand{\url}[1]{\href{#1}{#1}}
\providecommand{\dodoi}[1]{doi:~\href{http://doi.org/#1}{\nolinkurl{#1}}}
\providecommand{\doeprint}[1]{\href{http://ascl.net/#1}{\nolinkurl{http://ascl.net/#1}}}
\providecommand{\doarXiv}[1]{\href{https://arxiv.org/abs/#1}{\nolinkurl{https://arxiv.org/abs/#1}}}

\bibitem[{{Akhshik} {et~al.}(2020){Akhshik}, {Whitaker}, {Brammer}, {Mahler}, {Sharon}, {Leja}, {Bayliss}, {Bezanson}, {Gladders}, {Man}, {Nelson}, {Rigby}, {Rizzo}, {Toft}, {Wellons}, \& {Williams}}]{Akhshik2020}
{Akhshik}, M., {Whitaker}, K.~E., {Brammer}, G., {et~al.} 2020, arXiv e-prints, arXiv:2008.02276.
\newblock \doarXiv{2008.02276}

\bibitem[{{Akhshik} {et~al.}(2021){Akhshik}, {Whitaker}, {Leja}, {Mahler}, {Sharon}, {Brammer}, {Toft}, {Bezanson}, {Man}, {Nelson}, {Pacifici}, {Wellons}, \& {Williams}}]{Akhshik2021}
{Akhshik}, M., {Whitaker}, K.~E., {Leja}, J., {et~al.} 2021, \apjl, 907, L8, \dodoi{10.3847/2041-8213/abd416}

\bibitem[{{Akhshik} {et~al.}(2023){Akhshik}, {Whitaker}, {Leja}, {Richard}, {Spilker}, {Song}, {Brammer}, {Bezanson}, {Ebeling}, {Gallazzi}, {Mahler}, {Mowla}, {Nelson}, {Pacifici}, {Sharon}, {Toft}, {Williams}, {Wright}, \& {Zabl}}]{Akhshik2023}
---. 2023, \apj, 943, 179, \dodoi{10.3847/1538-4357/aca677}

\bibitem[{{Akins} {et~al.}(2023){Akins}, {Casey}, {Allen}, {Bagley}, {Dickinson}, {Finkelstein}, {Franco}, {Harish}, {Arrabal Haro}, {Ilbert}, {Kartaltepe}, {Koekemoer}, {Liu}, {Long}, {McCracken}, {Paquereau}, {Papovich}, {Pirzkal}, {Rhodes}, {Robertson}, {Shuntov}, {Toft}, {Yang}, {Barro}, {Bisigello}, {Buat}, {Champagne}, {Cooper}, {Costantin}, {de La Vega}, {Drakos}, {Faisst}, {Fontana}, {Fujimoto}, {Gillman}, {G{\'o}mez-Guijarro}, {Gozaliasl}, {Hathi}, {Hayward}, {Hirschmann}, {Holwerda}, {Jin}, {Kocevski}, {Kokorev}, {Lambrides}, {Lucas}, {Magdis}, {Magnelli}, {McKinney}, {Mobasher}, {P{\'e}rez-Gonz{\'a}lez}, {Rich}, {Seill{\'e}}, {Talia}, {Urry}, {Valentino}, {Whitaker}, {Yung}, {Zavala}, {Cosmos-Web Team}, \& {Ceers Team}}]{Akins2023}
{Akins}, H.~B., {Casey}, C.~M., {Allen}, N., {et~al.} 2023, \apj, 956, 61, \dodoi{10.3847/1538-4357/acef21}

\bibitem[{{Alberts} {et~al.}(2023){Alberts}, {Williams}, {Helton}, {Suess}, {Ji}, {Shivaei}, {Lyu}, {Rieke}, {Baker}, {Bonaventura}, {Bunker}, {Carniani}, {Charlot}, {Curtis-Lake}, {D'Eugenio}, {Eisenstein}, {de Graaff}, {Hainline}, {Hausen}, {Johnson}, {Maiolino}, {Parlanti}, {Rieke}, {Robertson}, {Sun}, {Tacchella}, {Willmer}, \& {Willott}}]{Alberts2023}
{Alberts}, S., {Williams}, C.~C., {Helton}, J.~M., {et~al.} 2023, arXiv e-prints, arXiv:2312.12207.
\newblock \doarXiv{2312.12207}

\bibitem[{{Antwi-Danso} {et~al.}(2023{\natexlab{a}}){Antwi-Danso}, {Papovich}, {Esdaile}, {Nanayakkara}, {Glazebrook}, {Hutchison}, {Whitaker}, {Marsan}, {Diaz}, {Marchesini}, {Muzzin}, {Tran}, {Setton}, {Kaushal}, {Speagle}, \& {Cole}}]{Antwi-Danso2023b}
{Antwi-Danso}, J., {Papovich}, C., {Esdaile}, J., {et~al.} 2023{\natexlab{a}}, arXiv e-prints, arXiv:2307.09590, \dodoi{10.48550/arXiv.2307.09590}

\bibitem[{{Antwi-Danso} {et~al.}(2023{\natexlab{b}}){Antwi-Danso}, {Papovich}, {Leja}, {Marchesini}, {Marsan}, {Martis}, {Labb{\'e}}, {Muzzin}, {Glazebrook}, {Straatman}, \& {Tran}}]{Antwi-Danso2023a}
{Antwi-Danso}, J., {Papovich}, C., {Leja}, J., {et~al.} 2023{\natexlab{b}}, \apj, 943, 166, \dodoi{10.3847/1538-4357/aca294}

\bibitem[{{Astropy Collaboration} {et~al.}(2013){Astropy Collaboration}, {Robitaille}, {Tollerud}, {Greenfield}, {Droettboom}, {Bray}, {Aldcroft}, {Davis}, {Ginsburg}, {Price-Whelan}, {Kerzendorf}, {Conley}, {Crighton}, {Barbary}, {Muna}, {Ferguson}, {Grollier}, {Parikh}, {Nair}, {Unther}, {Deil}, {Woillez}, {Conseil}, {Kramer}, {Turner}, {Singer}, {Fox}, {Weaver}, {Zabalza}, {Edwards}, {Azalee Bostroem}, {Burke}, {Casey}, {Crawford}, {Dencheva}, {Ely}, {Jenness}, {Labrie}, {Lim}, {Pierfederici}, {Pontzen}, {Ptak}, {Refsdal}, {Servillat}, \& {Streicher}}]{astropy2013}
{Astropy Collaboration}, {Robitaille}, T.~P., {Tollerud}, E.~J., {et~al.} 2013, \aap, 558, A33, \dodoi{10.1051/0004-6361/201322068}

\bibitem[{{Astropy Collaboration} {et~al.}(2018){Astropy Collaboration}, {Price-Whelan}, {Sip{\H{o}}cz}, {G{\"u}nther}, {Lim}, {Crawford}, {Conseil}, {Shupe}, {Craig}, {Dencheva}, {Ginsburg}, {Vand erPlas}, {Bradley}, {P{\'e}rez-Su{\'a}rez}, {de Val-Borro}, {Aldcroft}, {Cruz}, {Robitaille}, {Tollerud}, {Ardelean}, {Babej}, {Bach}, {Bachetti}, {Bakanov}, {Bamford}, {Barentsen}, {Barmby}, {Baumbach}, {Berry}, {Biscani}, {Boquien}, {Bostroem}, {Bouma}, {Brammer}, {Bray}, {Breytenbach}, {Buddelmeijer}, {Burke}, {Calderone}, {Cano Rodr{\'\i}guez}, {Cara}, {Cardoso}, {Cheedella}, {Copin}, {Corrales}, {Crichton}, {D'Avella}, {Deil}, {Depagne}, {Dietrich}, {Donath}, {Droettboom}, {Earl}, {Erben}, {Fabbro}, {Ferreira}, {Finethy}, {Fox}, {Garrison}, {Gibbons}, {Goldstein}, {Gommers}, {Greco}, {Greenfield}, {Groener}, {Grollier}, {Hagen}, {Hirst}, {Homeier}, {Horton}, {Hosseinzadeh}, {Hu}, {Hunkeler}, {Ivezi{\'c}}, {Jain}, {Jenness}, {Kanarek}, {Kendrew}, {Kern}, {Kerzendorf}, {Khvalko}, {King}, {Kirkby}, {Kulkarni},
  {Kumar}, {Lee}, {Lenz}, {Littlefair}, {Ma}, {Macleod}, {Mastropietro}, {McCully}, {Montagnac}, {Morris}, {Mueller}, {Mumford}, {Muna}, {Murphy}, {Nelson}, {Nguyen}, {Ninan}, {N{\"o}the}, {Ogaz}, {Oh}, {Parejko}, {Parley}, {Pascual}, {Patil}, {Patil}, {Plunkett}, {Prochaska}, {Rastogi}, {Reddy Janga}, {Sabater}, {Sakurikar}, {Seifert}, {Sherbert}, {Sherwood-Taylor}, {Shih}, {Sick}, {Silbiger}, {Singanamalla}, {Singer}, {Sladen}, {Sooley}, {Sornarajah}, {Streicher}, {Teuben}, {Thomas}, {Tremblay}, {Turner}, {Terr{\'o}n}, {van Kerkwijk}, {de la Vega}, {Watkins}, {Weaver}, {Whitmore}, {Woillez}, {Zabalza}, \& {Astropy Contributors}}]{astropy2018}
{Astropy Collaboration}, {Price-Whelan}, A.~M., {Sip{\H{o}}cz}, B.~M., {et~al.} 2018, \aj, 156, 123, \dodoi{10.3847/1538-3881/aabc4f}

\bibitem[{{Astropy Collaboration} {et~al.}(2022){Astropy Collaboration}, {Price-Whelan}, {Lim}, {Earl}, {Starkman}, {Bradley}, {Shupe}, {Patil}, {Corrales}, {Brasseur}, {N{\"o}the}, {Donath}, {Tollerud}, {Morris}, {Ginsburg}, {Vaher}, {Weaver}, {Tocknell}, {Jamieson}, {van Kerkwijk}, {Robitaille}, {Merry}, {Bachetti}, {G{\"u}nther}, {Aldcroft}, {Alvarado-Montes}, {Archibald}, {B{\'o}di}, {Bapat}, {Barentsen}, {Baz{\'a}n}, {Biswas}, {Boquien}, {Burke}, {Cara}, {Cara}, {Conroy}, {Conseil}, {Craig}, {Cross}, {Cruz}, {D'Eugenio}, {Dencheva}, {Devillepoix}, {Dietrich}, {Eigenbrot}, {Erben}, {Ferreira}, {Foreman-Mackey}, {Fox}, {Freij}, {Garg}, {Geda}, {Glattly}, {Gondhalekar}, {Gordon}, {Grant}, {Greenfield}, {Groener}, {Guest}, {Gurovich}, {Handberg}, {Hart}, {Hatfield-Dodds}, {Homeier}, {Hosseinzadeh}, {Jenness}, {Jones}, {Joseph}, {Kalmbach}, {Karamehmetoglu}, {Ka{\l}uszy{\'n}ski}, {Kelley}, {Kern}, {Kerzendorf}, {Koch}, {Kulumani}, {Lee}, {Ly}, {Ma}, {MacBride}, {Maljaars}, {Muna}, {Murphy}, {Norman},
  {O'Steen}, {Oman}, {Pacifici}, {Pascual}, {Pascual-Granado}, {Patil}, {Perren}, {Pickering}, {Rastogi}, {Roulston}, {Ryan}, {Rykoff}, {Sabater}, {Sakurikar}, {Salgado}, {Sanghi}, {Saunders}, {Savchenko}, {Schwardt}, {Seifert-Eckert}, {Shih}, {Jain}, {Shukla}, {Sick}, {Simpson}, {Singanamalla}, {Singer}, {Singhal}, {Sinha}, {Sip{\H{o}}cz}, {Spitler}, {Stansby}, {Streicher}, {{\v{S}}umak}, {Swinbank}, {Taranu}, {Tewary}, {Tremblay}, {de Val-Borro}, {Van Kooten}, {Vasovi{\'c}}, {Verma}, {de Miranda Cardoso}, {Williams}, {Wilson}, {Winkel}, {Wood-Vasey}, {Xue}, {Yoachim}, {Zhang}, {Zonca}, \& {Astropy Project Contributors}}]{astropy2022}
{Astropy Collaboration}, {Price-Whelan}, A.~M., {Lim}, P.~L., {et~al.} 2022, \apj, 935, 167, \dodoi{10.3847/1538-4357/ac7c74}

\bibitem[{{Baggen} {et~al.}(2023){Baggen}, {van Dokkum}, {Labb{\'e}}, {Brammer}, {Miller}, {Bezanson}, {Leja}, {Wang}, {Whitaker}, {Suess}, \& {Nelson}}]{Baggen2023}
{Baggen}, J. F.~W., {van Dokkum}, P., {Labb{\'e}}, I., {et~al.} 2023, \apjl, 955, L12, \dodoi{10.3847/2041-8213/acf5ef}

\bibitem[{{Baker} {et~al.}(2023){Baker}, {Tacchella}, {Johnson}, {Nelson}, {Suess}, {D'Eugenio}, {Curti}, {de Graaff}, {Ji}, {Maiolino}, {Robertson}, {Scholtz}, {Alberts}, {Arribas}, {Boyett}, {Bunker}, {Carniani}, {Charlot}, {Chen}, {Chevallard}, {Curtis-Lake}, {Danhaive}, {DeCoursey}, {Egami}, {Eisenstein}, {Endsley}, {Hausen}, {Helton}, {Kumari}, {Looser}, {Maseda}, {Pusk{\'a}s}, {Rieke}, {Sandles}, {Sun}, {{\"U}bler}, {Williams}, {Willmer}, \& {Witstok}}]{Baker2023}
{Baker}, W.~M., {Tacchella}, S., {Johnson}, B.~D., {et~al.} 2023, arXiv e-prints, arXiv:2306.02472, \dodoi{10.48550/arXiv.2306.02472}

\bibitem[{{Barro} {et~al.}(2017){Barro}, {Faber}, {Koo}, {Dekel}, {Fang}, {Trump}, {P{\'e}rez-Gonz{\'a}lez}, {Pacifici}, {Primack}, {Somerville}, {Yan}, {Guo}, {Liu}, {Ceverino}, {Kocevski}, \& {McGrath}}]{Barro2017}
{Barro}, G., {Faber}, S.~M., {Koo}, D.~C., {et~al.} 2017, \apj, 840, 47, \dodoi{10.3847/1538-4357/aa6b05}

\bibitem[{{Barrufet} {et~al.}(2023){Barrufet}, {Oesch}, {Weibel}, {Brammer}, {Bezanson}, {Bouwens}, {Fudamoto}, {Gonzalez}, {Gottumukkala}, {Illingworth}, {Heintz}, {Holden}, {Labbe}, {Magee}, {Naidu}, {Nelson}, {Stefanon}, {Smit}, {van Dokkum}, {Weaver}, \& {Williams}}]{Barrufet2023}
{Barrufet}, L., {Oesch}, P.~A., {Weibel}, A., {et~al.} 2023, \mnras, 522, 449, \dodoi{10.1093/mnras/stad947}

\bibitem[{{Barrufet} {et~al.}(2024){Barrufet}, {Oesch}, {Marques-Chaves}, {Arellano-Cordova}, {Baggen}, {Carnall}, {Cullen}, {Dunlop}, {Gottumukkala}, {Fudamoto}, {Illingworth}, {Magee}, {McLure}, {McLeod}, {Micha{\l}owski}, {Stefanon}, {van Dokkum}, \& {Weibel}}]{Barrufet2024}
{Barrufet}, L., {Oesch}, P., {Marques-Chaves}, R., {et~al.} 2024, arXiv e-prints, arXiv:2404.08052, \dodoi{10.48550/arXiv.2404.08052}

\bibitem[{{Belli} {et~al.}(2019){Belli}, {Newman}, \& {Ellis}}]{Belli2019}
{Belli}, S., {Newman}, A.~B., \& {Ellis}, R.~S. 2019, \apj, 874, 17, \dodoi{10.3847/1538-4357/ab07af}

\bibitem[{{Belli} {et~al.}(2021){Belli}, {Contursi}, {Genzel}, {Tacconi}, {F{\"o}rster-Schreiber}, {Lutz}, {Combes}, {Neri}, {Garc{\'\i}a-Burillo}, {Schuster}, {Herrera-Camus}, {Tadaki}, {Davies}, {Davies}, {Johnson}, {Lee}, {Leja}, {Nelson}, {Price}, {Shangguan}, {Shimizu}, {Tacchella}, \& {{\"U}bler}}]{Belli2021}
{Belli}, S., {Contursi}, A., {Genzel}, R., {et~al.} 2021, \apjl, 909, L11, \dodoi{10.3847/2041-8213/abe6a6}

\bibitem[{{Belli} {et~al.}(2023){Belli}, {Park}, {Davies}, {Mendel}, {Johnson}, {Conroy}, {Benton}, {Bugiani}, {Emami}, {Leja}, {Li}, {Maheson}, {Mathews}, {Naidu}, {Nelson}, {Tacchella}, {Terrazas}, \& {Weinberger}}]{Belli2023}
{Belli}, S., {Park}, M., {Davies}, R.~L., {et~al.} 2023, arXiv e-prints, arXiv:2308.05795, \dodoi{10.48550/arXiv.2308.05795}

\bibitem[{{Beverage} {et~al.}(2023){Beverage}, {Kriek}, {Suess}, {Conroy}, {Price}, {Barro}, {Bezanson}, {Franx}, {Lorenz}, {Ma}, {Mowla}, {Pasha}, {van Dokkum}, \& {Weisz}}]{Beverage2023}
{Beverage}, A.~G., {Kriek}, M., {Suess}, K.~A., {et~al.} 2023, arXiv e-prints, arXiv:2312.05307, \dodoi{10.48550/arXiv.2312.05307}

\bibitem[{{Bezanson} {et~al.}(2009){Bezanson}, {van Dokkum}, {Tal}, {Marchesini}, {Kriek}, {Franx}, \& {Coppi}}]{Bezanson2009}
{Bezanson}, R., {van Dokkum}, P.~G., {Tal}, T., {et~al.} 2009, \apj, 697, 1290, \dodoi{10.1088/0004-637X/697/2/1290}

\bibitem[{{Bezanson} {et~al.}(2022{\natexlab{a}}){Bezanson}, {Labbe}, {Whitaker}, {Leja}, {Price}, {Franx}, {Brammer}, {Marchesini}, {Zitrin}, {Wang}, {Weaver}, {Furtak}, {Atek}, {Coe}, {Cutler}, {Dayal}, {van Dokkum}, {Feldmann}, {Forster Schreiber}, {Fujimoto}, {Geha}, {Glazebrook}, {de Graaff}, {Greene}, {Juneau}, {Kassin}, {Kriek}, {Khullar}, {Maseda}, {Mowla}, {Muzzin}, {Nanayakkara}, {Nelson}, {Oesch}, {Pacifici}, {Pan}, {Papovich}, {Setton}, {Shapley}, {Smit}, {Stefanon}, {Taylor}, \& {Williams}}]{Bezanson2022b}
{Bezanson}, R., {Labbe}, I., {Whitaker}, K.~E., {et~al.} 2022{\natexlab{a}}, arXiv e-prints, arXiv:2212.04026, \dodoi{10.48550/arXiv.2212.04026}

\bibitem[{{Bezanson} {et~al.}(2022{\natexlab{b}}){Bezanson}, {Spilker}, {Suess}, {Setton}, {Feldmann}, {Greene}, {Kriek}, {Narayanan}, \& {Verrico}}]{Bezanson2022a}
{Bezanson}, R., {Spilker}, J.~S., {Suess}, K.~A., {et~al.} 2022{\natexlab{b}}, \apj, 925, 153, \dodoi{10.3847/1538-4357/ac3dfa}

\bibitem[{{Boizelle} {et~al.}(2017){Boizelle}, {Barth}, {Darling}, {Baker}, {Buote}, {Ho}, \& {Walsh}}]{Boizelle2017}
{Boizelle}, B.~D., {Barth}, A.~J., {Darling}, J., {et~al.} 2017, \apj, 845, 170, \dodoi{10.3847/1538-4357/aa8266}

\bibitem[{{Boyett} {et~al.}(2023){Boyett}, {Trenti}, {Leethochawalit}, {Calabr{\'o}}, {Metha}, {Roberts-Borsani}, {Dalmasso}, {Yang}, {Santini}, {Treu}, {Jones}, {Henry}, {Mason}, {Morishita}, {Nanayakkara}, {Roy}, {Wang}, {Fontana}, {Merlin}, {Castellano}, {Paris}, {Bradac}, {Marchesini}, {Mascia}, {Pentericci}, {Vanzella}, \& {Vulcani}}]{Boyett2023}
{Boyett}, K., {Trenti}, M., {Leethochawalit}, N., {et~al.} 2023, arXiv e-prints, arXiv:2303.00306, \dodoi{10.48550/arXiv.2303.00306}

\bibitem[{Brammer(2023)}]{grizli}
Brammer, G. 2023, grizli, 1.9.11,  Zenodo, \dodoi{10.5281/zenodo.8370018}

\bibitem[{{Brammer} {et~al.}(2022){Brammer}, {Strait}, {Matharu}, \& {Momcheva}}]{Brammer2022}
{Brammer}, G., {Strait}, V., {Matharu}, J., \& {Momcheva}, I. 2022, {grizli}, 1.5.0,  Zenodo, \dodoi{10.5281/zenodo.6672538}

\bibitem[{{Caliendo} {et~al.}(2021){Caliendo}, {Whitaker}, {Akhshik}, {Wilson}, {Williams}, {Spilker}, {Mahler}, {Pope}, {Sharon}, {Aguilar}, {Bezanson}, {Chavez Dagostino}, {G{\'o}mez-Ruiz}, {Monta{\~n}a}, {Toft}, {Velazquez de la Rosa}, \& {Zeballos}}]{Caliendo2021}
{Caliendo}, J.~N., {Whitaker}, K.~E., {Akhshik}, M., {et~al.} 2021, \apjl, 910, L7, \dodoi{10.3847/2041-8213/abe132}

\bibitem[{{Cappellari} {et~al.}(2013){Cappellari}, {McDermid}, {Alatalo}, {Blitz}, {Bois}, {Bournaud}, {Bureau}, {Crocker}, {Davies}, {Davis}, {de Zeeuw}, {Duc}, {Emsellem}, {Khochfar}, {Krajnovi{\'c}}, {Kuntschner}, {Morganti}, {Naab}, {Oosterloo}, {Sarzi}, {Scott}, {Serra}, {Weijmans}, \& {Young}}]{Cappellari2013}
{Cappellari}, M., {McDermid}, R.~M., {Alatalo}, K., {et~al.} 2013, \mnras, 432, 1862, \dodoi{10.1093/mnras/stt644}

\bibitem[{{Carnall} {et~al.}(2023{\natexlab{a}}){Carnall}, {McLeod}, {McLure}, {Dunlop}, {Begley}, {Cullen}, {Donnan}, {Hamadouche}, {Jewell}, {Jones}, {Pollock}, \& {Wild}}]{Carnall2023a}
{Carnall}, A.~C., {McLeod}, D.~J., {McLure}, R.~J., {et~al.} 2023{\natexlab{a}}, \mnras, 520, 3974, \dodoi{10.1093/mnras/stad369}

\bibitem[{{Carnall} {et~al.}(2023{\natexlab{b}}){Carnall}, {McLure}, {Dunlop}, {McLeod}, {Wild}, {Cullen}, {Magee}, {Begley}, {Cimatti}, {Donnan}, {Hamadouche}, {Jewell}, \& {Walker}}]{Carnall2023b}
{Carnall}, A.~C., {McLure}, R.~J., {Dunlop}, J.~S., {et~al.} 2023{\natexlab{b}}, arXiv e-prints, arXiv:2301.11413, \dodoi{10.48550/arXiv.2301.11413}

\bibitem[{Carnall {et~al.}(2024)Carnall, Cullen, McLure, McLeod, Begley, Donnan, Dunlop, Shapley, Rowlands, Almaini, Arellano-Córdova, Barrufet, Cimatti, Ellis, Grogin, Hamadouche, Illingworth, Koekemoer, Leung, Lovell, Pérez-González, Santini, Stanton, \& Wild}]{Carnall2024}
Carnall, A.~C., Cullen, F., McLure, R.~J., {et~al.} 2024, The JWST EXCELS survey: Too much, too young, too fast? Ultra-massive quiescent galaxies at 3 < z < 5.
\newblock \doarXiv{2405.02242}

\bibitem[{{Chabrier}(2003)}]{Chabrier2003}
{Chabrier}, G. 2003, \pasp, 115, 763, \dodoi{10.1086/376392}

\bibitem[{{Choi} {et~al.}(2016){Choi}, {Dotter}, {Conroy}, {Cantiello}, {Paxton}, \& {Johnson}}]{Choi2016}
{Choi}, J., {Dotter}, A., {Conroy}, C., {et~al.} 2016, \apj, 823, 102, \dodoi{10.3847/0004-637X/823/2/102}

\bibitem[{{Chworowsky} {et~al.}(2023){Chworowsky}, {Finkelstein}, {Spilker}, {Leung}, {Bagley}, {Casey}, {Gronwall}, {Jogee}, {Larson}, {Papovich}, {Somerville}, {Stevans}, {Wold}, \& {Yung}}]{Chworowsky2023}
{Chworowsky}, K., {Finkelstein}, S.~L., {Spilker}, J.~S., {et~al.} 2023, \apj, 951, 49, \dodoi{10.3847/1538-4357/acd1e3}

\bibitem[{{Cochrane} {et~al.}(2022){Cochrane}, {Hayward}, \& {Angl{\'e}s-Alc{\'a}zar}}]{Cochrane2022}
{Cochrane}, R.~K., {Hayward}, C.~C., \& {Angl{\'e}s-Alc{\'a}zar}, D. 2022, \apjl, 939, L27, \dodoi{10.3847/2041-8213/ac951d}

\bibitem[{{Conroy}(2013)}]{Conroy2013}
{Conroy}, C. 2013, \araa, 51, 393, \dodoi{10.1146/annurev-astro-082812-141017}

\bibitem[{{Conroy} \& {Gunn}(2010)}]{Conroy2010}
{Conroy}, C., \& {Gunn}, J.~E. 2010, \apj, 712, 833, \dodoi{10.1088/0004-637X/712/2/833}

\bibitem[{{Conroy} {et~al.}(2009){Conroy}, {Gunn}, \& {White}}]{Conroy2009}
{Conroy}, C., {Gunn}, J.~E., \& {White}, M. 2009, \apj, 699, 486, \dodoi{10.1088/0004-637X/699/1/486}

\bibitem[{{Costantin} {et~al.}(2021){Costantin}, {P{\'e}rez-Gonz{\'a}lez}, {M{\'e}ndez-Abreu}, {Huertas-Company}, {Dimauro}, {Alcalde-Pampliega}, {Buitrago}, {Ceverino}, {Daddi}, {Dom{\'\i}nguez-S{\'a}nchez}, {Espino-Briones}, {Hern{\'a}n-Caballero}, {Koekemoer}, \& {Rodighiero}}]{Costantin2021}
{Costantin}, L., {P{\'e}rez-Gonz{\'a}lez}, P.~G., {M{\'e}ndez-Abreu}, J., {et~al.} 2021, \apj, 913, 125, \dodoi{10.3847/1538-4357/abef72}

\bibitem[{{Costantin} {et~al.}(2022){Costantin}, {P{\'e}rez-Gonz{\'a}lez}, {M{\'e}ndez-Abreu}, {Huertas-Company}, {Pampliega}, {Balcells}, {Barro}, {Ceverino}, {Dimauro}, {S{\'a}nchez}, {Espino-Briones}, \& {Koekemoer}}]{Costantin2022}
---. 2022, \apj, 929, 121, \dodoi{10.3847/1538-4357/ac5a57}

\bibitem[{{Davis} {et~al.}(2013){Davis}, {Alatalo}, {Bureau}, {Cappellari}, {Scott}, {Young}, {Blitz}, {Crocker}, {Bayet}, {Bois}, {Bournaud}, {Davies}, {de Zeeuw}, {Duc}, {Emsellem}, {Khochfar}, {Krajnovi{\'c}}, {Kuntschner}, {Lablanche}, {McDermid}, {Morganti}, {Naab}, {Oosterloo}, {Sarzi}, {Serra}, \& {Weijmans}}]{Davis2013}
{Davis}, T.~A., {Alatalo}, K., {Bureau}, M., {et~al.} 2013, \mnras, 429, 534, \dodoi{10.1093/mnras/sts353}

\bibitem[{De~Cao {et~al.}(2020)De~Cao, Aziz, \& Titov}]{DeCao2020}
De~Cao, N., Aziz, W., \& Titov, I. 2020, in Uncertainty in artificial intelligence, PMLR, 1263--1273

\bibitem[{{de Graaff} {et~al.}(2023){de Graaff}, {Rix}, {Carniani}, {Suess}, {Charlot}, {Curtis-Lake}, {Arribas}, {Baker}, {Boyett}, {Bunker}, {Cameron}, {Chevallard}, {Curti}, {Eisenstein}, {Franx}, {Hainline}, {Hausen}, {Ji}, {Johnson}, {Jones}, {Maiolino}, {Maseda}, {Nelson}, {Parlanti}, {Rawle}, {Robertson}, {Tacchella}, {{\"U}bler}, {Williams}, {Willmer}, \& {Willott}}]{deGraaff2023}
{de Graaff}, A., {Rix}, H.-W., {Carniani}, S., {et~al.} 2023, arXiv e-prints, arXiv:2308.09742, \dodoi{10.48550/arXiv.2308.09742}

\bibitem[{{de Graaff} {et~al.}(2024){de Graaff}, {Setton}, {Brammer}, {Cutler}, {Suess}, {Labbe}, {Leja}, {Weibel}, {Maseda}, {Whitaker}, {Bezanson}, {Boogaard}, {Cleri}, {De Lucia}, {Franx}, {Greene}, {Hirschmann}, {Matthee}, {McConachie}, {Naidu}, {Oesch}, {Price}, {Rix}, {Valentino}, {Wang}, \& {Williams}}]{deGraaff2024}
{de Graaff}, A., {Setton}, D.~J., {Brammer}, G., {et~al.} 2024, arXiv e-prints, arXiv:2404.05683, \dodoi{10.48550/arXiv.2404.05683}

\bibitem[{{Dekel} \& {Burkert}(2014)}]{Dekel2014}
{Dekel}, A., \& {Burkert}, A. 2014, \mnras, 438, 1870, \dodoi{10.1093/mnras/stt2331}

\bibitem[{{D'Eugenio} {et~al.}(2020){D'Eugenio}, {Daddi}, {Gobat}, {Strazzullo}, {Lustig}, {Delvecchio}, {Jin}, {Puglisi}, {Calabr{\'o}}, {Mancini}, {Dickinson}, {Cimatti}, \& {Onodera}}]{DeugenioC2020}
{D'Eugenio}, C., {Daddi}, E., {Gobat}, R., {et~al.} 2020, \apjl, 892, L2, \dodoi{10.3847/2041-8213/ab7a96}

\bibitem[{{D'Eugenio} {et~al.}(2023){D'Eugenio}, {Perez-Gonzalez}, {Maiolino}, {Scholtz}, {Perna}, {Circosta}, {Uebler}, {Arribas}, {Boeker}, {Bunker}, {Carniani}, {Charlot}, {Chevallard}, {Cresci}, {Curtis-Lake}, {Jones}, {Kumari}, {Lamperti}, {Looser}, {Parlanti}, {Rix}, {Robertson}, {Rodriguez Del Pino}, {Tacchella}, {Venturi}, \& {Willott}}]{Deugenio2023}
{D'Eugenio}, F., {Perez-Gonzalez}, P., {Maiolino}, R., {et~al.} 2023, arXiv e-prints, arXiv:2308.06317, \dodoi{10.48550/arXiv.2308.06317}

\bibitem[{{Donevski} {et~al.}(2023){Donevski}, {Damjanov}, {Nanni}, {Man}, {Giulietti}, {Romano}, {Lapi}, {Narayanan}, {Dav{\'e}}, {Shivaei}, {Sohn}, {Junais}, {Pantoni}, \& {Li}}]{Donevski2023}
{Donevski}, D., {Damjanov}, I., {Nanni}, A., {et~al.} 2023, \aap, 678, A35, \dodoi{10.1051/0004-6361/202346066}

\bibitem[{{Dotter}(2016)}]{Dotter2016}
{Dotter}, A. 2016, \apjs, 222, 8, \dodoi{10.3847/0067-0049/222/1/8}

\bibitem[{{Draine} {et~al.}(2007){Draine}, {Dale}, {Bendo}, {Gordon}, {Smith}, {Armus}, {Engelbracht}, {Helou}, {Kennicutt}, {Li}, {Roussel}, {Walter}, {Calzetti}, {Moustakas}, {Murphy}, {Rieke}, {Bot}, {Hollenbach}, {Sheth}, \& {Teplitz}}]{Draine2007}
{Draine}, B.~T., {Dale}, D.~A., {Bendo}, G., {et~al.} 2007, \apj, 663, 866, \dodoi{10.1086/518306}

\bibitem[{{Ebneter} \& {Balick}(1985)}]{Ebneter1985}
{Ebneter}, K., \& {Balick}, B. 1985, \aj, 90, 183, \dodoi{10.1086/113724}

\bibitem[{{Ebneter} {et~al.}(1988){Ebneter}, {Djorgovski}, \& {Davis}}]{Ebneter1988}
{Ebneter}, K., {Djorgovski}, S., \& {Davis}, M. 1988, \aj, 95, 422, \dodoi{10.1086/114644}

\bibitem[{{Ellison} {et~al.}(2022){Ellison}, {Wilkinson}, {Woo}, {Leung}, {Wild}, {Bickley}, {Patton}, {Quai}, \& {Gwyn}}]{Ellison2022}
{Ellison}, S.~L., {Wilkinson}, S., {Woo}, J., {et~al.} 2022, arXiv e-prints, arXiv:2209.07613.
\newblock \doarXiv{2209.07613}

\bibitem[{{Fang} {et~al.}(2013){Fang}, {Faber}, {Koo}, \& {Dekel}}]{Fang2013}
{Fang}, J.~J., {Faber}, S.~M., {Koo}, D.~C., \& {Dekel}, A. 2013, \apj, 776, 63, \dodoi{10.1088/0004-637X/776/1/63}

\bibitem[{{Foreman-Mackey} {et~al.}(2013){Foreman-Mackey}, {Hogg}, {Lang}, \& {Goodman}}]{Foreman-Mackey2013}
{Foreman-Mackey}, D., {Hogg}, D.~W., {Lang}, D., \& {Goodman}, J. 2013, \pasp, 125, 306, \dodoi{10.1086/670067}

\bibitem[{{Forrest} {et~al.}(2020){Forrest}, {Marsan}, {Annunziatella}, {Wilson}, {Muzzin}, {Marchesini}, {Cooper}, {Chan}, {McConachie}, {Gomez}, {Kado-Fong}, {La Barbera}, {Lange-Vagle}, {Nantais}, {Nonino}, {Saracco}, {Stefanon}, \& {van der Burg}}]{Forrest2020b}
{Forrest}, B., {Marsan}, Z.~C., {Annunziatella}, M., {et~al.} 2020, \apj, 903, 47, \dodoi{10.3847/1538-4357/abb819}

\bibitem[{{French} {et~al.}(2015){French}, {Yang}, {Zabludoff}, {Narayanan}, {Shirley}, {Walter}, {Smith}, \& {Tremonti}}]{French2015}
{French}, K.~D., {Yang}, Y., {Zabludoff}, A., {et~al.} 2015, \apj, 801, 1, \dodoi{10.1088/0004-637X/801/1/1}

\bibitem[{{Fujimoto} {et~al.}(2023{\natexlab{a}}){Fujimoto}, {Kohno}, {Ouchi}, {Oguri}, {Kokorev}, {Brammer}, {Sun}, {Gonzalez-Lopez}, {Bauer}, {Caminha}, {Hatsukade}, {Richard}, {Smail}, {Tsujita}, {Ueda}, {Uematsu}, {Zitrin}, {Coe}, {Kneib}, {Postman}, {Umetsu}, {Lagos}, {Popping}, {Ao}, {Bradley}, {Caputi}, {Dessauges-Zavadsky}, {Egami}, {Espada}, {Ivison}, {Jauzac}, {Knudsen}, {Koekemoer}, {Magdis}, {Mahler}, {Munoz Arancibia}, {Rawle}, {Shimasaku}, {Toft}, {Umehata}, {Valentino}, {Wang}, \& {Wang}}]{Fujimoto2023a}
{Fujimoto}, S., {Kohno}, K., {Ouchi}, M., {et~al.} 2023{\natexlab{a}}, arXiv e-prints, arXiv:2303.01658, \dodoi{10.48550/arXiv.2303.01658}

\bibitem[{{Fujimoto} {et~al.}(2023{\natexlab{b}}){Fujimoto}, {Bezanson}, {Labbe}, {Brammer}, {Price}, {Wang}, {Weaver}, {Fudamoto}, {Oesch}, {Williams}, {Dayal}, {Feldmann}, {Greene}, {Leja}, {Whitaker}, {Zitrin}, {Cutler}, {Furtak}, {Pan}, {Chemerynska}, {Kokorev}, {Miller}, {Atek}, {van Dokkum}, {Juneau}, {Kassin}, {Khullar}, {Marchesini}, {Maseda}, {Nelson}, {Setton}, \& {Smit}}]{Fujimoto2023dualZ}
{Fujimoto}, S., {Bezanson}, R., {Labbe}, I., {et~al.} 2023{\natexlab{b}}, arXiv e-prints, arXiv:2309.07834, \dodoi{10.48550/arXiv.2309.07834}

\bibitem[{{Furtak} {et~al.}(2023){Furtak}, {Zitrin}, {Weaver}, {Atek}, {Bezanson}, {Labb{\'e}}, {Whitaker}, {Leja}, {Price}, {Brammer}, {Wang}, {Marchesini}, {Pan}, {Dayal}, {van Dokkum}, {Feldmann}, {Fujimoto}, {Franx}, {Khullar}, {Nelson}, \& {Mowla}}]{Furtak2023a}
{Furtak}, L.~J., {Zitrin}, A., {Weaver}, J.~R., {et~al.} 2023, \mnras, 523, 4568, \dodoi{10.1093/mnras/stad1627}

\bibitem[{{Glazebrook} {et~al.}(2017){Glazebrook}, {Schreiber}, {Labb{\'e}}, {Nanayakkara}, {Kacprzak}, {Oesch}, {Papovich}, {Spitler}, {Straatman}, {Tran}, \& {Yuan}}]{Glazebrook2017}
{Glazebrook}, K., {Schreiber}, C., {Labb{\'e}}, I., {et~al.} 2017, \nat, 544, 71, \dodoi{10.1038/nature21680}

\bibitem[{{Glazebrook} {et~al.}(2024){Glazebrook}, {Nanayakkara}, {Schreiber}, {Lagos}, {Kawinwanichakij}, {Jacobs}, {Chittenden}, {Brammer}, {Kacprzak}, {Labbe}, {Marchesini}, {Marsan}, {Oesch}, {Papovich}, {Remus}, {Tran}, {Esdaile}, \& {Chandro Gomez}}]{Glazebrook2024}
{Glazebrook}, K., {Nanayakkara}, T., {Schreiber}, C., {et~al.} 2024, Nature in press, arXiv:2308.05606, \dodoi{10.48550/arXiv.2308.05606}

\bibitem[{{Gobat} {et~al.}(2018){Gobat}, {Daddi}, {Magdis}, {Bournaud}, {Sargent}, {Martig}, {Jin}, {Finoguenov}, {B{\'e}thermin}, {Hwang}, {Renzini}, {Wilson}, {Aretxaga}, {Yun}, {Strazzullo}, \& {Valentino}}]{Gobat2018}
{Gobat}, R., {Daddi}, E., {Magdis}, G., {et~al.} 2018, Nature Astronomy, 2, 239, \dodoi{10.1038/s41550-017-0352-5}

\bibitem[{{Gould} {et~al.}(2023){Gould}, {Brammer}, {Valentino}, {Whitaker}, {Weaver}, {Lagos}, {Rizzo}, {Franco}, {Hsieh}, {Ilbert}, {Jin}, {Magdis}, {McCracken}, {Mobasher}, {Shuntov}, {Steinhardt}, {Strait}, \& {Toft}}]{Gould2023}
{Gould}, K. M.~L., {Brammer}, G., {Valentino}, F., {et~al.} 2023, \aj, 165, 248, \dodoi{10.3847/1538-3881/accadc}

\bibitem[{{Graves} {et~al.}(2007){Graves}, {Faber}, {Schiavon}, \& {Yan}}]{Graves2007}
{Graves}, G.~J., {Faber}, S.~M., {Schiavon}, R.~P., \& {Yan}, R. 2007, \apj, 671, 243, \dodoi{10.1086/522325}

\bibitem[{{Greene} {et~al.}(2013){Greene}, {Murphy}, {Graves}, {Gunn}, {Raskutti}, {Comerford}, \& {Gebhardt}}]{Greene2013}
{Greene}, J.~E., {Murphy}, J.~D., {Graves}, G.~J., {et~al.} 2013, \apj, 776, 64, \dodoi{10.1088/0004-637X/776/2/64}

\bibitem[{{Greve} {et~al.}(2012){Greve}, {Vieira}, {Wei{\ss}}, {Aguirre}, {Aird}, {Ashby}, {Benson}, {Bleem}, {Bradford}, {Brodwin}, {Carlstrom}, {Chang}, {Chapman}, {Crawford}, {de Breuck}, {de Haan}, {Dobbs}, {Downes}, {Fassnacht}, {Fazio}, {George}, {Gladders}, {Gonzalez}, {Halverson}, {Hezaveh}, {High}, {Holder}, {Holzapfel}, {Hoover}, {Hrubes}, {Johnson}, {Keisler}, {Knox}, {Lee}, {Leitch}, {Lueker}, {Luong-Van}, {Malkan}, {Marrone}, {McIntyre}, {McMahon}, {Mehl}, {Menten}, {Meyer}, {Montroy}, {Murphy}, {Natoli}, {Padin}, {Plagge}, {Pryke}, {Reichardt}, {Rest}, {Rosenman}, {Ruel}, {Ruhl}, {Schaffer}, {Sharon}, {Shaw}, {Shirokoff}, {Stalder}, {Stanford}, {Staniszewski}, {Stark}, {Story}, {Vanderlinde}, {Walsh}, {Welikala}, \& {Williamson}}]{Greve2012}
{Greve}, T.~R., {Vieira}, J.~D., {Wei{\ss}}, A., {et~al.} 2012, \apj, 756, 101, \dodoi{10.1088/0004-637X/756/1/101}

\bibitem[{{Hamadouche} {et~al.}(2022){Hamadouche}, {Carnall}, {McLure}, {Dunlop}, {McLeod}, {Cullen}, {Begley}, {Bolzonella}, {Buitrago}, {Castellano}, {Cucciati}, {Fontana}, {Gargiulo}, {Moresco}, {Pozzetti}, \& {Zamorani}}]{Hamadouche2022}
{Hamadouche}, M.~L., {Carnall}, A.~C., {McLure}, R.~J., {et~al.} 2022, arXiv e-prints, arXiv:2201.10576.
\newblock \doarXiv{2201.10576}

\bibitem[{{Hartley} {et~al.}(2023){Hartley}, {Nelson}, {Suess}, {Garcia}, {Park}, {Hernquist}, {Bezanson}, {Nevin}, {Pillepich}, {Schechter}, {Terrazas}, {Torrey}, {Wellons}, {Whitaker}, \& {Williams}}]{Hartley2023}
{Hartley}, A.~I., {Nelson}, E.~J., {Suess}, K.~A., {et~al.} 2023, \mnras, 522, 3138, \dodoi{10.1093/mnras/stad1162}

\bibitem[{{Hinshaw} {et~al.}(2013){Hinshaw}, {Larson}, {Komatsu}, {Spergel}, {Bennett}, {Dunkley}, {Nolta}, {Halpern}, {Hill}, {Odegard}, {Page}, {Smith}, {Weiland}, {Gold}, {Jarosik}, {Kogut}, {Limon}, {Meyer}, {Tucker}, {Wollack}, \& {Wright}}]{Hinshaw2013}
{Hinshaw}, G., {Larson}, D., {Komatsu}, E., {et~al.} 2013, \apjs, 208, 19, \dodoi{10.1088/0067-0049/208/2/19}

\bibitem[{Hoffman {et~al.}(2019)Hoffman, Sountsov, Dillon, Langmore, Tran, \& Vasudevan}]{Hoffman2019}
Hoffman, M., Sountsov, P., Dillon, J.~V., {et~al.} 2019, arXiv preprint arXiv:1903.03704

\bibitem[{Hoffman {et~al.}(2014)Hoffman, Gelman, {et~al.}}]{Hoffman2014}
Hoffman, M.~D., Gelman, A., {et~al.} 2014, J. Mach. Learn. Res., 15, 1593

\bibitem[{Hunter(2007)}]{Hunter:2007}
Hunter, J.~D. 2007, Computing in Science \& Engineering, 9, 90, \dodoi{10.1109/MCSE.2007.55}

\bibitem[{{Jafariyazani} {et~al.}(2020){Jafariyazani}, {Newman}, {Mobasher}, {Belli}, {Ellis}, \& {Patel}}]{Jafariyazani2020}
{Jafariyazani}, M., {Newman}, A.~B., {Mobasher}, B., {et~al.} 2020, \apjl, 897, L42, \dodoi{10.3847/2041-8213/aba11c}

\bibitem[{{Ji} {et~al.}(2024){Ji}, {Williams}, {Suess}, {Tacchella}, {Johnson}, {Robertson}, {Alberts}, {Baker}, {Baum}, {Bhatawdekar}, {Bonaventura}, {Boyett}, {Bunker}, {Carniani}, {Charlot}, {Chen}, {Chevallard}, {Curtis-Lake}, {D'Eugenio}, {de Graaff}, {DeCoursey}, {Egami}, {Eisenstein}, {Hainline}, {Hausen}, {Helton}, {Looser}, {Lyu}, {Maiolino}, {Maseda}, {Nelson}, {Rieke}, {Rieke}, {Rix}, {Sandles}, {Sun}, {{\"U}bler}, {Willmer}, {Willott}, \& {Witstok}}]{Ji2024}
{Ji}, Z., {Williams}, C.~C., {Suess}, K.~A., {et~al.} 2024, arXiv e-prints, arXiv:2401.00934, \dodoi{10.48550/arXiv.2401.00934}

\bibitem[{{Johnson} \& {Leja}(2017)}]{Johnson2017}
{Johnson}, B., \& {Leja}, J. 2017, {Bd-J/Prospector: Initial Release}, v0.1,  Zenodo, \dodoi{10.5281/zenodo.1116491}

\bibitem[{{Johnson} {et~al.}(2021){Johnson}, {Foreman-Mackey}, {Sick}, {Leja}, {Byler}, {Walmsley}, {Tollerud}, {Leung}, \& {Scott}}]{Johnson2021}
{Johnson}, B., {Foreman-Mackey}, D., {Sick}, J., {et~al.} 2021, {dfm/python-fsps: python-fsps v0.4.0}, v0.4.0,  Zenodo, \dodoi{10.5281/zenodo.4577191}

\bibitem[{{Johnson}(2019)}]{sedpy2019}
{Johnson}, B.~D. 2019, {SEDPY: Modules for storing and operating on astronomical source spectral energy distribution}, Astrophysics Source Code Library, record ascl:1905.026

\bibitem[{{Kakimoto} {et~al.}(2024){Kakimoto}, {Tanaka}, {Onodera}, {Shimakawa}, {Wu}, {Gould}, {Ito}, {Jin}, {Kubo}, {Suzuki}, {Toft}, {Valentino}, \& {Yabe}}]{Kakimoto2024}
{Kakimoto}, T., {Tanaka}, M., {Onodera}, M., {et~al.} 2024, \apj, 963, 49, \dodoi{10.3847/1538-4357/ad1ff1}

\bibitem[{{Kalita} {et~al.}(2021){Kalita}, {Daddi}, {D'Eugenio}, {Valentino}, {Rich}, {G{\'o}mez-Guijarro}, {Coogan}, {Delvecchio}, {Elbaz}, {Neill}, {Puglisi}, \& {Strazzullo}}]{Kalita2021}
{Kalita}, B.~S., {Daddi}, E., {D'Eugenio}, C., {et~al.} 2021, \apjl, 917, L17, \dodoi{10.3847/2041-8213/ac16dc}

\bibitem[{{Kawinwanichakij} {et~al.}(2021){Kawinwanichakij}, {Silverman}, {Ding}, {George}, {Damjanov}, {Sawicki}, {Tanaka}, {Taranu}, {Birrer}, {Huang}, {Li}, {Onodera}, {Shibuya}, \& {Yasuda}}]{Kawinwanichakij2021}
{Kawinwanichakij}, L., {Silverman}, J.~D., {Ding}, X., {et~al.} 2021, \apj, 921, 38, \dodoi{10.3847/1538-4357/ac1f21}

\bibitem[{{Kewley} {et~al.}(2006){Kewley}, {Groves}, {Kauffmann}, \& {Heckman}}]{Kewley2006}
{Kewley}, L.~J., {Groves}, B., {Kauffmann}, G., \& {Heckman}, T. 2006, \mnras, 372, 961, \dodoi{10.1111/j.1365-2966.2006.10859.x}

\bibitem[{{Kokorev} {et~al.}(2023){Kokorev}, {Jin}, {Magdis}, {Caputi}, {Valentino}, {Dayal}, {Trebitsch}, {Brammer}, {Fujimoto}, {Bauer}, {Iani}, {Kohno}, {Bl{\'a}nquez Ses{\'e}}, {G{\'o}mez-Guijarro}, {Rinaldi}, \& {Navarro-Carrera}}]{Kokorev2023a}
{Kokorev}, V., {Jin}, S., {Magdis}, G.~E., {et~al.} 2023, \apjl, 945, L25, \dodoi{10.3847/2041-8213/acbd9d}

\bibitem[{{Kriek} \& {Conroy}(2013)}]{Kriek2013}
{Kriek}, M., \& {Conroy}, C. 2013, \apjl, 775, L16, \dodoi{10.1088/2041-8205/775/1/L16}

\bibitem[{{Kriek} {et~al.}(2016){Kriek}, {Conroy}, {van Dokkum}, {Shapley}, {Choi}, {Reddy}, {Siana}, {van de Voort}, {Coil}, \& {Mobasher}}]{Kriek2016}
{Kriek}, M., {Conroy}, C., {van Dokkum}, P.~G., {et~al.} 2016, \nat, 540, 248, \dodoi{10.1038/nature20570}

\bibitem[{{Kubo} {et~al.}(2021){Kubo}, {Umehata}, {Matsuda}, {Kajisawa}, {Steidel}, {Yamada}, {Tanaka}, {Hatsukade}, {Tamura}, {Nakanishi}, {Kohno}, {Lee}, \& {Matsuda}}]{Kubo2021}
{Kubo}, M., {Umehata}, H., {Matsuda}, Y., {et~al.} 2021, \apj, 919, 6, \dodoi{10.3847/1538-4357/ac0cf8}

\bibitem[{{Labb{\'e}} {et~al.}(2023){Labb{\'e}}, {van Dokkum}, {Nelson}, {Bezanson}, {Suess}, {Leja}, {Brammer}, {Whitaker}, {Mathews}, {Stefanon}, \& {Wang}}]{Labbe2023}
{Labb{\'e}}, I., {van Dokkum}, P., {Nelson}, E., {et~al.} 2023, \nat, 616, 266, \dodoi{10.1038/s41586-023-05786-2}

\bibitem[{{Laine} {et~al.}(2003){Laine}, {van der Marel}, {Lauer}, {Postman}, {O'Dea}, \& {Owen}}]{Laine2003}
{Laine}, S., {van der Marel}, R.~P., {Lauer}, T.~R., {et~al.} 2003, \aj, 125, 478, \dodoi{10.1086/345823}

\bibitem[{{Lauer} {et~al.}(2005){Lauer}, {Faber}, {Gebhardt}, {Richstone}, {Tremaine}, {Ajhar}, {Aller}, {Bender}, {Dressler}, {Filippenko}, {Green}, {Grillmair}, {Ho}, {Kormendy}, {Magorrian}, {Pinkney}, \& {Siopis}}]{Lauer2005}
{Lauer}, T.~R., {Faber}, S.~M., {Gebhardt}, K., {et~al.} 2005, \aj, 129, 2138, \dodoi{10.1086/429565}

\bibitem[{{Lee} {et~al.}(2024){Lee}, {Steidel}, {Brammer}, {F{\"o}rster-Schreiber}, {Renzini}, {Liu}, {Herrera-Camus}, {Naab}, {Price}, {{\"U}bler}, {Arriagada-Neira}, \& {Magdis}}]{Lee2024}
{Lee}, M.~M., {Steidel}, C.~C., {Brammer}, G., {et~al.} 2024, \mnras, 527, 9529, \dodoi{10.1093/mnras/stad3718}

\bibitem[{{Leitherer} {et~al.}(1999){Leitherer}, {Schaerer}, {Goldader}, {Delgado}, {Robert}, {Kune}, {de Mello}, {Devost}, \& {Heckman}}]{Leitherer1999}
{Leitherer}, C., {Schaerer}, D., {Goldader}, J.~D., {et~al.} 1999, \apjs, 123, 3, \dodoi{10.1086/313233}

\bibitem[{{Leja} {et~al.}(2017){Leja}, {Johnson}, {Conroy}, {van Dokkum}, \& {Byler}}]{Leja2017}
{Leja}, J., {Johnson}, B.~D., {Conroy}, C., {van Dokkum}, P.~G., \& {Byler}, N. 2017, \apj, 837, 170, \dodoi{10.3847/1538-4357/aa5ffe}

\bibitem[{{Leja} {et~al.}(2019){Leja}, {Johnson}, {Conroy}, {van Dokkum}, {Speagle}, {Brammer}, {Momcheva}, {Skelton}, {Whitaker}, {Franx}, \& {Nelson}}]{Leja2019}
{Leja}, J., {Johnson}, B.~D., {Conroy}, C., {et~al.} 2019, \apj, 877, 140, \dodoi{10.3847/1538-4357/ab1d5a}

\bibitem[{{Li} {et~al.}(2023){Li}, {Jiang}, {He}, {Guo}, \& {Wang}}]{Li2023}
{Li}, K., {Jiang}, Z., {He}, P., {Guo}, Q., \& {Wang}, J. 2023, Research in Astronomy and Astrophysics, 23, 015010, \dodoi{10.1088/1674-4527/ac9e90}

\bibitem[{{Long} {et~al.}(2023){Long}, {Casey}, {del P. Lagos}, {Lambrides}, {Zavala}, {Champagne}, {Cooper}, \& {Cooray}}]{Long2023}
{Long}, A.~S., {Casey}, C.~M., {del P. Lagos}, C., {et~al.} 2023, \apj, 953, 11, \dodoi{10.3847/1538-4357/acddde}

\bibitem[{{Ma} {et~al.}(2015){Ma}, {Smail}, {Swinbank}, {Simpson}, {Thomson}, {Chen}, {Danielson}, {Hilton}, {Tadaki}, {Stott}, \& {Kodama}}]{Ma2015}
{Ma}, C.~J., {Smail}, I., {Swinbank}, A.~M., {et~al.} 2015, \apj, 806, 257, \dodoi{10.1088/0004-637X/806/2/257}

\bibitem[{{Man} {et~al.}(2021){Man}, {Zabl}, {Brammer}, {Richard}, {Toft}, {Stockmann}, {Gallazzi}, {Zibetti}, \& {Ebeling}}]{Man2021}
{Man}, A. W.~S., {Zabl}, J., {Brammer}, G.~B., {et~al.} 2021, arXiv e-prints, arXiv:2106.08338.
\newblock \doarXiv{2106.08338}

\bibitem[{{Marchesini} {et~al.}(2014){Marchesini}, {Muzzin}, {Stefanon}, {Franx}, {Brammer}, {Marsan}, {Vulcani}, {Fynbo}, {Milvang-Jensen}, {Dunlop}, \& {Buitrago}}]{Marchesini2014}
{Marchesini}, D., {Muzzin}, A., {Stefanon}, M., {et~al.} 2014, \apj, 794, 65, \dodoi{10.1088/0004-637X/794/1/65}

\bibitem[{{Marsan} {et~al.}(2015){Marsan}, {Marchesini}, {Brammer}, {Stefanon}, {Muzzin}, {Fern{\'a}ndez-Soto}, {Geier}, {Hainline}, {Intema}, {Karim}, {Labb{\'e}}, {Toft}, \& {van Dokkum}}]{Marsan2015}
{Marsan}, Z.~C., {Marchesini}, D., {Brammer}, G.~B., {et~al.} 2015, \apj, 801, 133, \dodoi{10.1088/0004-637X/801/2/133}

\bibitem[{{Martis} {et~al.}(2019){Martis}, {Marchesini}, {Muzzin}, {Stefanon}, {Brammer}, {da Cunha}, {Sajina}, \& {Labbe}}]{Martis2019}
{Martis}, N.~S., {Marchesini}, D.~M., {Muzzin}, A., {et~al.} 2019, \apj, 882, 65, \dodoi{10.3847/1538-4357/ab32f1}

\bibitem[{{McDermid} {et~al.}(2015){McDermid}, {Alatalo}, {Blitz}, {Bournaud}, {Bureau}, {Cappellari}, {Crocker}, {Davies}, {Davis}, {de Zeeuw}, {Duc}, {Emsellem}, {Khochfar}, {Krajnovi{\'c}}, {Kuntschner}, {Morganti}, {Naab}, {Oosterloo}, {Sarzi}, {Scott}, {Serra}, {Weijmans}, \& {Young}}]{McDermid2015}
{McDermid}, R.~M., {Alatalo}, K., {Blitz}, L., {et~al.} 2015, \mnras, 448, 3484, \dodoi{10.1093/mnras/stv105}

\bibitem[{{McKinney} {et~al.}(2023){McKinney}, {Manning}, {Cooper}, {Long}, {Akins}, {Casey}, {Faisst}, {Franco}, {Hayward}, {Lambrides}, {Magdis}, {Whitaker}, {Yun}, {Champagne}, {Drakos}, {Gentile}, {Gillman}, {Gozaliasl}, {Ilbert}, {Jin}, {Koekemoer}, {Kokorev}, {Liu}, {Rich}, {Robertson}, {Valentino}, {Weaver}, {Zavala}, {Allen}, {Kartaltepe}, {McCracken}, {Paquereau}, {Rhodes}, {Shuntov}, \& {Toft}}]{McKinney2023}
{McKinney}, J., {Manning}, S.~M., {Cooper}, O.~R., {et~al.} 2023, \apj, 956, 72, \dodoi{10.3847/1538-4357/acf614}

\bibitem[{{Miller} {et~al.}(2023){Miller}, {van Dokkum}, \& {Mowla}}]{Miller2023}
{Miller}, T.~B., {van Dokkum}, P., \& {Mowla}, L. 2023, \apj, 945, 155, \dodoi{10.3847/1538-4357/acbc74}

\bibitem[{{Miller} {et~al.}(2022){Miller}, {Whitaker}, {Nelson}, {van Dokkum}, {Bezanson}, {Brammer}, {Heintz}, {Leja}, {Suess}, \& {Weaver}}]{Miller2022}
{Miller}, T.~B., {Whitaker}, K.~E., {Nelson}, E.~J., {et~al.} 2022, \apjl, 941, L37, \dodoi{10.3847/2041-8213/aca675}

\bibitem[{{Morishita} {et~al.}(2022){Morishita}, {Abdurro'uf}, {Hirashita}, {Newman}, {Stiavelli}, \& {Chiaberge}}]{Morishita2022}
{Morishita}, T., {Abdurro'uf}, {Hirashita}, H., {et~al.} 2022, \apj, 938, 144, \dodoi{10.3847/1538-4357/ac9055}

\bibitem[{{Mosleh} {et~al.}(2017){Mosleh}, {Tacchella}, {Renzini}, {Carollo}, {Molaeinezhad}, {Onodera}, {Khosroshahi}, \& {Lilly}}]{Mosleh2017}
{Mosleh}, M., {Tacchella}, S., {Renzini}, A., {et~al.} 2017, \apj, 837, 2, \dodoi{10.3847/1538-4357/aa5f14}

\bibitem[{{Mowla} {et~al.}(2019){Mowla}, {van Dokkum}, {Brammer}, {Momcheva}, {van der Wel}, {Whitaker}, {Nelson}, {Bezanson}, {Muzzin}, {Franx}, {MacKenty}, {Leja}, {Kriek}, \& {Marchesini}}]{Mowla2019}
{Mowla}, L.~A., {van Dokkum}, P., {Brammer}, G.~B., {et~al.} 2019, \apj, 880, 57, \dodoi{10.3847/1538-4357/ab290a}

\bibitem[{{Mu{\~n}oz Arancibia} {et~al.}(2023){Mu{\~n}oz Arancibia}, {Gonz{\'a}lez-L{\'o}pez}, {Ibar}, {Bauer}, {Anguita}, {Aravena}, {Demarco}, {Kneissl}, {Koekemoer}, {Troncoso-Iribarren}, \& {Zitrin}}]{MunozArancibia2023}
{Mu{\~n}oz Arancibia}, A.~M., {Gonz{\'a}lez-L{\'o}pez}, J., {Ibar}, E., {et~al.} 2023, \aap, 675, A85, \dodoi{10.1051/0004-6361/202243528}

\bibitem[{{Nanayakkara} {et~al.}(2024){Nanayakkara}, {Glazebrook}, {Jacobs}, {Kawinwanichakij}, {Schreiber}, {Brammer}, {Esdaile}, {Kacprzak}, {Labbe}, {Lagos}, {Marchesini}, {Marsan}, {Oesch}, {Papovich}, {Remus}, \& {Tran}}]{Nanayakkara2024}
{Nanayakkara}, T., {Glazebrook}, K., {Jacobs}, C., {et~al.} 2024, Nature Scientific Reports in press, arXiv:2212.11638, \dodoi{10.48550/arXiv.2212.11638}

\bibitem[{{Nelson} {et~al.}(2023{\natexlab{a}}){Nelson}, {Suess}, {Bezanson}, {Price}, {van Dokkum}, {Leja}, {Wang}, {Whitaker}, {Labb{\'e}}, {Barrufet}, {Brammer}, {Eisenstein}, {Gibson}, {Hartley}, {Johnson}, {Heintz}, {Mathews}, {Miller}, {Oesch}, {Sandles}, {Setton}, {Speagle}, {Tacchella}, {Tadaki}, {{\"U}bler}, \& {Weaver}}]{Nelson2022}
{Nelson}, E.~J., {Suess}, K.~A., {Bezanson}, R., {et~al.} 2023{\natexlab{a}}, \apjl, 948, L18, \dodoi{10.3847/2041-8213/acc1e1}

\bibitem[{{Nelson} {et~al.}(2023{\natexlab{b}}){Nelson}, {Brammer}, {Gimenez-Arteaga}, {Oesch}, {Ubler}, {de Graaff}, {Matharu}, {Naidu}, {Shapley}, {Whitaker}, {Wisnioski}, {Forster Schreiber}, {Smit}, {van Dokkum}, {Chisholm}, {Endsley}, {Hartley}, {Gibson}, {Giovinazzo}, {Illingworth}, {Labbe}, {Maseda}, {Matthee}, {Covelo Paz}, {Price}, {Reddy}, {Shivaei}, {Weibel}, {Wuyts}, {Xiao}, {Alberts}, {Baker}, {Bunker}, {Cameron}, {Charlot}, {Eisenstein}, {Ji}, {Johnson}, {Jones}, {Maiolino}, {Robertson}, {Sandles}, {Suess}, {Tacchella}, {Williams}, \& {Witstok}}]{Nelson2023}
{Nelson}, E.~J., {Brammer}, G., {Gimenez-Arteaga}, C., {et~al.} 2023{\natexlab{b}}, arXiv e-prints, arXiv:2310.06887, \dodoi{10.48550/arXiv.2310.06887}

\bibitem[{{Otter} {et~al.}(2022){Otter}, {Rowlands}, {Alatalo}, {Leung}, {Wild}, {Luo}, {Petric}, {Sazonova}, {Stark}, {Heckman}, {Davis}, {Ellison}, {French}, {Baker}, {Bluck}, {Lanz}, {Lin}, {Liu}, {L{\'o}pez Cob{\'a}}, {Masters}, {Nair}, {Pan}, {Riffel}, {Scudder}, {Smercina}, {van de Voort}, \& {Weaver}}]{Otter2022}
{Otter}, J.~A., {Rowlands}, K., {Alatalo}, K., {et~al.} 2022, arXiv e-prints, arXiv:2210.12199.
\newblock \doarXiv{2210.12199}

\bibitem[{{Pacifici} {et~al.}(2016){Pacifici}, {Kassin}, {Weiner}, {Holden}, {Gardner}, {Faber}, {Ferguson}, {Koo}, {Primack}, {Bell}, {Dekel}, {Gawiser}, {Giavalisco}, {Rafelski}, {Simons}, {Barro}, {Croton}, {Dav{\'e}}, {Fontana}, {Grogin}, {Koekemoer}, {Lee}, {Salmon}, {Somerville}, \& {Behroozi}}]{Pacifici2016}
{Pacifici}, C., {Kassin}, S.~A., {Weiner}, B.~J., {et~al.} 2016, \apj, 832, 79, \dodoi{10.3847/0004-637X/832/1/79}

\bibitem[{{Park} {et~al.}(2022){Park}, {Belli}, {Conroy}, {Tacchella}, {Leja}, {Cutler}, {Johnson}, {Nelson}, \& {Emami}}]{Park2022}
{Park}, M., {Belli}, S., {Conroy}, C., {et~al.} 2022, arXiv e-prints, arXiv:2210.03747.
\newblock \doarXiv{2210.03747}

\bibitem[{{Park} {et~al.}(2024){Park}, {Belli}, {Conroy}, {Johnson}, {Davies}, {Leja}, {Tacchella}, {Mendel}, {Benton}, {Bugiani}, {Emami}, {Khoram}, {Li}, {Maheson}, {Mathews}, {Naidu}, {Nelson}, {Terrazas}, \& {Weinberger}}]{Park2024}
---. 2024, arXiv e-prints, arXiv:2404.17945, \dodoi{10.48550/arXiv.2404.17945}

\bibitem[{{Pasha} \& {Miller}(2023)}]{Pasha2023}
{Pasha}, I., \& {Miller}, T.~B. 2023, The Journal of Open Source Software, 8, 5703, \dodoi{10.21105/joss.05703}

\bibitem[{{Pattarakijwanich} {et~al.}(2016){Pattarakijwanich}, {Strauss}, {Ho}, \& {Ross}}]{Pattarakijwanich2016}
{Pattarakijwanich}, P., {Strauss}, M.~A., {Ho}, S., \& {Ross}, N.~P. 2016, \apj, 833, 19, \dodoi{10.3847/0004-637X/833/1/19}

\bibitem[{{Pawlik} {et~al.}(2016){Pawlik}, {Wild}, {Walcher}, {Johansson}, {Villforth}, {Rowlands}, {Mendez-Abreu}, \& {Hewlett}}]{Pawlik2016}
{Pawlik}, M.~M., {Wild}, V., {Walcher}, C.~J., {et~al.} 2016, \mnras, 456, 3032, \dodoi{10.1093/mnras/stv2878}

\bibitem[{{P{\'e}rez-Gonz{\'a}lez} {et~al.}(2023){P{\'e}rez-Gonz{\'a}lez}, {Barro}, {Annunziatella}, {Costantin}, {Garc{\'\i}a-Argum{\'a}nez}, {McGrath}, {M{\'e}rida}, {Zavala}, {Arrabal Haro}, {Bagley}, {Backhaus}, {Behroozi}, {Bell}, {Bisigello}, {Buat}, {Calabr{\`o}}, {Casey}, {Cleri}, {Coogan}, {Cooper}, {Cooray}, {Dekel}, {Dickinson}, {Elbaz}, {Ferguson}, {Finkelstein}, {Fontana}, {Franco}, {Gardner}, {Giavalisco}, {G{\'o}mez-Guijarro}, {Grazian}, {Grogin}, {Guo}, {Huertas-Company}, {Jogee}, {Kartaltepe}, {Kewley}, {Kirkpatrick}, {Kocevski}, {Koekemoer}, {Long}, {Lotz}, {Lucas}, {Papovich}, {Pirzkal}, {Ravindranath}, {Somerville}, {Tacchella}, {Trump}, {Wang}, {Wilkins}, {Wuyts}, {Yang}, \& {Yung}}]{PerezGonzalez2023}
{P{\'e}rez-Gonz{\'a}lez}, P.~G., {Barro}, G., {Annunziatella}, M., {et~al.} 2023, \apjl, 946, L16, \dodoi{10.3847/2041-8213/acb3a5}

\bibitem[{Phan {et~al.}(2019)Phan, Pradhan, \& Jankowiak}]{Phan2019}
Phan, D., Pradhan, N., \& Jankowiak, M. 2019, arXiv preprint arXiv:1912.11554

\bibitem[{{Price} {et~al.}(2023){Price}, {Suess}, {Williams}, {Bezanson}, {Khullar}, {Nelson}, {Wang}, {Weaver}, {Fujimoto}, {Kokorev}, {Greene}, {Brammer}, {Cutler}, {Dayal}, {Furtak}, {Labbe}, {Leja}, {Miller}, {Nanayakkara}, {Pan}, \& {Whitaker}}]{Price2023}
{Price}, S.~H., {Suess}, K.~A., {Williams}, C.~C., {et~al.} 2023, arXiv e-prints, arXiv:2310.02500, \dodoi{10.48550/arXiv.2310.02500}

\bibitem[{{Rowlands} {et~al.}(2018){Rowlands}, {Wild}, {Bourne}, {Bremer}, {Brough}, {Driver}, {Hopkins}, {Owers}, {Phillipps}, {Pimbblet}, {Sansom}, {Wang}, {Alpaslan}, {Bland-Hawthorn}, {Colless}, {Holwerda}, \& {Taylor}}]{Rowlands2018a}
{Rowlands}, K., {Wild}, V., {Bourne}, N., {et~al.} 2018, \mnras, 473, 1168, \dodoi{10.1093/mnras/stx1903}

\bibitem[{{S{\'a}nchez-Bl{\'a}zquez} {et~al.}(2006{\natexlab{a}}){S{\'a}nchez-Bl{\'a}zquez}, {Gorgas}, {Cardiel}, \& {Gonz{\'a}lez}}]{Sanchez-Blazquez2006b}
{S{\'a}nchez-Bl{\'a}zquez}, P., {Gorgas}, J., {Cardiel}, N., \& {Gonz{\'a}lez}, J.~J. 2006{\natexlab{a}}, \aap, 457, 787, \dodoi{10.1051/0004-6361:20064842}

\bibitem[{{S{\'a}nchez-Bl{\'a}zquez} {et~al.}(2006{\natexlab{b}}){S{\'a}nchez-Bl{\'a}zquez}, {Peletier}, {Jim{\'e}nez-Vicente}, {Cardiel}, {Cenarro}, {Falc{\'o}n-Barroso}, {Gorgas}, {Selam}, \& {Vazdekis}}]{Sanchez-Blazquez2006a}
{S{\'a}nchez-Bl{\'a}zquez}, P., {Peletier}, R.~F., {Jim{\'e}nez-Vicente}, J., {et~al.} 2006{\natexlab{b}}, \mnras, 371, 703, \dodoi{10.1111/j.1365-2966.2006.10699.x}

\bibitem[{{Sanders} {et~al.}(2016){Sanders}, {Shapley}, {Kriek}, {Reddy}, {Freeman}, {Coil}, {Siana}, {Mobasher}, {Shivaei}, {Price}, \& {de Groot}}]{Sanders2016}
{Sanders}, R.~L., {Shapley}, A.~E., {Kriek}, M., {et~al.} 2016, \apj, 816, 23, \dodoi{10.3847/0004-637X/816/1/23}

\bibitem[{{Santini} {et~al.}(2019){Santini}, {Merlin}, {Fontana}, {Magnelli}, {Paris}, {Castellano}, {Grazian}, {Pentericci}, {Pilo}, \& {Torelli}}]{Santini2019}
{Santini}, P., {Merlin}, E., {Fontana}, A., {et~al.} 2019, \mnras, 486, 560, \dodoi{10.1093/mnras/stz801}

\bibitem[{{Saracco} {et~al.}(2020){Saracco}, {Marchesini}, {La Barbera}, {Gargiulo}, {Annunziatella}, {Forrest}, {Lange Vagle}, {Marsan}, {Muzzin}, {Stefanon}, \& {Wilson}}]{Saracco2020}
{Saracco}, P., {Marchesini}, D., {La Barbera}, F., {et~al.} 2020, \apj, 905, 40, \dodoi{10.3847/1538-4357/abc7c4}

\bibitem[{{Sargent} {et~al.}(2015){Sargent}, {Daddi}, {Bournaud}, {Onodera}, {Feruglio}, {Martig}, {Gobat}, {Dannerbauer}, \& {Schinnerer}}]{Sargent2015}
{Sargent}, M.~T., {Daddi}, E., {Bournaud}, F., {et~al.} 2015, \apjl, 806, L20, \dodoi{10.1088/2041-8205/806/1/L20}

\bibitem[{{Sazonova} {et~al.}(2021){Sazonova}, {Alatalo}, {Rowlands}, {Deustua}, {French}, {Heckman}, {Lanz}, {Lisenfeld}, {Luo}, {Medling}, {Nyland}, {Otter}, {Petric}, {Snyder}, \& {Urry}}]{Sazonova2021}
{Sazonova}, E., {Alatalo}, K., {Rowlands}, K., {et~al.} 2021, \apj, 919, 134, \dodoi{10.3847/1538-4357/ac0f7f}

\bibitem[{{Schreiber} {et~al.}(2018{\natexlab{a}}){Schreiber}, {Labb{\'e}}, {Glazebrook}, {Bekiaris}, {Papovich}, {Costa}, {Elbaz}, {Kacprzak}, {Nanayakkara}, {Oesch}, {Pannella}, {Spitler}, {Straatman}, {Tran}, \& {Wang}}]{Schreiber2018a}
{Schreiber}, C., {Labb{\'e}}, I., {Glazebrook}, K., {et~al.} 2018{\natexlab{a}}, \aap, 611, A22, \dodoi{10.1051/0004-6361/201731917}

\bibitem[{{Schreiber} {et~al.}(2018{\natexlab{b}}){Schreiber}, {Glazebrook}, {Nanayakkara}, {Kacprzak}, {Labb{\'e}}, {Oesch}, {Yuan}, {Tran}, {Papovich}, {Spitler}, \& {Straatman}}]{Schreiber2018b}
{Schreiber}, C., {Glazebrook}, K., {Nanayakkara}, T., {et~al.} 2018{\natexlab{b}}, \aap, 618, A85, \dodoi{10.1051/0004-6361/201833070}

\bibitem[{{Scoville} {et~al.}(2016){Scoville}, {Sheth}, {Aussel}, {Vanden Bout}, {Capak}, {Bongiorno}, {Casey}, {Murchikova}, {Koda}, {{\'A}lvarez-M{\'a}rquez}, {Lee}, {Laigle}, {McCracken}, {Ilbert}, {Pope}, {Sanders}, {Chu}, {Toft}, {Ivison}, \& {Manohar}}]{Scoville2016}
{Scoville}, N., {Sheth}, K., {Aussel}, H., {et~al.} 2016, \apj, 820, 83, \dodoi{10.3847/0004-637X/820/2/83}

\bibitem[{{Setton} {et~al.}(2023){Setton}, {Dey}, {Khullar}, {Bezanson}, {Newman}, {Aguilar}, {Ahlen}, {Andrews}, {Brooks}, {de la Macorra}, {Dey}, {Eftekharzadeh}, {Font-Ribera}, {A Gontcho}, {Kremin}, {Juneau}, {Landriau}, {Meisner}, {Miquel}, {Moustakas}, {Pearl}, {Prada}, {Tarl{\'e}}, {Siudek}, {Weaver}, {Zhou}, \& {Zou}}]{Setton2023}
{Setton}, D.~J., {Dey}, B., {Khullar}, G., {et~al.} 2023, \apjl, 947, L31, \dodoi{10.3847/2041-8213/acc9b5}

\bibitem[{{Shapley} {et~al.}(2023){Shapley}, {Sanders}, {Reddy}, {Topping}, \& {Brammer}}]{Shapley2023}
{Shapley}, A.~E., {Sanders}, R.~L., {Reddy}, N.~A., {Topping}, M.~W., \& {Brammer}, G.~B. 2023, \apj, 954, 157, \dodoi{10.3847/1538-4357/acea5a}

\bibitem[{{Shen} {et~al.}(2003){Shen}, {Mo}, {White}, {Blanton}, {Kauffmann}, {Voges}, {Brinkmann}, \& {Csabai}}]{Shen2003}
{Shen}, S., {Mo}, H.~J., {White}, S. D.~M., {et~al.} 2003, \mnras, 343, 978, \dodoi{10.1046/j.1365-8711.2003.06740.x}

\bibitem[{{Slob} {et~al.}(2024){Slob}, {Kriek}, {Beverage}, {Suess}, {Barro}, {Bezanson}, {Cheng}, {Conroy}, {de Graaff}, {F{\"o}rster Schreiber}, {Franx}, {Lorenz}, {Mancera Pi{\~n}a}, {Marchesini}, {Muzzin}, {Newman}, {Price}, {Shapley}, {Stefanon}, {van Dokkum}, \& {Weisz}}]{Slob2024}
{Slob}, M., {Kriek}, M., {Beverage}, A.~G., {et~al.} 2024, arXiv e-prints, arXiv:2404.12432, \dodoi{10.48550/arXiv.2404.12432}

\bibitem[{{Smail} {et~al.}(2023){Smail}, {Dudzevi{\v{c}}i{\={u}}t{\.{e}}}, {Gurwell}, {Fazio}, {Willner}, {Swinbank}, {Arumugam}, {Summers}, {Cohen}, {Jansen}, {Windhorst}, {Meena}, {Zitrin}, {Keel}, {Cheng}, {Coe}, {Conselice}, {D'Silva}, {Driver}, {Frye}, {Grogin}, {Koekemoer}, {Marshall}, {Nonino}, {Pirzkal}, {Robotham}, {Rutkowski}, {Ryan}, {Tompkins}, {Willmer}, {Yan}, {Broadhurst}, {Diego}, {Kamieneski}, \& {Yun}}]{Smail2023}
{Smail}, I., {Dudzevi{\v{c}}i{\={u}}t{\.{e}}}, U., {Gurwell}, M., {et~al.} 2023, \apj, 958, 36, \dodoi{10.3847/1538-4357/acf931}

\bibitem[{{Smercina} {et~al.}(2022){Smercina}, {Smith}, {French}, {Bell}, {Dale}, {Medling}, {Nyland}, {Privon}, {Rowlands}, {Walter}, \& {Zabludoff}}]{Smercina2022}
{Smercina}, A., {Smith}, J.-D.~T., {French}, K.~D., {et~al.} 2022, \apj, 929, 154, \dodoi{10.3847/1538-4357/ac5d5f}

\bibitem[{{Speagle}(2020)}]{Speagle2020}
{Speagle}, J.~S. 2020, \mnras, 493, 3132, \dodoi{10.1093/mnras/staa278}

\bibitem[{{Speagle} {et~al.}(2014){Speagle}, {Steinhardt}, {Capak}, \& {Silverman}}]{Speagle2014}
{Speagle}, J.~S., {Steinhardt}, C.~L., {Capak}, P.~L., \& {Silverman}, J.~D. 2014, \apjs, 214, 15, \dodoi{10.1088/0067-0049/214/2/15}

\bibitem[{{Spilker} {et~al.}(2022){Spilker}, {Suess}, {Setton}, {Bezanson}, {Feldmann}, {Greene}, {Kriek}, {Lower}, {Narayanan}, \& {Verrico}}]{Spilker2022}
{Spilker}, J.~S., {Suess}, K.~A., {Setton}, D.~J., {et~al.} 2022, \apjl, 936, L11, \dodoi{10.3847/2041-8213/ac75ea}

\bibitem[{{Straatman} {et~al.}(2014){Straatman}, {Labb{\'e}}, {Spitler}, {Allen}, {Altieri}, {Brammer}, {Dickinson}, {van Dokkum}, {Inami}, {Glazebrook}, {Kacprzak}, {Kawinwanichakij}, {Kelson}, {McCarthy}, {Mehrtens}, {Monson}, {Murphy}, {Papovich}, {Persson}, {Quadri}, {Rees}, {Tomczak}, {Tran}, \& {Tilvi}}]{Straatman2014}
{Straatman}, C. M.~S., {Labb{\'e}}, I., {Spitler}, L.~R., {et~al.} 2014, \apjl, 783, L14, \dodoi{10.1088/2041-8205/783/1/L14}

\bibitem[{{Straatman} {et~al.}(2016){Straatman}, {Spitler}, {Quadri}, {Labb{\'e}}, {Glazebrook}, {Persson}, {Papovich}, {Tran}, {Brammer}, {Cowley}, {Tomczak}, {Nanayakkara}, {Alcorn}, {Allen}, {Broussard}, {van Dokkum}, {Forrest}, {van Houdt}, {Kacprzak}, {Kawinwanichakij}, {Kelson}, {Lee}, {McCarthy}, {Mehrtens}, {Monson}, {Murphy}, {Rees}, {Tilvi}, \& {Whitaker}}]{Straatman2016}
{Straatman}, C. M.~S., {Spitler}, L.~R., {Quadri}, R.~F., {et~al.} 2016, \apj, 830, 51, \dodoi{10.3847/0004-637X/830/1/51}

\bibitem[{{Suess} {et~al.}(2017){Suess}, {Bezanson}, {Spilker}, {Kriek}, {Greene}, {Feldmann}, {Hunt}, \& {Narayanan}}]{Suess2017}
{Suess}, K.~A., {Bezanson}, R., {Spilker}, J.~S., {et~al.} 2017, \apjl, 846, L14, \dodoi{10.3847/2041-8213/aa85dc}

\bibitem[{{Suess} {et~al.}(2019){Suess}, {Kriek}, {Price}, \& {Barro}}]{Suess2019a}
{Suess}, K.~A., {Kriek}, M., {Price}, S.~H., \& {Barro}, G. 2019, \apj, 877, 103, \dodoi{10.3847/1538-4357/ab1bda}

\bibitem[{{Suess} {et~al.}(2021){Suess}, {Kriek}, {Price}, \& {Barro}}]{Suess2021}
---. 2021, \apj, 915, 87, \dodoi{10.3847/1538-4357/abf1e4}

\bibitem[{{Suess} {et~al.}(2024){Suess}, {Weaver}, {Price}, {Pan}, {Wang}, {Bezanson}, {Brammer}, {Cutler}, {Labbe}, {Leja}, {Williams}, {Whitaker}, {Dayal}, {de Graaff}, {Feldmann}, {Franx}, {Fudamoto}, {Fujimoto}, {Furtak}, {Goulding}, {Greene}, {Khullar}, {Kokorev}, {Kriek}, {Lorenz}, {Marchesini}, {Maseda}, {Matthee}, {Miller}, {Mitsuhashi}, {Mowla}, {Muzzin}, {Naidu}, {Nanayakkara}, {Nelson}, {Oesch}, {Setton}, {Shipley}, {Smit}, {Spilker}, {van Dokkum}, \& {Zitrin}}]{Suess2024}
{Suess}, K.~A., {Weaver}, J.~R., {Price}, S.~H., {et~al.} 2024, arXiv e-prints, arXiv:2404.13132, \dodoi{10.48550/arXiv.2404.13132}

\bibitem[{{Suzuki} {et~al.}(2022){Suzuki}, {Glazebrook}, {Schreiber}, {Kodama}, {Kacprzak}, {Leiton}, {Nanayakkara}, {Oesch}, {Papovich}, {Spitler}, {Straatman}, {Tran}, \& {Wang}}]{Suzuki2022}
{Suzuki}, T.~L., {Glazebrook}, K., {Schreiber}, C., {et~al.} 2022, \apj, 936, 61, \dodoi{10.3847/1538-4357/ac7ce3}

\bibitem[{{Swinbank} {et~al.}(2006){Swinbank}, {Chapman}, {Smail}, {Lindner}, {Borys}, {Blain}, {Ivison}, \& {Lewis}}]{Swinbank2006}
{Swinbank}, A.~M., {Chapman}, S.~C., {Smail}, I., {et~al.} 2006, \mnras, 371, 465, \dodoi{10.1111/j.1365-2966.2006.10673.x}

\bibitem[{{Tacchella} {et~al.}(2016){Tacchella}, {Dekel}, {Carollo}, {Ceverino}, {DeGraf}, {Lapiner}, {Mand elker}, \& {Primack}}]{Tacchella2016}
{Tacchella}, S., {Dekel}, A., {Carollo}, C.~M., {et~al.} 2016, \mnras, 458, 242, \dodoi{10.1093/mnras/stw303}

\bibitem[{{Tacchella} {et~al.}(2015){Tacchella}, {Carollo}, {Renzini}, {Schreiber}, {Lang}, {Wuyts}, {Cresci}, {Dekel}, {Genzel}, {Lilly}, {Mancini}, {Newman}, {Onodera}, {Shapley}, {Tacconi}, {Woo}, \& {Zamorani}}]{Tacchella2015a}
{Tacchella}, S., {Carollo}, C.~M., {Renzini}, A., {et~al.} 2015, Science, 348, 314, \dodoi{10.1126/science.1261094}

\bibitem[{{Tacchella} {et~al.}(2022){Tacchella}, {Conroy}, {Faber}, {Johnson}, {Leja}, {Barro}, {Cunningham}, {Deason}, {Guhathakurta}, {Guo}, {Hernquist}, {Koo}, {McKinnon}, {Rockosi}, {Speagle}, {van Dokkum}, \& {Yesuf}}]{Tacchella2022}
{Tacchella}, S., {Conroy}, C., {Faber}, S.~M., {et~al.} 2022, \apj, 926, 134, \dodoi{10.3847/1538-4357/ac449b}

\bibitem[{{Tacchella} {et~al.}(2023){Tacchella}, {Eisenstein}, {Hainline}, {Johnson}, {Baker}, {Helton}, {Robertson}, {Suess}, {Chen}, {Nelson}, {Pusk{\'a}s}, {Sun}, {Alberts}, {Egami}, {Hausen}, {Rieke}, {Rieke}, {Shivaei}, {Williams}, {Willmer}, {Bunker}, {Cameron}, {Carniani}, {Charlot}, {Curti}, {Curtis-Lake}, {Looser}, {Maiolino}, {Maseda}, {Rawle}, {Rix}, {Smit}, {{\"U}bler}, {Willott}, {Witstok}, {Baum}, {Bhatawdekar}, {Boyett}, {Danhaive}, {de Graaff}, {Endsley}, {Ji}, {Lyu}, {Sandles}, {Saxena}, {Scholtz}, {Topping}, \& {Whitler}}]{Tacchella2023}
{Tacchella}, S., {Eisenstein}, D.~J., {Hainline}, K., {et~al.} 2023, \apj, 952, 74, \dodoi{10.3847/1538-4357/acdbc6}

\bibitem[{{Tadaki} {et~al.}(2020){Tadaki}, {Belli}, {Burkert}, {Dekel}, {F{\"o}rster Schreiber}, {Genzel}, {Hayashi}, {Herrera-Camus}, {Kodama}, {Kohno}, {Koyama}, {Lee}, {Lutz}, {Mowla}, {Nelson}, {Renzini}, {Suzuki}, {Tacconi}, {{\"U}bler}, {Wisnioski}, \& {Wuyts}}]{Tadaki2020}
{Tadaki}, K.-i., {Belli}, S., {Burkert}, A., {et~al.} 2020, \apj, 901, 74, \dodoi{10.3847/1538-4357/abaf4a}

\bibitem[{{Tanaka} {et~al.}(2019){Tanaka}, {Valentino}, {Toft}, {Onodera}, {Shimakawa}, {Ceverino}, {Faisst}, {Gallazzi}, {G{\'o}mez-Guijarro}, {Kubo}, {Magdis}, {Steinhardt}, {Stockmann}, {Yabe}, \& {Zabl}}]{Tanaka2019}
{Tanaka}, M., {Valentino}, F., {Toft}, S., {et~al.} 2019, \apjl, 885, L34, \dodoi{10.3847/2041-8213/ab4ff3}

\bibitem[{{Tanaka} {et~al.}(2023){Tanaka}, {Onodera}, {Shimakawa}, {Ito}, {Kakimoto}, {Kubo}, {Morishita}, {Toft}, {Valentino}, \& {Wu}}]{Tanaka2023}
{Tanaka}, M., {Onodera}, M., {Shimakawa}, R., {et~al.} 2023, arXiv e-prints, arXiv:2311.11569, \dodoi{10.48550/arXiv.2311.11569}

\bibitem[{{Thomas} {et~al.}(2005){Thomas}, {Maraston}, {Bender}, \& {Mendes de Oliveira}}]{Thomas2005}
{Thomas}, D., {Maraston}, C., {Bender}, R., \& {Mendes de Oliveira}, C. 2005, \apj, 621, 673, \dodoi{10.1086/426932}

\bibitem[{{Tomita} {et~al.}(2000){Tomita}, {Aoki}, {Watanabe}, {Takata}, \& {Ichikawa}}]{Tomita2000}
{Tomita}, A., {Aoki}, K., {Watanabe}, M., {Takata}, T., \& {Ichikawa}, S.-i. 2000, \aj, 120, 123, \dodoi{10.1086/301440}

\bibitem[{{Tortora} {et~al.}(2010){Tortora}, {Napolitano}, {Cardone}, {Capaccioli}, {Jetzer}, \& {Molinaro}}]{Tortora2010}
{Tortora}, C., {Napolitano}, N.~R., {Cardone}, V.~F., {et~al.} 2010, \mnras, 407, 144, \dodoi{10.1111/j.1365-2966.2010.16938.x}

\bibitem[{{Trager} {et~al.}(2000){Trager}, {Faber}, {Worthey}, \& {Gonz{\'a}lez}}]{Trager2000}
{Trager}, S.~C., {Faber}, S.~M., {Worthey}, G., \& {Gonz{\'a}lez}, J.~J. 2000, \aj, 119, 1645, \dodoi{10.1086/301299}

\bibitem[{{Tran} {et~al.}(2001){Tran}, {Tsvetanov}, {Ford}, {Davies}, {Jaffe}, {van den Bosch}, \& {Rest}}]{Tran2001}
{Tran}, H.~D., {Tsvetanov}, Z., {Ford}, H.~C., {et~al.} 2001, \aj, 121, 2928, \dodoi{10.1086/321072}

\bibitem[{{Trujillo} {et~al.}(2004){Trujillo}, {Erwin}, {Asensio Ramos}, \& {Graham}}]{Trujillo2004}
{Trujillo}, I., {Erwin}, P., {Asensio Ramos}, A., \& {Graham}, A.~W. 2004, \aj, 127, 1917, \dodoi{10.1086/382712}

\bibitem[{{Urbano Stawinski} {et~al.}(2024){Urbano Stawinski}, {Cooper}, {Forrest}, {Muzzin}, {Marchesini}, {Wilson}, {Gomez}, {McConachie}, {Marsan}, {Annuziatella}, \& {Chang}}]{UrbanoStawinski2024}
{Urbano Stawinski}, S.~M., {Cooper}, M.~C., {Forrest}, B., {et~al.} 2024, arXiv e-prints, arXiv:2404.16036, \dodoi{10.48550/arXiv.2404.16036}

\bibitem[{{Valentino} {et~al.}(2020){Valentino}, {Tanaka}, {Davidzon}, {Toft}, {G{\'o}mez-Guijarro}, {Stockmann}, {Onodera}, {Brammer}, {Ceverino}, {Faisst}, {Gallazzi}, {Hayward}, {Ilbert}, {Kubo}, {Magdis}, {Selsing}, {Shimakawa}, {Sparre}, {Steinhardt}, {Yabe}, \& {Zabl}}]{Valentino2020}
{Valentino}, F., {Tanaka}, M., {Davidzon}, I., {et~al.} 2020, \apj, 889, 93, \dodoi{10.3847/1538-4357/ab64dc}

\bibitem[{{Valentino} {et~al.}(2023){Valentino}, {Brammer}, {Gould}, {Kokorev}, {Fujimoto}, {Jespersen}, {Vijayan}, {Weaver}, {Ito}, {Tanaka}, {Ilbert}, {Magdis}, {Whitaker}, {Faisst}, {Gallazzi}, {Gillman}, {Gim{\'e}nez-Arteaga}, {G{\'o}mez-Guijarro}, {Kubo}, {Heintz}, {Hirschmann}, {Oesch}, {Onodera}, {Rizzo}, {Lee}, {Strait}, \& {Toft}}]{Valentino2023}
{Valentino}, F., {Brammer}, G., {Gould}, K. M.~L., {et~al.} 2023, \apj, 947, 20, \dodoi{10.3847/1538-4357/acbefa}

\bibitem[{{van der Wel} {et~al.}(2014){van der Wel}, {Franx}, {van Dokkum}, {Skelton}, {Momcheva}, {Whitaker}, {Brammer}, {Bell}, {Rix}, {Wuyts}, {Ferguson}, {Holden}, {Barro}, {Koekemoer}, {Chang}, {McGrath}, {H{\"a}ussler}, {Dekel}, {Behroozi}, {Fumagalli}, {Leja}, {Lundgren}, {Maseda}, {Nelson}, {Wake}, {Patel}, {Labb{\'e}}, {Faber}, {Grogin}, \& {Kocevski}}]{VanDerWel2014}
{van der Wel}, A., {Franx}, M., {van Dokkum}, P.~G., {et~al.} 2014, \apj, 788, 28, \dodoi{10.1088/0004-637X/788/1/28}

\bibitem[{{van Dokkum} {et~al.}(2023){van Dokkum}, {Brammer}, {Wang}, {Leja}, \& {Conroy}}]{VanDokkum2023}
{van Dokkum}, P., {Brammer}, G., {Wang}, B., {Leja}, J., \& {Conroy}, C. 2023, Nature Astronomy, \dodoi{10.1038/s41550-023-02103-9}

\bibitem[{{van Dokkum} \& {Franx}(1995)}]{VanDokkum1995}
{van Dokkum}, P.~G., \& {Franx}, M. 1995, \aj, 110, 2027, \dodoi{10.1086/117667}

\bibitem[{{van Dokkum} {et~al.}(2015){van Dokkum}, {Nelson}, {Franx}, {Oesch}, {Momcheva}, {Brammer}, {F{\"o}rster Schreiber}, {Skelton}, {Whitaker}, {van der Wel}, {Bezanson}, {Fumagalli}, {Illingworth}, {Kriek}, {Leja}, \& {Wuyts}}]{VanDokkum2015}
{van Dokkum}, P.~G., {Nelson}, E.~J., {Franx}, M., {et~al.} 2015, \apj, 813, 23, \dodoi{10.1088/0004-637X/813/1/23}

\bibitem[{Vehtari {et~al.}(2021)Vehtari, Gelman, Simpson, Carpenter, \& B{\"u}rkner}]{Vehtari2021}
Vehtari, A., Gelman, A., Simpson, D., Carpenter, B., \& B{\"u}rkner, P.-C. 2021, Bayesian analysis, 16, 667

\bibitem[{{Verrico} {et~al.}(2023){Verrico}, {Setton}, {Bezanson}, {Greene}, {Suess}, {Goulding}, {Spilker}, {Kriek}, {Feldmann}, {Narayanan}, {Donofrio}, \& {Khullar}}]{Verrico2023}
{Verrico}, M.~E., {Setton}, D.~J., {Bezanson}, R., {et~al.} 2023, \apj, 949, 5, \dodoi{10.3847/1538-4357/acc38b}

\bibitem[{{Wang} {et~al.}(2023){Wang}, {Leja}, {Bezanson}, {Johnson}, {Khullar}, {Labb{\'e}}, {Price}, {Weaver}, \& {Whitaker}}]{Wang2023}
{Wang}, B., {Leja}, J., {Bezanson}, R., {et~al.} 2023, \apjl, 944, L58, \dodoi{10.3847/2041-8213/acba99}

\bibitem[{{Wang} {et~al.}(2024{\natexlab{a}}){Wang}, {Leja}, {Labb{\'e}}, {Bezanson}, {Whitaker}, {Brammer}, {Furtak}, {Weaver}, {Price}, {Zitrin}, {Atek}, {Coe}, {Cutler}, {Dayal}, {van Dokkum}, {Feldmann}, {Marchesini}, {Franx}, {F{\"o}rster Schreiber}, {Fujimoto}, {Geha}, {Glazebrook}, {de Graaff}, {Greene}, {Juneau}, {Kassin}, {Kriek}, {Khullar}, {Maseda}, {Mowla}, {Muzzin}, {Nanayakkara}, {Nelson}, {Oesch}, {Pacifici}, {Pan}, {Papovich}, {Setton}, {Shapley}, {Smit}, {Stefanon}, {Suess}, {Taylor}, \& {Williams}}]{WangB2024_uncoverpops}
{Wang}, B., {Leja}, J., {Labb{\'e}}, I., {et~al.} 2024{\natexlab{a}}, \apjs, 270, 12, \dodoi{10.3847/1538-4365/ad0846}

\bibitem[{{Wang} {et~al.}(2024{\natexlab{b}}){Wang}, {Leja}, {de Graaff}, {Brammer}, {Weibel}, {van Dokkum}, {Baggen}, {Suess}, {Greene}, {Bezanson}, {Cleri}, {Hirschmann}, {Labbe}, {Matthee}, {McConachie}, {Naidu}, {Nelson}, {Oesch}, {Setton}, \& {Williams}}]{WangB2024_UB}
{Wang}, B., {Leja}, J., {de Graaff}, A., {et~al.} 2024{\natexlab{b}}, arXiv e-prints, arXiv:2405.01473, \dodoi{10.48550/arXiv.2405.01473}

\bibitem[{{Weaver} {et~al.}(2018){Weaver}, {Husemann}, {Kuntschner}, {Mart{\'\i}n-Navarro}, {Bournaud}, {Duc}, {Emsellem}, {Krajnovi{\'c}}, {Lyubenova}, \& {McDermid}}]{Weaver2018}
{Weaver}, J., {Husemann}, B., {Kuntschner}, H., {et~al.} 2018, \aap, 614, A32, \dodoi{10.1051/0004-6361/201732448}

\bibitem[{{Weaver} {et~al.}(2024){Weaver}, {Cutler}, {Pan}, {Whitaker}, {Labb{\'e}}, {Price}, {Bezanson}, {Brammer}, {Marchesini}, {Leja}, {Wang}, {Furtak}, {Zitrin}, {Atek}, {Chemerynska}, {Coe}, {Dayal}, {van Dokkum}, {Feldmann}, {F{\"o}rster Schreiber}, {Franx}, {Fujimoto}, {Fudamoto}, {Glazebrook}, {de Graaff}, {Greene}, {Juneau}, {Kassin}, {Kriek}, {Khullar}, {Maseda}, {Mowla}, {Muzzin}, {Nanayakkara}, {Nelson}, {Oesch}, {Pacifici}, {Papovich}, {Setton}, {Shapley}, {Shipley}, {Smit}, {Stefanon}, {Taylor}, {Weibel}, \& {Williams}}]{Weaver2024}
{Weaver}, J.~R., {Cutler}, S.~E., {Pan}, R., {et~al.} 2024, \apjs, 270, 7, \dodoi{10.3847/1538-4365/ad07e0}

\bibitem[{{Wellons} {et~al.}(2015){Wellons}, {Torrey}, {Ma}, {Rodriguez-Gomez}, {Vogelsberger}, {Kriek}, {van Dokkum}, {Nelson}, {Genel}, {Pillepich}, {Springel}, {Sijacki}, {Snyder}, {Nelson}, {Sales}, \& {Hernquist}}]{Wellons2015}
{Wellons}, S., {Torrey}, P., {Ma}, C.-P., {et~al.} 2015, \mnras, 449, 361, \dodoi{10.1093/mnras/stv303}

\bibitem[{{Whitaker} \& {van Dokkum}(2008)}]{Whitaker2008}
{Whitaker}, K.~E., \& {van Dokkum}, P.~G. 2008, \apjl, 676, L105, \dodoi{10.1086/587516}

\bibitem[{{Whitaker} {et~al.}(2011){Whitaker}, {Labb{\'e}}, {van Dokkum}, {Brammer}, {Kriek}, {Marchesini}, {Quadri}, {Franx}, {Muzzin}, {Williams}, {Bezanson}, {Illingworth}, {Lee}, {Lundgren}, {Nelson}, {Rudnick}, {Tal}, \& {Wake}}]{Whitaker2011}
{Whitaker}, K.~E., {Labb{\'e}}, I., {van Dokkum}, P.~G., {et~al.} 2011, \apj, 735, 86, \dodoi{10.1088/0004-637X/735/2/86}

\bibitem[{{Whitaker} {et~al.}(2017){Whitaker}, {Bezanson}, {van Dokkum}, {Franx}, {van der Wel}, {Brammer}, {F{\"o}rster-Schreiber}, {Giavalisco}, {Labb{\'e}}, {Momcheva}, {Nelson}, \& {Skelton}}]{Whitaker2017}
{Whitaker}, K.~E., {Bezanson}, R., {van Dokkum}, P.~G., {et~al.} 2017, \apj, 838, 19, \dodoi{10.3847/1538-4357/aa6258}

\bibitem[{{Whitaker} {et~al.}(2021{\natexlab{a}}){Whitaker}, {Narayanan}, {Williams}, {Li}, {Spilker}, {Dav{\'e}}, {Akhshik}, {Akins}, {Bezanson}, {Katz}, {Leja}, {Magdis}, {Mowla}, {Nelson}, {Pope}, {Privon}, {Toft}, \& {Valentino}}]{Whitaker2021b}
{Whitaker}, K.~E., {Narayanan}, D., {Williams}, C.~C., {et~al.} 2021{\natexlab{a}}, \apjl, 922, L30, \dodoi{10.3847/2041-8213/ac399f}

\bibitem[{{Whitaker} {et~al.}(2021{\natexlab{b}}){Whitaker}, {Williams}, {Mowla}, {Spilker}, {Toft}, {Narayanan}, {Pope}, {Magdis}, {van Dokkum}, {Akhshik}, {Bezanson}, {Brammer}, {Leja}, {Man}, {Nelson}, {Richard}, {Pacifici}, {Sharon}, \& {Valentino}}]{Whitaker2021a}
{Whitaker}, K.~E., {Williams}, C.~C., {Mowla}, L., {et~al.} 2021{\natexlab{b}}, \nat, 597, 485, \dodoi{10.1038/s41586-021-03806-7}

\bibitem[{{Wild} {et~al.}(2016){Wild}, {Almaini}, {Dunlop}, {Simpson}, {Rowlands}, {Bowler}, {Maltby}, \& {McLure}}]{Wild2016}
{Wild}, V., {Almaini}, O., {Dunlop}, J., {et~al.} 2016, \mnras, 463, 832, \dodoi{10.1093/mnras/stw1996}

\bibitem[{{Wild} {et~al.}(2020){Wild}, {Taj Aldeen}, {Carnall}, {Maltby}, {Almaini}, {Werle}, {Wilkinson}, {Rowlands}, {Bolzonella}, {Castellano}, {Gargiulo}, {McLure}, {Pentericci}, \& {Pozzetti}}]{Wild2020}
{Wild}, V., {Taj Aldeen}, L., {Carnall}, A., {et~al.} 2020, \mnras, 494, 529, \dodoi{10.1093/mnras/staa674}

\bibitem[{{Williams} {et~al.}(2021){Williams}, {Spilker}, {Whitaker}, {Dav{\'e}}, {Woodrum}, {Brammer}, {Bezanson}, {Narayanan}, \& {Weiner}}]{Williams2021}
{Williams}, C.~C., {Spilker}, J.~S., {Whitaker}, K.~E., {et~al.} 2021, \apj, 908, 54, \dodoi{10.3847/1538-4357/abcbf6}

\bibitem[{{Williams} {et~al.}(2023){Williams}, {Alberts}, {Ji}, {Hainline}, {Lyu}, {Rieke}, {Endsley}, {Suess}, {Johnson}, {Florian}, {Shivaei}, {Rujopakarn}, {Baker}, {Bhatawdekar}, {Boyett}, {Bunker}, {Carniani}, {Charlot}, {Curtis-Lake}, {DeCoursey}, {de Graaff}, {Egami}, {Eisenstein}, {Gibson}, {Hausen}, {Helton}, {Maiolino}, {Maseda}, {Nelson}, {Perez-Gonzalez}, {Rieke}, {Robertson}, {Sun}, {Tacchella}, {Willmer}, \& {Willott}}]{Williams2023}
{Williams}, C.~C., {Alberts}, S., {Ji}, Z., {et~al.} 2023, arXiv e-prints, arXiv:2311.07483.
\newblock \doarXiv{2311.07483}

\bibitem[{{Williams} {et~al.}(2009){Williams}, {Quadri}, {Franx}, {van Dokkum}, \& {Labb{\'e}}}]{Williams2009}
{Williams}, R.~J., {Quadri}, R.~F., {Franx}, M., {van Dokkum}, P., \& {Labb{\'e}}, I. 2009, \apj, 691, 1879, \dodoi{10.1088/0004-637X/691/2/1879}

\bibitem[{{Woodrum} {et~al.}(2022){Woodrum}, {Williams}, {Rieke}, {Leja}, {Johnson}, {Bezanson}, {Kennicutt}, {Spilker}, \& {Tacchella}}]{Woodrum2022}
{Woodrum}, C., {Williams}, C.~C., {Rieke}, M., {et~al.} 2022, \apj, 940, 39, \dodoi{10.3847/1538-4357/ac9af7}

\bibitem[{{Wu} {et~al.}(2023){Wu}, {Bezanson}, {D'Eugenio}, {Gallazzi}, {Greene}, {Maseda}, {Suess}, \& {van der Wel}}]{Wu2023}
{Wu}, P.-F., {Bezanson}, R., {D'Eugenio}, F., {et~al.} 2023, \apj, 955, 75, \dodoi{10.3847/1538-4357/acf0bd}

\bibitem[{{Wuyts} {et~al.}(2010){Wuyts}, {Cox}, {Hayward}, {Franx}, {Hernquist}, {Hopkins}, {Jonsson}, \& {van Dokkum}}]{Wuyts2010}
{Wuyts}, S., {Cox}, T.~J., {Hayward}, C.~C., {et~al.} 2010, \apj, 722, 1666, \dodoi{10.1088/0004-637X/722/2/1666}

\bibitem[{{Xie} {et~al.}(2024){Xie}, {De Lucia}, {Fontanot}, {Hirschmann}, {Bah{\'e}}, {Balogh}, {Muzzin}, {Vulcani}, {Baxter}, {Forrest}, {Wilson}, {Rudnick}, {Cooper}, \& {Rescigno}}]{Xie2024}
{Xie}, L., {De Lucia}, G., {Fontanot}, F., {et~al.} 2024, arXiv e-prints, arXiv:2402.01314.
\newblock \doarXiv{2402.01314}

\bibitem[{{Zaidi} {et~al.}(2024){Zaidi}, {Marchesini}, {Papovich}, {Antwi-Danso}, {Nonino}, {Annunziatella}, {Brammer}, {Glazebrook}, {Iyer}, {Labb{\'e}}, {Marsan}, {Muzzin}, \& {Wake}}]{Zaidi2024}
{Zaidi}, K., {Marchesini}, D., {Papovich}, C., {et~al.} 2024, arXiv e-prints, arXiv:2401.03107, \dodoi{10.48550/arXiv.2401.03107}

\bibitem[{{Zolotov} {et~al.}(2015){Zolotov}, {Dekel}, {Mandelker}, {Tweed}, {Inoue}, {DeGraf}, {Ceverino}, {Primack}, {Barro}, \& {Faber}}]{Zolotov2015}
{Zolotov}, A., {Dekel}, A., {Mandelker}, N., {et~al.} 2015, \mnras, 450, 2327, \dodoi{10.1093/mnras/stv740}

\end{thebibliography}

\end{document}